\documentclass[twocolumn]{aastex631}

\usepackage{xspace}
\usepackage{graphicx}
\usepackage{subfigure}

\newcommand\TESS{\textit{TESS}\xspace}
\newcommand\exofast{$\tt{EXOFASTv2}$\xspace}
\newcommand\mj{M$_{\rm J}$\xspace}

\newcommand\rj{R$_{\rm J}$\xspace}

\newcommand{\bjdtdb}{\ensuremath{\rm {BJD_{TDB}}}}
\newcommand{\msol}{M$_\odot$\xspace}
\newcommand{\teff}{T$_{\rm eff}$\xspace}

\newenvironment{rotatepage}%
    {\clearpage\pagebreak[4]\global\pdfpageattr\expandafter{\the\pdfpageattr/Rotate 90}}%
    {\clearpage\pagebreak[4]\global\pdfpageattr\expandafter{\the\pdfpageattr/Rotate 0}}%

\shorttitle{MEEP I: Nine New Hot Jupiters}
\shortauthors{Schulte et al.}

\graphicspath{{./}{figures/}}

\begin{document}

\title{Migration and Evolution of giant ExoPlanets (MEEP) I: Nine Newly Confirmed Hot Jupiters from the \TESS Mission}

\newcommand{\cfa}{Center for Astrophysics \textbar \ Harvard \& Smithsonian, 60 Garden St, Cambridge, MA 02138, USA}
\newcommand{\msu}{Center for Data Intensive and Time Domain Astronomy, Department of Physics and Astronomy, Michigan State University, East Lansing, MI 48824, USA}
\newcommand{\umich}{Astronomy Department, University of Michigan, 1085 S University Avenue, Ann Arbor, MI 48109, USA}
\newcommand{\utaustin}{Department of Astronomy, The University of Texas at Austin, Austin, TX 78712, USA}
\newcommand{\MIT}{Department of Physics and Kavli Institute for Astrophysics and Space Research, Massachusetts Institute of Technology, Cambridge, MA 02139, USA}
\newcommand{\MITEPS}{Department of Earth, Atmospheric and Planetary Science, Massachusetts Institute of Technology, 77 Massachusetts Avenue, Cambridge, MA 02139, USA}
\newcommand{\uflorida}{Department of Astronomy, University of Florida, 211 Bryant Space Science Center, Gainesville, FL, 32611, USA}
\newcommand{\riverside}{Department of Earth and Planetary Sciences, University of California, Riverside, CA 92521, USA}
\newcommand{\usq}{Centre for Astrophysics, University of Southern Queensland, West Street, Toowoomba, QLD 4350, Australia}
\newcommand{\ames}{NASA Ames Research Center, Moffett Field, CA 94035, USA}
\newcommand{\geneva}{Geneva Observatory, University of Geneva, Chemin des Mailettes 51, 1290 Versoix, Switzerland}
\newcommand{\uw}{Astronomy Department, University of Washington, Seattle, WA 98195, USA}
\newcommand{\warwick}{Department of Physics, University of Warwick, Gibbet Hill Road, Coventry CV4 7AL, UK}
\newcommand{\warwickceh}{Centre for Exoplanets and Habitability, University of Warwick, Gibbet Hill Road, Coventry CV4 7AL, UK}
\newcommand{\princeton}{Department of Astrophysical Sciences, Princeton University, 4 Ivy Lane, Princeton, NJ 08544, USA}
\newcommand{\liege}{Astrobiology Research Unit, Universit\'e de Li\`ege, 19C All\'ee du 6 Ao\^ut, 4000 Li\`ege, Belgium}
\newcommand{\liegestar}{Space Sciences, Technologies and Astrophysics Research (STAR) Institute, Universit\'e de Li\`ege, All\'ee du 6 Ao\^ut 19C, B-4000 Li\`ege, Belgium}
\newcommand{\vanderbilt}{Department of Physics and Astronomy, Vanderbilt University, Nashville, TN 37235, USA}
\newcommand{\fisk}{Department of Physics, Fisk University, 1000 17th Avenue North, Nashville, TN 37208, USA}
\newcommand{\columbia}{Department of Astronomy, Columbia University, 550 West 120th Street, New York, NY 10027, USA}
\newcommand{\toronto}{Dunlap Institute for Astronomy and Astrophysics, University of Toronto, Ontario M5S 3H4, Canada}
\newcommand{\unc}{Department of Physics and Astronomy, The University of North Carolina at Chapel Hill, Chapel Hill, NC 27599-3255, USA}
\newcommand{\iac}{Instituto de Astrof\'isica de Canarias (IAC), 38200 La Laguna, Tenerife, Spain}
\newcommand{\lalaguna}{Deptartamento de Astrof\'isica, Universidad de La Laguna (ULL), 38206 La Laguna, Tenerife, Spain}
\newcommand{\louisville}{Department of Physics and Astronomy, University of Louisville, Louisville, KY 40292, USA}
\newcommand{\aavso}{American Association of Variable Star Observers, 185 Alewife Brook Parkway, Suite 410, Cambridge, MA 02138, USA}
\newcommand{\tokyo}{Komaba Institute for Science, The University of Tokyo, 3-8-1 Komaba, Meguro, Tokyo 153-8902, Japan}
\newcommand{\naoj}{National Astronomical Observatory of Japan, 2-21-1 Osawa, Mitaka, Tokyo 181-8588, Japan}
\newcommand{\jstpresto}{JST, PRESTO, 7-3-1 Hongo, Bunkyo-ku, Tokyo 113-0033, Japan}
\newcommand{\astrobiojapan}{Astrobiology Center, 2-21-1 Osawa, Mitaka, Tokyo 181-8588, Japan}
\newcommand{\tokyomultidisc}{Department of Multi-Disciplinary Sciences, Graduate School of Arts and Sciences, The University of Tokyo, 3-8-1 Komaba, Meguro, Tokyo 153-8902, Japan}
\newcommand{\ctio}{Cerro Tololo Inter-American Observatory/NSF’s NOIRLab, Casilla 603, La Serena, Chile}
\newcommand{\noirlab}{NSF’s National Optical-Infrared Astronomy Research Laboratory, 950 N. Cherry Ave., Tucson, AZ 85719, USA}
\newcommand{\nexsci}{NASA Exoplanet Science Institute-Caltech/IPAC, Pasadena, CA 91125 USA}
\newcommand{\ucsc}{Department of Astronomy and Astrophysics, University of
California, Santa Cruz, CA 95064, USA}
\newcommand{\gsfc}{Exoplanets and Stellar Astrophysics Laboratory, Code 667, NASA Goddard Space Flight Center, Greenbelt, MD 20771, USA}
\newcommand{\sgtinc}{SGT, Inc./NASA AMES Research Center, Mailstop 269-3, Bldg T35C, P.O. Box 1, Moffett Field, CA 94035, USA}
\newcommand{\chile}{Center of Astro-Engineering UC, Pontificia Universidad Cat\'olica de Chile, Av. Vicu\~{n}a Mackenna 4860, 7820436 Macul, Santiago, Chile}
\newcommand{\Pontificia}{Facultad de Ingeniería y Ciencias, Universidad Adolfo Ib\'a\~nez, Av. Diagonal las Torres 2640, Pe\~nalol\'en, Santiago, Chile}
\newcommand{\Millennium}{Millennium Institute for Astrophysics, Chile}
\newcommand{\maxplank}{Max-Planck-Institut f\"ur Astronomie, K\"onigstuhl 17, Heidelberg 69117, Germany}
\newcommand{\utdallas}{Department of Physics, The University of Texas at Dallas, 800 West
Campbell Road, Richardson, TX 75080-3021, USA}
\newcommand{\MauryLewin}{Maury Lewin Astronomical Observatory, Glendora, CA 91741, USA}
\newcommand{\umbc}{University of Maryland, Baltimore County, 1000 Hilltop Circle, Baltimore, MD 21250, USA}
\newcommand{\osu}{Department of Astronomy, The Ohio State University, 140 West 18th Avenue, Columbus, OH 43210, USA}
\newcommand{\MITAA}{Department of Aeronautics and Astronautics, MIT, 77 Massachusetts Avenue, Cambridge, MA 02139, USA}
\newcommand{\openu}{School of Physical Sciences, The Open University, Milton Keynes MK7 6AA, UK}
\newcommand{\swarthmore}{Department of Physics and Astronomy, Swarthmore College, Swarthmore, PA 19081, USA}
\newcommand{\seti}{SETI Institute, Mountain View, CA 94043, USA}
\newcommand{\lehigh}{Department of Physics, Lehigh University, 16 Memorial Drive East, Bethlehem, PA 18015, USA}
\newcommand{\utah}{Department of Physics and Astronomy, University of Utah, 115 South 1400 East, Salt Lake City, UT 84112, USA}
\newcommand{\USNA}{Department of Physics, United States Naval Academy, 572C Holloway Rd., Annapolis, MD 21402, USA}
\newcommand{\DTM}{Department of Terrestrial Magnetism, Carnegie Institution for Science, 5241 Broad Branch Road, NW, Washington, DC 20015, USA}
\newcommand{\UPenn}{The University of Pennsylvania, Department of Physics and Astronomy, Philadelphia, PA, 19104, USA}
\newcommand{\montana}{Department of Physics and Astronomy, University of Montana, 32 Campus Drive, No. 1080, Missoula, MT 59812, USA}
\newcommand{\psu}{Department of Astronomy \& Astrophysics, The Pennsylvania State University, 525 Davey Lab, University Park, PA 16802, USA}
\newcommand{\psust}{Center for Exoplanets and Habitable Worlds, The Pennsylvania State University, 525 Davey Lab, University Park, PA 16802, USA}
\newcommand{\Kutztown}{Department of Physical Sciences, Kutztown University, Kutztown, PA 19530, USA}
\newcommand{\udel}{Department of Physics \& Astronomy, University of Delaware, Newark, DE 19716, USA}
\newcommand{\Westminster}{Department of Physics, Westminster College, New Wilmington, PA 16172, USA}
\newcommand{\steward}{Department of Astronomy and Steward Observatory, University of Arizona, Tucson, AZ 85721, USA}
\newcommand{\saao}{South African Astronomical Observatory, PO Box 9, Observatory, 7935, Cape Town, South Africa}
\newcommand{\salt}{Southern African Large Telescope, PO Box 9, Observatory, 7935, Cape Town, South Africa}
\newcommand{\ssl}{Societ\`{a} Astronomica Lunae, Italy}
\newcommand{\spot}{Spot Observatory, Nashville, TN 37206, USA}
\newcommand{\txamGP}{George P.\ and Cynthia Woods Mitchell Institute for Fundamental Physics and Astronomy, Texas A\&M University, College Station, TX77843, USA}
\newcommand{\txam}{Department of Physics and Astronomy, Texas A\&M university, College Station, TX 77843, USA}
\newcommand{\txammul}{Munnerlyn Astronomical Instrumentation Laboratory, Department of Physics \& Astronomy, Texas A\&M university, College Station, TX 77843 USA}
\newcommand{\wellesley}{Department of Astronomy, Wellesley College, Wellesley, MA 02481, USA}
\newcommand{\Wesleyan}{Astronomy Department and Van Vleck Observatory, Wesleyan University, Middletown, CT 06459, USA}
\newcommand{\inaf}{INAF -- Osservatorio Astronomico di Padova, Vicolo dell'Osservatorio 5, I-35122, Padova, Italy}
\newcommand{\byu}{Department of Physics and Astronomy, Brigham Young University, Provo, UT 84602, USA}
\newcommand{\Hazelwood}{Hazelwood Observatory, Churchill, Victoria, Australia}
\newcommand{\pest}{Perth Exoplanet Survey Telescope}
\newcommand{\Winer}{Winer Observatory, PO Box 797, Sonoita, AZ 85637, USA}
\newcommand{\icpo}{Ivan Curtis Private Observatory}
\newcommand{\elsauce}{El Sauce Observatory, Coquimbo Province, Chile}
\newcommand{\crow}{Atalaia Group \& CROW Observatory, Portalegre, Portugal}
\newcommand{\dfus}{Dipartimento di Fisica ``E.R.Caianiello'', Universit\`a di Salerno, Via Giovanni Paolo II 132, Fisciano 84084, Italy}
\newcommand{\indfn}{Istituto Nazionale di Fisica Nucleare, Napoli, Italy}
\newcommand{\sotes}{Gabriel Murawski Private Observatory (SOTES)}
\newcommand{\lco}{Las Cumbres Observatory Global Telescope, 6740 Cortona Dr., Suite 102, Goleta, CA 93111, USA}
\newcommand{\ucsb}{Department of Physics, University of California, Santa Barbara, CA 93106-9530, USA}
\newcommand{\yale}{Department of Astronomy, Yale University, 52 Hillhouse Avenue, New Haven, CT 06511, USA}
\newcommand{\eso}{European Southern Observatory, D-85748 Garching, Germany}
\newcommand{\stsci}{Space Telescope Science Institute, Baltimore, MD 21218, USA}
\newcommand{\keele}{Astrophysics Group, Keele University, Staffordshire ST5 5BG, UK}
\newcommand{\gsfcsellers}{GSFC Sellers Exoplanet Environments Collaboration, NASA Goddard Space Flight Center, Greenbelt, MD 20771, USA}
\newcommand{\usno}{U.S. Naval Observatory, Washington, DC 20392, USA}
\newcommand{\kansas}{Department of Physics and Astronomy, University of Kansas, 1251 Wescoe Hall Dr., Lawrence, KS 66045, USA}
\newcommand{\gmu}{George Mason University, 4400 University Drive, Fairfax, VA, 22030 USA}
\newcommand{\unsw}{Exoplanetary Science at UNSW, School of Physics, UNSW Sydney, NSW 2052, Australia}
\newcommand{\sifa}{School of Physics, Sydney Institute for Astronomy (SIfA), The University of Sydney, NSW 2006, Australia}
\newcommand{\nanjing}{School of Astronomy and Space Science, Key Laboratory of Modern Astronomy and Astrophysics in Ministry of Education, Nanjing University, Nanjing 210046, Jiangsu, China}
\newcommand{\berkeley}{Department of Astronomy, University of California Berkeley, Berkeley, CA 94720-3411, USA}
\newcommand{\bhicfa}{Black Hole Initiative at Harvard University, 20 Garden Street, Cambridge, MA 02138, USA}
\newcommand{\Silesian}{Department of Electronics, Electronical Engineering and Microelectronics, Silesian University of Techhnology Akademicka 16, 44-100 Gliwice, Poland}
\newcommand{\Patashnick}{Patashnick Voorheesville Observatory, Voorheesville, NY 12186, USA}
\newcommand{\austincollege}{Physics Department, Austin College, 900 North Grand Avenue, Sherman TX 75090, USA}
\newcommand{\Tsinghua}{Department of Astronomy, Tsinghua University, Beijing 100084, China}
\newcommand{\Tsinghuaschool}{Tsinghua International School, Beijing 100084, China}
\newcommand{\chinaNAO}{National Astronomical Observatories, Chinese Academy of Sciences, 20A Datun Road, Chaoyang District, Beijing 100012, China}
\newcommand{\Tautenburg}{Th{\"u}ringer Landessternwarte Tautenburg, Sternwarte 5, 07778 Tautenburg, Germany}
\newcommand{\brierfield}{Brierfield Observatory, New South Wales, Australia}
\newcommand{\Indiana}{Indiana University Department of Astronomy, SW319, 727 E 3rd Street, Bloomington, IN 47405, USA}
\newcommand{\wisconsin}{Department of Astronomy, University of Wisconsin-Madison, 475 N Charter St., Madison, WI 53706, USA}
\newcommand{\protologic}{Proto-Logic Consulting LLC, Washington, DC 20009, USA}
\newcommand{\ASTRAVEO}{ASTRAVEO LLC, PO Box 1668, MA 01931, USA}
\newcommand{\TJHS}{Thomas Jefferson High School, 6560 Braddock Rd, Alexandria, VA 22312 USA}
\newcommand{\ucatchile}{Instituto de Astrof\'isica, Facultad de F\'isica, Pontificia Universidad Cat\'olica de Chile}
\newcommand{\lasa}{Liberal Arts and Science Academy, Austin, Texas 78724, USA}
\newcommand{\gemini}{Gemini Observatory/NSF’s NOIRLab, 670 N. A’ohoku Place, Hilo, HI, 96720, USA}
\newcommand{\umd}{Department of Astronomy, University of Maryland, College Park, College Park, MD 20742, USA}
\newcommand{\ucscchile}{Departamento de Matem\'atica y F\i'sica Aplicadas, Universidad Cat\'olica de la Sant\'isima Concepci\'on, Alonso de Rivera 2850, Concepci\'on, Chile}
\newcommand{\austinstate}{Department of Physics, Engineering and Astronomy, Stephen F. Austin State University, 1936 North St, Nacogdoches, TX 75962, USA}
\newcommand{\carleton}{Department of Physics and Astronomy, Carleton College, 215 Goodsell Circle, Northfield, MN 55057, USA}
\newcommand{\newmexico}{Department of Physics and Astronomy, University of New Mexico, 210 Yale Blvd NE, Albuquerque, NM 87106, USA}
\newcommand{\caltech}{Department of Astronomy, California Institute of Technology, Pasadena, CA 91125, USA}
\newcommand{\chicago}{Department of Astronomy and Astrophysics, 5640 South Ellis Avenue, University of Chicago, IL 60637, USA}
\newcommand{\baeri}{Bay Area Environmental Research Institute, Moffett Field, CA 94035, USA}
\newcommand{\iaacsic}{Instituto de Astrof\'isica de Andaluc\'ia (IAA-CSIC), Glorieta de la Astronom\'ia s/n, 18008 Granada, Spain}
\newcommand{\lomonosov}{Sternberg Astronomical Institute, Lomonosov Moscow State University,  Universitetsky pr. 13, Moscow 119234, Russia}
\newcommand{\slovakia}{Astronomical Institute, Slovak Academy of Sciences, 059 60 Tatransk\'a Lomnica, Slovakia}
\newcommand{\jpl}{Jet Propulsion Laboratory, California Institute of Technology, Pasadena, CA 91109, USA}
\newcommand{\supa}{SUPA Physics and Astronomy, University of St. Andrews, Fife, KY16 9SS Scotland, UK}
\newcommand{\calou}{Observatori de Ca l'Ou, Carrer de dalt 18, Sant Martí Sesgueioles 08282, Barcelona, Spain}
\newcommand{\okayama}{Okayama Observatory, Kyoto University, 3037-5 Honjo, Kamogatacho,
Asakuchi, Okayama 719-0232, Japan}
\newcommand{\ondrejov}{Astronomical Institute of the Czech Academy of Sciences, Fri\v{c}ova 298, 25165 Ond\v{r}ejov, Czech Republic}


\newcommand{\eberly}{\altaffiliation{Eberly Research Fellow}}
\newcommand{\torres}{\altaffiliation{Juan Carlos Torres Fellow}}
\newcommand{\sagan}{\altaffiliation{NASA Sagan Fellow}}
\newcommand{\bernoulli}{\altaffiliation{Bernoulli fellow}}
\newcommand{\gruber}{\altaffiliation{Gruber fellow}}
\newcommand{\kavli}{\altaffiliation{Kavli Fellow}}
\newcommand{\peg}{\altaffiliation{51 Pegasi b Fellow}}
\newcommand{\pappalardo}{\altaffiliation{Pappalardo Fellow}}
\newcommand{\hubble}{\altaffiliation{NASA Hubble Fellow}}
\newcommand{\nsf}{\altaffiliation{National Science Foundation Graduate Research Fellow}}

\correspondingauthor{Jack Schulte}
\email{jschulte@msu.edu}

\author[0000-0002-7382-0160]{Jack Schulte}
\affiliation{\msu}

\author[0000-0001-8812-0565]{Joseph E. Rodriguez} 
\affiliation{\msu}

\author[0000-0001-6637-5401]{Allyson Bieryla} 
\affiliation{\cfa}

\author[0000-0002-8964-8377]{Samuel~N.~Quinn}
\affiliation{\cfa}

\author[0000-0001-6588-9574]{Karen A.\ Collins} 
\affiliation{\cfa}

\author[0000-0001-7961-3907]{Samuel W. Yee} 
\altaffiliation{51 Pegasi b Fellow}
\affiliation{\princeton}
\affiliation{\cfa}

\author[0000-0002-6478-0611]{Andrew C. Nine} 
\affiliation{\wisconsin}
\affiliation{\carleton}

\author[0000-0001-7493-7419]{Melinda Soares-Furtado} 
\affiliation{\wisconsin}
\affiliation{\MIT}

\author[0000-0001-9911-7388]{David~W.~Latham} 
\affiliation{\cfa}

\author[0000-0003-3773-5142]{Jason~D.~Eastman} 
\affiliation{\cfa}


\author[0000-0003-1464-9276]{Khalid Barkaoui} 
\affiliation{\liege}
\affiliation{\MITEPS}
\affiliation{\iac}

\author[0000-0002-5741-3047]{David R. Ciardi}
\affiliation{\nexsci}


\author[0000-0003-2313-467X]{Diana~Dragomir} 
\affiliation{\newmexico}

\author[0000-0002-0885-7215]{Mark~E.~Everett}
\affiliation{\noirlab}



\author[0000-0002-8965-3969]{Steven Giacalone} 
\altaffiliation{NSF Astronomy and Astrophysics Postdoctoral Fellow}
\affiliation{\caltech}

\author[0000-0002-4510-2268]{Ismael~Mireles} 
\affiliation{\newmexico}

\author{Felipe Murgas}
\affiliation{\iac}
\affiliation{\lalaguna}

\author[0000-0001-8511-2981]{Norio Narita} 
\affiliation{\tokyo}
\affiliation{\astrobiojapan}
\affiliation{\iac}

\author[0000-0002-1836-3120]{Avi~Shporer} 
\affiliation{\MIT}

\author[0000-0003-0647-6133]{Ivan A. Strakhov} 
\affiliation{\lomonosov}

\author[0009-0008-5145-0446]{Stephanie~Striegel} 
\affiliation{\seti}
\affiliation{\ames}

\author[0000-0002-2798-6944]{Martin Va\v{n}ko} 
\affil{\slovakia}

\author[0000-0002-0701-4005]{Noah Vowell} 
\affiliation{\msu}

\author[0000-0003-3092-4418]{Gavin Wang} 
\affiliation{\Tsinghuaschool}

\author[0000-0002-0619-7639]{Carl Ziegler}
\affiliation{\austinstate}



\author{Michael Bellaver} 
\affiliation{\msu}

\author[0000-0001-6981-8722]{Paul Benni} 
\affiliation{Acton Sky Portal (private observatory), Acton, MA USA}

\author[0000-0003-0846-1744]{Serge Bergeron} 
\affiliation{\aavso}

\author[0000-0002-9486-4840]{Henri M. J. Boffin} 
\affiliation{\eso}

\author[0000-0001-7124-4094]{C\'{e}sar Brice\~{n}o} 
\affiliation{\ctio}


\author[0000-0002-2361-5812]{Catherine A. Clark} 
\affiliation{\nexsci}
\affiliation{\jpl}

\author[0000-0003-2781-3207]{Kevin I.\ Collins} 
\affiliation{\gmu}

\author[0000-0002-6424-3410]{Jerome P. de Leon} 
\affiliation{\tokyomultidisc}

\author[0000-0001-8189-0233]{Courtney D. Dressing} 
\affiliation{\berkeley}

\author[0000-0002-5674-2404]{Phil Evans} 
\affiliation{\elsauce}

\author[0000-0002-2341-3233]{Emma Esparza-Borges} 
\affiliation{\iac}
\affiliation{\lalaguna}

\author{Jeremy Fedewa} 
\affiliation{\msu}

\author[0000-0002-4909-5763]{Akihiko Fukui} 
\affiliation{\tokyo}
\affiliation{\iac}

\author[0000-0002-4503-9705]{Tianjun Gan} 
\affiliation{\Tsinghua}

\author[0000-0001-7113-8152]{Ivan S. Gerasimov} 
\affiliation{\lomonosov}


\author[0000-0001-8732-6166]{Joel D.\ Hartman} 
\affiliation{\princeton}

\author[0000-0001-6171-7951]{Holden Gill} 
\affil{\berkeley}

\author{Micha\"el Gillon} 
\affiliation{\liege}

\author[0000-0003-1728-0304]{Keith Horne} 
\affiliation{\supa}

\author[0000-0001-9927-7269]{Ferran Grau Horta} 
\affiliation{\calou}

\author[0000-0002-2532-2853]{Steve~B.~Howell} 
\affil{\ames}

\author[0000-0002-6480-3799]{Keisuke Isogai} 
\affiliation{\okayama}
\affiliation{\tokyomultidisc}

\author[0000-0001-8923-488X]{Emmanu\"el Jehin} 
\affiliation{\liegestar}

\author[0000-0002-4715-9460]{Jon~M.~Jenkins} 
\affiliation{\ames}

\author[0000-0002-2656-909X]{Raine Karjalainen} 
\affil{\ondrejov}


\author[0000-0003-0497-2651]{John F.\ Kielkopf} 
\affiliation{\louisville}


\author[0000-0002-9903-9911]{Kathryn V. Lester} 
\affil{\ames}

\author[0000-0001-7746-5795]{Colin Littlefield} 
\affiliation{\baeri}
\affiliation{\ames}

\author[0000-0003-2527-1598]{Michael B. Lund} 
\affiliation{\nexsci}

\author[0000-0003-3654-1602]{Andrew W. Mann} 
\affiliation{\unc}

\author[0000-0002-1463-9847]{Mason McCormack} 
\affiliation{\chicago}

\author{Edward J. Michaels} 
\affiliation{Waffelow Creek Observatory}

\author{Shane Painter} 
\affiliation{\msu}

\author[0000-0003-0987-1593]{Enric Palle} 
\affiliation{\iac}
\affiliation{\lalaguna}

\author[0000-0001-5519-1391]{Hannu Parviainen} 
\affiliation{\lalaguna}
\affiliation{\iac}

\author[0009-0009-1439-3101]{David-Michael Peterson} 
\affiliation{\msu}

\author[0000-0003-1572-7707]{Francisco J. Pozuelos} 
\affiliation{\iaacsic}
\affiliation{\liege}


\author[0000-0003-4278-6374]{Zachary Raup} 
\affiliation{\Kutztown}

\author[0000-0002-5005-1215]{Phillip Reed} 
\affiliation{\Kutztown}

\author[0009-0009-5132-9520]{Howard M. Relles} 
\affiliation{\cfa}

\author[0000-0003-2058-6662]{George~R.~Ricker} 
\affiliation{\MIT}

\author[0000-0002-2454-768X]{Arjun B. Savel} 
\affiliation{\umd}

\author[0000-0001-8227-1020]{Richard P. Schwarz} 
\affiliation{\cfa}

\author[0000-0002-6892-6948]{S.~Seager} 
\affiliation{\MIT}
\affiliation{\MITEPS}
\affiliation{\MITAA}

\author[0000-0003-3904-6754]{Ramotholo Sefako} 
\affiliation{\saao}

\author{Gregor Srdoc} 
\affiliation{Kotizarovci Observatory, Sarsoni 90, 51216 Viskovo, Croatia}

\author[0000-0003-2163-1437]{Chris Stockdale} 
\affiliation{Hazelwood Observatory, Australia}

\author{Hannah Sullivan} 
\affiliation{\msu}

\author[0009-0008-2214-5039]{Mathilde Timmermans} 
\affiliation{\liege}


\author[0000-0002-4265-047X]{Joshua N.\ Winn} 
\affiliation{\princeton}

\begin{abstract}

Hot Jupiters were many of the first exoplanets discovered in the 1990s, but in the decades since their discovery, the mysteries surrounding their origins remain. Here, we present nine new hot Jupiters (TOI-1855 b, TOI-2107 b, TOI-2368 b, TOI-3321 b, TOI-3894 b, TOI-3919 b, TOI-4153 b, TOI-5232 b, and TOI-5301 b) discovered by NASA's \TESS mission and confirmed using ground-based imaging and spectroscopy. These discoveries are the first in a series of papers named the Migration and Evolution of giant ExoPlanets (MEEP) survey and are part of an ongoing effort to build a complete sample of hot Jupiters orbiting FGK stars, with a limiting Gaia $G$-band magnitude of 12.5. This effort aims to use homogeneous detection and analysis techniques to generate a set of precisely measured stellar and planetary properties that is ripe for statistical analysis. The nine planets presented in this work occupy a range of masses (0.55 Jupiter masses (M$_{\rm{J}}$) $<$ M$_{\rm{P}}$ $<$ 3.88 M$_{\rm{J}}$) and sizes (0.967 Jupiter radii (R$_{\rm{J}}$) $<$ R$_{\rm{P}}$ $<$ 1.438 R$_{\rm{J}}$) and orbit stars that range in temperature from 5360 K $<$ \teff $<$ 6860 K with Gaia $G$-band magnitudes ranging from 11.1 to 12.7. Two of the planets in our sample have detectable orbital eccentricity: TOI-3919 b ($e = 0.259^{+0.033}_{-0.036}$) and TOI-5301 b ($e = 0.33^{+0.11}_{-0.10}$). These eccentric planets join a growing sample of eccentric hot Jupiters that are consistent with high-eccentricity tidal migration, one of the three most prominent theories explaining hot Jupiter formation and evolution.

\end{abstract}

\keywords{Exoplanet astronomy (486) --- Exoplanet migration (2205) --- Exoplanet detection methods (489) --- Exoplanets (498) --- Transits (1711) --- Radial velocity (1332) --- Direct imaging (387)}

\section{Introduction} \label{sec:intro}

The first exoplanet discovered around a main sequence star, 51 Pegasi b \citep{Mayor:1995}, immediately had an impact on the fundamental theories of planet formation and evolution. 51 Pegasi b, like many of the planets that followed, is a hot Jupiter (HJ), or a gas giant planet ($>$ 0.25 \mj) orbiting its host star with a period less than 10 days \citep{Dawson:2018}. Now, close to thirty years later, over 5,500 additional planets have been confirmed, approximately 10\% of which are HJs. The rapid discovery of HJs has prompted questions about how these giant planets, with orbits far more compact than any in our Solar System, came to be. Several theories have been proposed to explain the preponderance of short-period giant planets, which we group into three categories: in situ formation \citep{Batygin:2016}, gas-disk migration \citep{Goldreich:1980, Lin:1986}, and high-eccentricity tidal migration \citep[e.g.,][]{Kozai:1962, Lidov:1962, Rasio:1996, Naoz:2016}.

The plausibility of in situ formation, the idea that giant planets can form in their present-day orbital configurations and not require large-scale migration, has been debated throughout the last two decades \citep{Rafikov:2005, Rafikov:2006, Lee:2014, Boley:2016, Poon:2021}. The majority of the currently accepted literature suggests that in situ formation is unlikely to be the dominant mechanism to form HJs, as it would require a rapid build-up of $\sim 10$ Earth masses ($\rm{ M}_{\rm{E}}$) of solids in a region of the protoplanetary disk where the feeding zones are tiny \citep{Dawson:2018}. Additionally, the cores that form would not likely be able to accrete enough gas to form a HJ before the dissipation of the gas disk \citep{Lee:2014, Dawson:2018}. \cite{Batygin:2016} argued that rapid gas accretion onto planets with masses between 10 and 20 M$_{\rm E}$ is possible, but that HJs that form in situ should be accompanied by low-mass planets on short periods, a scenario which has proven to be rare \citep{Huang:2016}.

If in situ formation is not the dominant mechanism to produce HJs, then there must be some mechanisms for giant planets to undergo large-scale migration from orbits beyond $\sim 3$ au to less than $\sim 0.1$ au. The currently favored pathways for this large-scale migration are gas-disk migration, where the planet transfers its orbital energy and angular momentum to the protoplanetary disk \citep{Goldreich:1980, Lin:1986}, and high-eccentricity tidal migration, where the planet exchanges angular momentum with another planet or star, exciting the planet's eccentricity to be later circularized on a much smaller orbit as a consequence of tidal interactions with the host star \citep{Rasio:1996, Fabrycky:2007}. Each of these mechanisms is expected to generate different observable outcomes in planetary systems that have undergone migration. Gas-disk migration does not typically excite the orbital eccentricities or misalign the orbits of the migrating planets, whereas planets that have undergone high-eccentricity tidal migration may still have observable eccentric or misaligned orbits that are remnants of their migration \citep{Dawson:2018}. Additionally, high-eccentricity tidal migration typically destabilizes the orbits of nearby planets, leading to the preponderance of isolated HJs \citep{Huang:2016}, while gas-disk migration is more quiescent, enabling nearby companions to survive \citep{Becker:2015, Vanderburg:2017}.

This article is the first in connection with a survey intended to construct a magnitude-limited, complete sample of hot and warm Jupiters with precisely measured masses, radii, semi-major axes, and eccentricities. This set of parameters will be useful for future efforts to understand the HJ population and constrain the migration pathways of giant planets. Our survey, which we name the Migration and Evolution of giant ExoPlanets (MEEP) survey, is joining other efforts \citep{Rodriguez:2019, Rodriguez:2021, Ikwut-Ukwa:2022, Yee:2022, Yee:2023, Rodriguez:2023} to build such a catalog of parameters using data from the Transiting Exoplanet Survey Satellite \citep[\TESS;][]{Ricker:2015} and homogeneous analysis techniques, to ensure that statistical studies can be performed on the sample with a good understanding of the selection biases. This sample will be constrained by the host star brightness to ensure that radial velocity (RV) observations yield a precise mass and eccentricity, and so only planets with host star Gaia $G$-band magnitudes brighter than 12.5 will be included. \cite{Yee:2021} finds that there are likely roughly 300-400 discoverable transiting HJs orbiting FGK stars brighter than $G = 12.5$ mag and, while $< 300$ of these have been confirmed so far, \TESS should be able to detect nearly all of the remaining undiscovered HJs. With a complete sample of HJs orbiting bright FGK stars, we will be able to confidently explore the distribution of planetary and stellar parameters of HJ-hosting systems, nailing down the occurrence rate of HJs around Sun-like stars and probing the likelihood of each formation and evolution mechanism by comparing population synthesis models to the observed distribution of orbital eccentricity and orbital period. \cite{Ikwut-Ukwa:2022} found tentative evidence for multiple, separate distributions in mass-period space in the incomplete HJ population. A complete population of HJs could enable follow-up investigations of this and other tentative trends and connect them to the physics of planet formation and evolution.

In this article, we present nine new HJs discovered by \TESS, eight of which orbit an FGK star brighter than $ G = 12.5$ mag. These planets were first detected via transit photometry by \TESS and classified as \TESS Objects of Interest (TOIs) before being confirmed by ground-based photometry, radial velocities, and high-resolution imaging. In \S \ref{sec:obs}, we discuss these observations in detail. In \S \ref{sec:exofast}, we discuss the use of Markov Chain Monte Carlo to globally fit each planetary system and obtain relevant stellar and planetary parameters. In \S \ref{sec:results}, we present the results of these global fits and discuss the future of our survey. Finally, in \S \ref{sec:summary}, we summarize the key arguments made in this article.

\section{Observations} \label{sec:obs}

We used a combination of space-based and ground-based photometric, high-resolution imaging, and spectroscopic observations to characterize each planetary system and rule out possible false-positive scenarios, such as eclipsing binaries, blended nearby eclipsing binaries, and stellar activity. Table \ref{tab:lit} shows the relevant parameters of each system from archival observations and ground-based spectroscopy.

\subsection{\TESS Photometry}\label{subsec:TESS}

All nine of the planets confirmed in this work were first classified as candidates after transits were detected by \TESS. The \TESS spacecraft observes $24^\circ \times 96^\circ$ sectors of the sky for 27 days at a time, with a pixel scale of 21 arcseconds per pixel. During \TESS's Prime Mission (Sectors 1 -- 26), more than 200,000 bright stars were selected for 2-minute cadence, high-precision photometry, while the remainder of the stars in the fields were observed in the 30-minute cadence full frame images. In \TESS's first extended mission (Sectors 27 -- 55), 2-minute cadence photometry was obtained for $\sim$15,000 stars per sector, and the cadence of the full frame images was reduced to 10 minutes. In addition, approximately 600 targets per sector received 20-second cadence data. In the ongoing second extended mission (Sectors 56 -- 97), the full frame images have 200-second sampling, $\sim$8,000 targets per sector will be selected for 2-minute cadence, and $\sim$2,000 targets per sector will be selected for 20-second cadence.

The systems studied in this work were observed in at least two sectors each, between Sector 8 and Sector 63. These observations were collected in 2-minute cadence, 10-minute cadence, and 30-minute cadence. The \TESS data were reduced using both the \TESS Science Processing Operations Center (SPOC) Pipeline \citep{Jenkins:2016, Caldwell:2020} located at NASA Ames Research Center and the MIT Quick-Look Pipeline (QLP; \cite{Huang:2020a, Huang:2020b, Kunimoto:2021}). Both pipelines account for contamination from other stars in the pixel mask. Where available, the SPOC lightcurves were chosen over the QLP lightcurves. See Table \ref{tab:tess} for a full list of the sectors and cadences in which each system was observed and the pipelines used to process the data. In the table, SPOC lightcurves generated from the full-frame images have their source listed as TESS-SPOC, as opposed to SPOC, which refers to the SPOC lightcurves from the 2-minute data.

\begin{rotatepage}
\movetableright=-1in
\movetabledown=2.3in
\begin{rotatetable}
\begin{deluxetable*}{l>{\centering}cccccccc}
\tablecaption{Measured Properties from Literature \label{tab:lit}}
\tabletypesize{\scriptsize}
\providecommand{\bjdtdb}{\ensuremath{\rm {BJD_{TDB}}}}
\providecommand{\feh}{\ensuremath{\left[{\rm Fe}/{\rm H}\right]}}
\providecommand{\teff}{\ensuremath{T_{\rm eff}}}
\providecommand{\teq}{\ensuremath{T_{\rm eq}}}
\providecommand{\ecosw}{\ensuremath{e\cos{\omega_*}}}
\providecommand{\esinw}{\ensuremath{e\sin{\omega_*}}}
\providecommand{\msun}{\ensuremath{\,M_\Sun}}
\providecommand{\rsun}{\ensuremath{\,R_\Sun}}
\providecommand{\lsun}{\ensuremath{\,L_\Sun}}
\providecommand{\mj}{\ensuremath{\,M_{\rm J}}}
\providecommand{\rj}{\ensuremath{\,R_{\rm J}}}
\providecommand{\me}{\ensuremath{\,M_{\rm E}}}
\providecommand{\re}{\ensuremath{\,R_{\rm E}}}
\providecommand{\fave}{\langle F \rangle}
\providecommand{\fluxcgs}{10$^9$ erg s$^{-1}$ cm$^{-2}$}
\providecommand{\tess}{\textit{TESS}\xspace}
\tablecolumns{8}
\tablehead{ &  & \colhead{TOI-1855} & \colhead{TOI-2107} & \colhead{TOI-2368} & \colhead{TOI-3321} & \colhead{TOI-3894} & \colhead{Source}}
\startdata
\multicolumn{8}{l}{\textbf{Other identifiers}:} \\
& \tess Input Catalog & TIC 81247740 & TIC 446549906 & TIC 401125028 & TIC 306648160 & TIC 165464482\\
& TYCHO-2 & TYC 1463-150-1 & --- & --- & TYC 9062-2690-1 & TYC 3850-280-1\\
& 2MASS & J13412503+1741141 & J19071116-5841485 & J08594984-4829182 & J17430248-6500297 & J13252998+5246050\\
& Gaia DR3 & 1247603719345703808 & 6633448311454447232 & 5325498111172934784 & 5909085279371278336 & 1562935820172427392\\
\hline
\multicolumn{8}{l}{\textbf{Astrometric Parameters}:} \\
$\alpha_{J2000}\ddagger$ & Right Ascension (h:m:s) & 13:41:25.028 & 19:07:11.162 & 08:59:49.838 & 17:43:02.481 & 13:25:29.974  & 1 \\
$\delta_{J2000}\ddagger$ & Declination (d:m:s) & 17:41:13.99 & -58:41:48.48 & -48:29:18.269 & -65:00:29.677 & 52:46:05.039  & 1 \\
$\mu_{\alpha}$ & Gaia DR3 proper motion in RA (mas yr$^{-1}$)& $-73.265 \pm 0.017$ & $3.703 \pm 0.012$ & $-5.084 \pm 0.011$ & $-0.048 \pm 0.014$ & $-8.895 \pm 0.012$  & 1 \\
$\mu_{\delta}$ & Gaia DR3 proper motion in Dec (mas yr$^{-1}$)& $-80.743 \pm 0.011$ & $-8.099 \pm 0.011$ & $-19.694 \pm 0.01$ & $-1.832 \pm 0.016$ & $8.65 \pm 0.011$  & 1 \\
$\pi$ & Gaia DR3 Parallax (mas) & $5.6659 \pm 0.0174$ & $4.2151 \pm 0.0146$ & $4.7511 \pm 0.0093$ & $3.4852 \pm 0.0213$ & $2.3814 \pm 0.0114$  & 1 \\
$v\sin{i_\star}$ & Projected rotational velocity (km s$^{-1}$) & $5.74 \pm 0.15$ & $9.85 \pm 0.34$ & $10.6 \pm 0.19$ & $5.19 \pm 0.013$ & $5.27 \pm 0.17$  & 2 \\
\multicolumn{8}{l}{\textbf{Photometric Parameters}:} \\
${\rm G}$ & Gaia $G$ mag. & $11.176755 \pm 0.02$ & $11.857414 \pm 0.02$ & $12.2656 \pm 0.02$ & $11.098307 \pm 0.02$ & $11.687243 \pm 0.02$  & 1 \\
$G_{\rm BP}$ & Gaia $G_{\rm BP}$ mag. & $11.578712 \pm 0.02$ & $12.230471 \pm 0.02$ & $12.737476 \pm 0.02$ & $11.412926 \pm 0.02$ & $11.956965 \pm 0.02$  & 1 \\
$G_{\rm RP}$ & Gaia $G_{\rm RP}$ mag. & $10.606661 \pm 0.02$ & $11.318671 \pm 0.02$ & $11.637787 \pm 0.02$ & $10.606308 \pm 0.02$ & $11.258775 \pm 0.02$  & 1 \\
${\rm T}$ & TESS mag. & $10.6503 \pm 0.007$ & $11.3819 \pm 0.006$ & $11.6955 \pm 0.006$ & $10.6651 \pm 0.006$ & $11.323 \pm 0.007$  & 3 \\
$J$ & 2MASS $J$ mag. & $9.979 \pm 0.022$ & $10.737 \pm 0.026$ & $10.916 \pm 0.021$ & $10.046 \pm 0.021$ & $10.78 \pm 0.021$  & 4 \\
$H$ & 2MASS $H$ mag. & $9.62 \pm 0.023$ & $10.435 \pm 0.026$ & $10.509 \pm 0.024$ & $9.718 \pm 0.021$ & $10.51 \pm 0.02$  & 4 \\
$K$ & 2MASS $K$ mag. & $9.523 \pm 0.02$ & $10.328 \pm 0.021$ & $10.342 \pm 0.02$ & $9.67 \pm 0.021$ & $10.479 \pm 0.02$  & 4 \\
$W1$ & WISE $W1$ mag. & $9.354 \pm 0.03$ & $10.269 \pm 0.031$ & $10.355 \pm 0.03$ & $9.616 \pm 0.03$ & $10.447 \pm 0.03$  & 5 \\
$W2$ & WISE $W2$ mag. & $9.404 \pm 0.03$ & $10.302 \pm 0.031$ & $10.445 \pm 0.03$ & $9.64 \pm 0.03$ & $10.487 \pm 0.03$  & 5 \\
$W3$ & WISE $W3$ mag. & $9.376 \pm 0.036$ & $10.293 \pm 0.075$ & $10.364 \pm 0.069$ & $9.588 \pm 0.04$ & $10.449 \pm 0.059$  & 5 \\
\enddata

\end{deluxetable*}
\end{rotatetable}
\addtocounter{table}{-1} 
\end{rotatepage}

\begin{rotatepage}
\movetableright=-1in
\movetabledown=1.7in
\begin{rotatetable}
\begin{deluxetable*}{l>{\centering}cccccccc}
\tablecaption{\textit{(Continued)}}
\tabletypesize{\scriptsize}
\providecommand{\bjdtdb}{\ensuremath{\rm {BJD_{TDB}}}}
\providecommand{\feh}{\ensuremath{\left[{\rm Fe}/{\rm H}\right]}}
\providecommand{\teff}{\ensuremath{T_{\rm eff}}}
\providecommand{\teq}{\ensuremath{T_{\rm eq}}}
\providecommand{\ecosw}{\ensuremath{e\cos{\omega_*}}}
\providecommand{\esinw}{\ensuremath{e\sin{\omega_*}}}
\providecommand{\msun}{\ensuremath{\,M_\Sun}}
\providecommand{\rsun}{\ensuremath{\,R_\Sun}}
\providecommand{\lsun}{\ensuremath{\,L_\Sun}}
\providecommand{\mj}{\ensuremath{\,M_{\rm J}}}
\providecommand{\rj}{\ensuremath{\,R_{\rm J}}}
\providecommand{\me}{\ensuremath{\,M_{\rm E}}}
\providecommand{\re}{\ensuremath{\,R_{\rm E}}}
\providecommand{\fave}{\langle F \rangle}
\providecommand{\fluxcgs}{10$^9$ erg s$^{-1}$ cm$^{-2}$}
\providecommand{\tess}{\textit{TESS}\xspace}
\tablecolumns{7}
\tablehead{ &  & \colhead{TOI-3919} & \colhead{TOI-4153} & \colhead{TOI-5232} & \colhead{TOI-5301} & \colhead{Source}}
\startdata
\multicolumn{7}{l}{\textbf{Other identifiers}:} \\
& \tess Input Catalog & TIC 23769326 & TIC 470171739 & TIC 69356857 & TIC 58825110\\
& TYCHO-2 & --- & TYC 4612-244-1 & --- & ---\\
& 2MASS & J13554673+4023304 & J22213433+8212578 & J19372100+3516360 & ---\\
& Gaia DR3 & 1497132660589966976 & 2299246592282070400 & 2048174967505679488 & 2776823148593566592\\
\hline
\multicolumn{7}{l}{\textbf{Astrometric Parameters}:} \\
$\alpha_{J2000}\ddagger$ & Right Ascension (h:m:s) & 13:55:46.738 & 22:21:34.307 & 19:37:21.0 & 00:51:57.604  & 1 \\
$\delta_{J2000}\ddagger$ & Declination (d:m:s) & 40:23:30.487 & 82:12:57.759 & 35:16:35.933 & 13:04:41.835  & 1 \\
$\mu_{\alpha}$ & Gaia DR3 proper motion in RA (mas yr$^{-1}$)& $2.236 \pm 0.01$ & $16.409 \pm 0.015$ & $-14.071 \pm 0.01$ & $5.273 \pm 0.259$  & 1 \\
$\mu_{\delta}$ & Gaia DR3 proper motion in Dec (mas yr$^{-1}$)& $1.004 \pm 0.011$ & $8.158 \pm 0.013$ & $-35.962 \pm 0.011$ & $-5.756 \pm 0.314$  & 1 \\
$\pi$ & Gaia DR3 Parallax (mas) & $1.6301 \pm 0.012$ & $2.3424 \pm 0.0109$ & $1.6203 \pm 0.0099$ & $0.8934 \pm 0.1932$  & 1 \\
$v\sin{i_\star}$ & Projected rotational velocity (km s$^{-1}$) & $7.73 \pm 0.09$ & $9.44 \pm 0.22$ & $9.39 \pm 0.23$ & $11.8 \pm 0.19$  & 2 \\
\multicolumn{7}{l}{\textbf{Photometric Parameters}:} \\
${\rm G}$ & Gaia $G$ mag. & $12.701433 \pm 0.02$ & $11.559911 \pm 0.02$ & $12.160616 \pm 0.02$ & $11.565127 \pm 0.02$  & 1 \\
$G_{\rm BP}$ & Gaia $G_{\rm BP}$ mag. & $12.979944 \pm 0.02$ & $11.866878 \pm 0.02$ & $12.438007 \pm 0.02$ & $11.842548 \pm 0.02$  & 1 \\
$G_{\rm RP}$ & Gaia $G_{\rm RP}$ mag. & $12.268707 \pm 0.02$ & $11.080535 \pm 0.02$ & $11.706043 \pm 0.02$ & $11.124998 \pm 0.02$  & 1 \\
${\rm T}$ & TESS mag. & $12.3342 \pm 0.007$ & $11.1477 \pm 0.006$ & $11.7791 \pm 0.009$ & $11.1893 \pm 0.007$  & 3 \\
$J$ & 2MASS $J$ mag. & $11.819 \pm 0.022$ & $10.559 \pm 0.023$ & $11.182 \pm 0.022$ & $10.616 \pm 0.023$  & 4 \\
$H$ & 2MASS $H$ mag. & $11.545 \pm 0.02$ & $10.359 \pm 0.031$ & $10.906 \pm 0.021$ & $10.405 \pm 0.021$  & 4 \\
$K$ & 2MASS $K$ mag. & $11.502 \pm 0.02$ & $10.296 \pm 0.021$ & $10.857 \pm 0.02$ & $10.336 \pm 0.02$  & 4 \\
$W1$ & WISE $W1$ mag. & $11.476 \pm 0.03$ & $10.228 \pm 0.03$ & $10.804 \pm 0.03$ & $10.288 \pm 0.03$  & 5 \\
$W2$ & WISE $W2$ mag. & $11.51 \pm 0.03$ & $10.234 \pm 0.03$ & $10.83 \pm 0.03$ & $10.298 \pm 0.03$  & 5 \\
$W3$ & WISE $W3$ mag. & $11.507 \pm 0.158$ & $10.182 \pm 0.054$ & $10.753 \pm 0.084$ & $10.349 \pm 0.094$  & 5 \\
\enddata

\begin{flushleft}
 \footnotesize{ \textbf{\textsc{NOTES:}}
 The uncertainties of the photometric measurements have a systematic floor applied that is usually larger than the reported catalog errors. \\
 $\ddagger$ Right Ascension and Declination are in epoch J2000. The coordinates come from Vizier where the Gaia RA and Dec have been precessed and corrected to J2000 from epoch J2016.\\
 Sources are: (1) \cite{GaiaEDR3}; (2) \S \ref{subsubsec:tres} \& \S \ref{subsubsec:chiron}; (3) \cite{Stassun:2019}; (4) \cite{Cutri:2003, Skrutskie:2006}; (5) \cite{Wright:2010, Cutri:2012} \\
}
\end{flushleft}
\end{deluxetable*}
\end{rotatetable}
\end{rotatepage}

\startlongtable
\begin{deluxetable}{l l l l l}
\tabletypesize{\scriptsize}
\tablecaption{Summary of Observations from TESS \label{tab:tess}}
\tablewidth{0pt}
\tablehead{
\colhead{Target} & \colhead{TESS Sector} & \colhead{Cadence (s)} & \colhead{Source}
}
\startdata
TOI-1855 & 23 & 1800 & TESS-SPOC \\
--- & 50 & 120 & SPOC \\
TOI-2107 & 13 & 1800 & QLP \\
--- & 27 & 600 & QLP \\
TOI-2368 & 8 & 1800 & TESS-SPOC \\
--- & 9 & 1800 & TESS-SPOC \\
--- & 35 & 600 & TESS-SPOC \\
--- & 36 & 600 & TESS-SPOC \\
--- & 62 & 120 & SPOC \\
--- & 63 & 120 & SPOC \\
TOI-3321 & 12 & 1800 & TESS-SPOC \\
--- & 13 & 1800 & TESS-SPOC \\
TOI-3894 & 15 & 1800 & TESS-SPOC \\
--- & 16 & 1800 & TESS-SPOC \\
--- & 22 & 1800 & TESS-SPOC \\
--- & 49 & 120 & SPOC \\
TOI-3919 & 16 & 1800 & TESS-SPOC \\
--- & 23 & 1800 & TESS-SPOC \\
--- & 49 & 120 & SPOC \\
--- & 50 & 120 & SPOC \\
TOI-4153 & 18 & 1800 & TESS-SPOC \\
--- & 19 & 1800 & TESS-SPOC \\
--- & 24 & 1800 & TESS-SPOC \\
--- & 25 & 1800 & TESS-SPOC \\
--- & 26 & 1800 & TESS-SPOC \\
--- & 53 & 120 & SPOC \\
--- & 58 & 120 & SPOC \\
--- & 59 & 120 & SPOC \\
--- & 60 & 120 & SPOC \\
TOI-5232 & 14 & 1800 & QLP \\
--- & 40 & 600 & QLP \\
--- & 41 & 600 & QLP \\
--- & 54 & 600 & QLP \\
--- & 55 & 600 & QLP \\
TOI-5301 & 17 & 1800 & TESS-SPOC \\
--- & 42 & 600 & TESS-SPOC \\
--- & 43 & 600 & TESS-SPOC \\
\hline
\enddata
\end{deluxetable}

Both the SPOC and QLP pipelines search for and identify transit-like signals, which are then vetted by the \TESS Science Office. The SPOC conducted its transit searches with an adaptive, noise-compensating matched filter \citep{Jenkins:2002, Jenkins:2010, Jenkins:2020}, producing a threshold crossing event for which an initial limb-darkened transit model was fitted \citep{Li:2019} and a suite of diagnostic tests were conducted to help make or break the planetary nature of the signal \citep{Twicken:2018}. Those that survive this process are identified as candidates, labeled \TESS Objects of Interest \citep[TOIs;][]{Guerrero:2021}, and are released to the public. The lightcurves were downloaded using the $\tt{lightkurve}\footnote{\url{https://github.com/lightkurve/lightkurve}}$ package \citep{LightkurveCollab:2018}, which accesses the Mikulski Archive for Space Telescopes (MAST\footnote{\url{https://archive.stsci.edu/}}). The selected lightcurves were then flattened using $\tt{keplersplinev2}\footnote{\url{https://github.com/avanderburg/keplersplinev2}}$, which fits a spline to any out-of-transit stellar variability in the lightcurve, which can then be divided out to flatten the lightcurve without significantly affecting the transit \citep{Vanderburg:2014}. In order to optimally choose the spline breakpoints, we adopted the methodology from \cite{Shallue:2018} and used the $\tt{choosekeplersplinev2}$ feature, which chooses the breakpoint spacing that minimizes the Bayesian Information Criterion \citep[BIC;][]{Schwarz:1978}. We then chop the flattened lightcurves to ensure that only a baseline of one transit duration (T$_{14}$) is kept on either side of each transit to reduce the computational cost of using 27-day, nearly uninterrupted, lightcurves in our global fits.

\subsection{Ground-based Photometry}\label{subsec:followup}

To confirm and better characterize each system, we collected ground-based photometric follow-up of all nine targets through the \TESS Follow-up Observing Program \citep[TFOP;][]{Collins:2018} Sub Group 1 \citep[SG1;][]{Collins:2019}. This ground-based follow-up serves multiple purposes: to rule out contamination from nearby eclipsing binaries that were blended in \TESS's larger pixels, to collect observations in multiple different bandpasses and rule out that the target star itself is an eclipsing binary, and to extend the observation baselines to provide better constraints on the orbital period and transit epoch. The follow-up photometry from SG1 is included in our global analysis (see \S\ref{sec:exofast}) and shown in Table \ref{tab:followup} and Figures \ref{fig:toi1855}-\ref{fig:toi5301}. Additionally, for more detailed information on the follow-up observations collected for each of these planets, see Table \ref{tab:followup}. All 40 follow-up lightcurves are available to the public and can be downloaded from ExoFOP\footnote{\url{https://exofop.ipac.caltech.edu/tess/}}.

\begin{table*}[ht]
\movetableright=-1.5in
\centering
\caption{Follow-up observations}
\label{tab:followup}
\resizebox{0.95\textwidth}{!}{%
\begin{tabular}{llllllllll}
\hline \hline
TIC ID & TOI & Telescope & Camera & Observation Date (UT) & Telescope Size (m) & Filter & Pixel Scale (arcsec) & Exposure Time (s) & Detrend params \\
\hline
81247740 & 1855 & Astrosberge (Bergeron) & SBIG ST10xme & 2020 May 13 & 0.305 & Rc & 1.85 & 38 & Airmass, tot\_C\_cnts \\
 &  & NAOJ-Okayama & MuSCAT & 2021 April 05 & 1.88 & g' & 0.358 & 8 & Airmass \\
 &  & NAOJ-Okayama & MuSCAT & 2021 April 05 & 1.88 & r' & 0.358 & 6 & Airmass \\
 &  & NAOJ-Okayama & MuSCAT & 2021 April 05 & 1.88 & z' & 0.358 & 9 & Airmass \\
 &  & TCS & MuSCAT2 & 2020 May 19 & 1.52 & g' & 0.44 & 8 & Airmass \\
 &  & TCS & MuSCAT2 & 2020 May 19 & 1.52 & r' & 0.44 & 6 & Airmass \\
 &  & TCS & MuSCAT2 & 2020 May 19 & 1.52 & i' & 0.44 & 12-15$^\dagger$ & Airmass \\
 &  & TCS & MuSCAT2 & 2020 May 19 & 1.52 & z' & 0.44 & 9-12$^\dagger$ & Airmass \\
 &  & Deep Sky West & Apogee Ultra 16m & 2020 May 21 & 0.5 & g' & 1.09 & 90 & BJD\_TDB, FWHM\_T1, Airmass \\
 &  & LCO-CTIO & Sinistro & 2021 January 27 & 1 & z' & 0.39 & 40 & Airmass \\
 &  & SOAR & Goodman & 2022 April 20 & 4.1 & i' & 0.15 & 25 & Airmass, Y(FITS)\_T1 \\
446549906 & 2107 & LCO-SSO & SBIG STX6303 & 2020 May 06 & 0.4 & i' & 0.57 & 300 & Width\_T1, Y(FITS)\_T1 \\
 &  & LCO-SAAO & SINISTRO & 2020 July 16 & 1 & z' & 0.389 & 70 & Airmass \\
 &  & LCO-SAAO & SINISTRO & 2020 July 16 & 1 & g' & 0.389 & 26 & Airmass \\
 &  & TRAPPIST-South & FLI ProLine PL3041-BB & 2023 May 10 & 0.6 & B & 0.64 & 50 & None \\
401125028 & 2368 & CDK14 (El Sauce) & STT-1603 & 2020 November 14 & 0.36 & Rc & 1.47 & 120 & Airmass \\
 &  & Hazelwood Observatory & STT3200 & 2021 January 20 & 0.32 & Rc & 0.556 & 180 & Airmass \\
 &  & LCO-SSO & Sinistro & 2021 February 20 & 1 & g' & 0.39 & 47 & Airmass \\
 &  & LCO-SSO & Sinistro & 2021 February 20 & 1 & i' & 0.39 & 27 & Airmass \\
 &  & SOAR & Goodman & 2022 April 20 & 4.1 & i' & 0.15 & 40 & Airmass \\
 &  & LCO-CTIO & SINISTRO & 2023 April 12 & 1 & i' & 0.389 & 27 & Airmass \\
 &  & LCO-CTIO & SINISTRO & 2023 April 12 & 1 & i' & 0.389 & 40 & Airmass \\
306648160 & 3321 & LCO-SAAO & Sinistro & 2022 April 30 & 1 & z' & 0.389 & 44 & None \\
165464482 & 3894 & KeplerCam & KeplerCam & 2022 January 26 & 1.2 & i' & 0.672 & 10 & Airmass \\
 &  & KeplerCam & KeplerCam & 2022 March 19 & 1.2 & B & 0.672 & 20 & Airmass, X(FITS)\_T1 \\
 &  & Acton Sky Portal & SBIG A4710 & 2023 April 09 & 0.36 & r' & 1 & 20 & Airmass \\
23769326 & 3919 & CALOU & FLI PL1001 & 2022 January 14 & 0.4 & R & 1.11 & 120 & Airmass \\
 &  & WCO & SBIG STXL-6303E & 2022 April 28 & 0.36 & r' & 0.66 & 90 & Meridian Flip \\
 &  & KeplerCam & KeplerCam & 2022 April 28 & 1.2 & i' & 0.672 & 25 & Airmass, X(FITS)\_T1 \\
 &  & LCO-McD & Sinistro & 2023 March 06 & 1 & i' & 0.389 & 46 & Y(FITS)\_T1 \\
470171739 & 4153 & MSU & Apogee Alta U47 & 2022 July 03 & 0.6 & V & 0.55 & 60 & Airmass \\
 &  & MSU & Apogee Alta U47 & 2022 August 09 & 0.6 & V & 0.55 & 60 & Airmass \\
 &  & CRCAO-KU-OGS-RC & SBIG STXL-6303E & 2022 September 16 & 0.61 & B & 0.76 & 120 & Airmass, X(FITS)\_T1 \\
 &  & CRCAO-KU-OGS-RC & SBIG STXL-6303E & 2022 September 16 & 0.61 & Ic & 0.76 & 90 & None \\
69356857 & 5232 & ULMT & ASI 6200 & 2023 June 12 & 0.6 & R & 0.167 & 30 & None \\
 &  & LCO-McD & SINISTRO & 2023 June 12 & 1 & i' & 0.389 & 27 & Airmass \\
58825110 & 5301 & KeplerCam & KeplerCam & 2022 October 28 & 1.2 & i' & 0.672 & 7 & Airmass \\
\hline
\end{tabular}%
}
\begin{flushleft}
    \textbf{NOTE:} All lightcurves are available on ExoFOP.\\
    $\dagger$ Exposure times were adjusted to avoid target saturation. \S \ref{subsubsec:muscat} provides further details.
\end{flushleft}
\end{table*}

The Las Cumbres Observatory Global Telescope (LCOGT; \citealt{Brown:2013}) generated twelve of the follow-up lightcurves of our targets using observations from the following LCOGT sites: Cerro Tololo Inter-American Observatory (CTIO), the Siding Spring Observatory (SSO), the South African Astronomical Observatory (SAAO), and the McDonald Observatory (McD). Additional ground-based observations of our targets were collected using the following facilities: the National Astronomical Observatory of Japan (NAOJ) \citep{Narita2015}, Teide Observatory, Deep Sky West, the Southern Astrophysical Research (SOAR) telescope at CTIO, TRAPPIST-South at the La Silla Observatory \citep{Jehin:2011}, El Sauce Observatory, Hazelwood Observatory, Brierfield Observatory, KeplerCam at the Fred Lawrence Whipple Observatory (FLWO), the Acton Sky Portal, Observatori de Ca l’Ou (CALOU), Waffelow Creek Observatory (WCO), the Michigan State University (MSU) Observatory, Carlson R. Chambliss Astronomical Observatory (CRCAO) at Kutztown University, and the University of Louisville Manner Telescope (ULMT) at Mt. Lemmon Observatory.

The data calibration and reduction for each follow-up observation (with the exception of the observations from MuSCAT and MuSCAT2; see \S \ref{subsubsec:muscat}) were done using {\tt AstroImageJ} (AIJ) \citep{Collins_2017}, a tool designed to enable point-and-click photometry and data reduction. The exposures from the observation are first calibrated using flat field images, dark images, and bias images by the observer or facility that collected the observations. The general procedure\footnote{For more details and observing tips, see Dennis Conti's guide: \url{https://astrodennis.com/Guide.pdf}} that is then used to generate a reduced lightcurve for each target is as follows:

\begin{itemize}
    \item Generate calibrated images using AIJ's CCD Data Processor tool, median combining the biases, darks, and flats, before subtracting the master bias and dark and dividing the master flat.
    \item Locate the target star in the first image and develop an aperture that has a diameter of roughly twice the full width at half maximum (FWHM) of the target. The background annulus should have roughly the same area as the target aperture.
    \item Place apertures on the target star and at least 5-10 comparison stars of similar brightness. The number of comparison stars varies significantly depending on the field of view of the telescope and the density of stars in the surrounding area of the sky.
    \item Run AIJ's multi-aperture photometry, which populates AIJ's measurement table and plot.
    \item Assess the quality of the data and search for relevant parameters that change significantly throughout the observation and correlate with changes in the relative flux of the host star. Likely culprits usually include the airmass, the target star's X and Y pixel location (due to imperfect guiding), and the width or FWHM of the target source (due to evolving weather conditions).
    \item Use AIJ's built-in transit-fitting tool to fit the transit with detrending enabled. Choose the detrending parameters from the previous step that minimize the Bayesian Information Criterion.
    \item Normalize the relative flux and save a lightcurve file containing the Barycentric Julian Date in Barycentric Dynamical Time (\bjdtdb), normalized relative flux, relative flux error, and the selected detrending parameters.
\end{itemize}

This reduced lightcurve is then ready to be included in our global fit (see \S \ref{sec:exofast}).

\subsubsection{MuSCAT and MuSCAT2 observations of TOI-1855} \label{subsubsec:muscat}

Two of the follow-up light curves that we obtained for TOI-1855 b were reduced using a reduction strategy different than the one presented above.

We observed TOI-1855 b using MuSCAT, a simultaneous multi-band photometer installed on the 188cm telescope of the National Astronomical Observatory of Japan (NAOJ) in Okayama, Japan \citep{Narita2015}. It has three 1k CCDs each with $6.1\arcmin \times 6.1\arcmin$ field-of-view (FOV) and $0.358\arcsec$ pix$^{-1}$ pixel scale, enabling simultaneous photometry in the $g$ (400–550 nm), $r$ (550–700 nm), and $z_s$ (820–920 nm) bands. The data reduction and differential photometry were performed using the pipeline described in \citet{Fukui2011}. We optimized both the aperture radii and the set of comparison stars by minimizing the dispersion of the resulting relative light curves. However, there are only two useful comparison stars within the FOV, one of which is a faint star 12.8\arcsec from the target. Using the brighter comparison star and uncontaminated aperture radius of 7.2\arcsec yielded the optimum light curves with a typical rms of 0.5 ppt/10 min across all bands.

We additionally observed TOI-1855 b with the 1.5-m Telescopio Carlos S\'{a}nchez (TCS) instrument MuSCAT2 \citep{Narita2019} from Teide Observatory, Tenerife, Spain. MuSCAT2 is another multi-band imager, equipped with four cameras with a field of view of $7.4 \times 7.4$ arcmin$^2$ (pixel scale of 0.44 arcsec per pixel) and is capable of obtaining simultaneous observations in $g'$, $r'$, $i'$, and $z_s$ bands. The exposure times were initially set to 8 s, 6 s, 15 s, and 12 s in $g'$, $r'$, $i'$, and $z_s$, respectively. During the observations, the $i'$, and $z_s$ band exposure times were changed to 12 s and 9 s respectively to avoid the saturation of the target star. The dedicated MuSCAT2 pipeline \citep{Parviainen2019} reduced the raw science frames using standard procedures (i.e., dark and flat field corrections) and computed the photometry for the stars in the field using a set of circular apertures. Finally, the pipeline obtained the best light curve by fitting a transit model that includes instrumental effect noise components and choosing the aperture that delivered the lowest scatter in the time series.



\subsection{Spectroscopy}

In order to rule out false-positive scenarios and measure the mass and orbital eccentricity of each planet, we collected spectroscopic observations of each host star to measure the radial velocity (RV) shift imparted on the star by the orbiting planet. The spectra that we used in our final fits of the nine systems in this sample were collected by the Tillinghast Reflector \'Echelle Spectrograph (TRES) on the 1.5-meter Tillinghast Reflector at the Fred Lawrence Whipple Observatory \citep{gaborthesis}, the CHIRON spectrograph on the SMARTS 1.5-meter telescope at the Cerro Tololo Inter-American Observatory \citep{Tokovinin:2013}, and the NEID spectrograph on the 3.5-meter WIYN Telescope at the Kitt Peak National Observatory \citep{Schwab:2016}. The RV of each star as a function of time and orbital phase are shown in Figures \ref{fig:toi1855}-\ref{fig:toi5301}. An example RV measurement is shown for each target and instrument in Table \ref{tab:rv}. Additionally, the spectroscopic metallicity measurements from TRES and CHIRON were averaged and used as a wide Gaussian prior in our \exofast global fits, with a prior width equal to twice the standard deviation of the measured metallicities.

\begin{deluxetable}{l l l l l}[bt]
\tabletypesize{\scriptsize}
\tablecaption{The first RV measurement of each system, per instrument used. \label{tab:rv}}
\tablewidth{0pt}
\tablehead{
\colhead{Target} & \colhead{Spectrograph} & \colhead{\bjdtdb} & \colhead{RV (m s$^{-1}$)}  & \colhead{$\sigma_{RV}$ (m s$^{-1}$)}
}
\startdata
TOI-1855 & CHIRON & 2459281.82160 & -34358.0 & 29.0 \\
--- & OES & 2459659.41157 & -7998.8 & 20 \\
--- & TCES & 2459657.51218 & 253.996 & 62.3 \\
--- & MUSICOS & 2459657.55728 & -34400 & 80 \\
TOI-2107 & CHIRON & 2459320.88131 & 2413.0 & 44.0 \\
TOI-2368 & CHIRON & 2459299.60262 & 16039.0 & 64.0 \\
TOI-3321 & CHIRON & 2459375.74691 & 12710.0 & 28.0 \\
TOI-3894 & TRES & 2459405.67847 & 231 & 52.6 \\
--- & NEID & 2459709.79500 & 8084.2 & 10.4 \\
TOI-3919 & TRES & 2459622.02268 & 679 & 39.9 \\
TOI-4153 & TRES & 2459421.95883 & 251 & 48.5 \\
TOI-5232 & TRES & 2459684.92784 & 409 & 53.2 \\
TOI-5301 & TRES & 2459823.79933 & -478 & 46.0 \\
\hline
\enddata
\begin{flushleft}
    \textbf{NOTE:} The full table of RVs for each system is available in machine-readable form in the online journal.
\end{flushleft}
\end{deluxetable}

\subsubsection{TRES Spectroscopy} \label{subsubsec:tres}

Five of the nine host stars presented in this paper (TOI-3894, TOI-3919, TOI-4153, TOI-5232, and TOI-5301) were observed using the Tillinghast Reflector \'Echelle Spectrograph (TRES) to obtain precise radial velocities. TRES is a fiber-fed, high-resolution \'echelle spectrograph mounted on a 1.5-meter telescope with a resolving power (R) of 44,000. The reduction of the spectra was done following the work of \cite{Buchhave:2010}. Multi-order velocities were derived by cross-correlating the spectrum with the highest signal-to-noise per resolution element against each observation, order by order. To measure the metallicity, effective temperature, surface gravity, and projected rotational velocity of the star, we analyzed the spectra with the Stellar Parameter Classification (SPC) tool \citep{Buchhave:2012}. Finally, we calculated the bisector spans of the TRES spectra following the work of \cite{Torres:2007} to hunt for correlations between the bisector spans and RVs, an indication that the RVs are being induced by an eclipsing binary. For each of the five targets with TRES spectra, we found no evidence of correlation between the bisector spans and the RVs.


\subsubsection{CHIRON Spectroscopy} \label{subsubsec:chiron}

The other four host stars in this work (TOI-1855, TOI-2107, TOI-2368, and TOI-3321) were observed using CHIRON, a fiber-fed, high-resolution \'echelle spectrograph. For the majority of the observations in this paper, we used an image slicer that offers a resolving power of $\sim$ 80,000. The exception to this is TOI-2107, of which we observed 23 spectra in the fiber mode (R $\sim$ 25,000) before switching to the slicer mode for the remaining 13 observations. In both cases, we bracketed each of the spectra we collected with ThAr calibration spectra. To obtain estimates of the metallicity, effective temperature, surface gravity, and projected rotational velocity, each spectrum is matched to an interpolated catalog of $\sim$ 10,000 spectra classified by SPC \citep{Buchhave:2012}. RVs were derived by the least-squares deconvolution \citep{Donati:1997, Zhou:2021} of the observed spectra against non-rotating synthetic templates generated using ATLAS9 model atmospheres \citep{Kurucz:1992}, before fitting the resulting line profile with a rotational broadening kernel prescribed by \cite{Gray:2005}.

\subsubsection{NEID Spectroscopy}

We obtained six observations of TOI-3894 with the NEID spectrograph, between UT 2022 April 29 and UT 2022 June 09. NEID is a highly-stabilized, fiber-fed optical spectrograph \citep{NEID_Schwab2016,NEID_Halverson2016} on the WIYN 3.5m telescope at Kitt Peak National Observatory (KPNO). We used NEID in high-resolution (HR) mode ($R \sim 110{,}000$), and the data were reduced using versions 1.1.2--1.1.4 of the NEID Data Reduction Pipeline (DRP). The DRP extracts precise RVs by cross-correlating the observed spectra with a stellar line mask \citep{Baranne1996,Pepe2002}. Each observation was made with exposure times between 300s and 420s, achieving a typical RV precision of $\approx 10$~m\,s$^{-1}$. On the nights of 9 and 10 May 2022, the NEID data were affected by a failure of the Fabry-Perot etalon laser, which is used to calibrate nightly offsets. We used observations of standard stars to determine those offsets ($35.8\pm1.2$~m\,s$^{-1}$ and $31.5\pm1.3$~m\,s$^{-1}$ respectively, see \citealt{Yee:2023}) and correct the NEID RVs taken on the affected nights.

\subsection{Supplementary Spectroscopy}

Three additional spectra of TOI-1855 were collected by the Skalnat\'e Pleso Observatory in the High Tatras (Slovak Republic), using the 1.3 m f/8.36 Astelco Alt-azimuthal Nasmyth-Cassegrain reflecting telescope, equipped with a fiber-fed \'echelle spectrograph of MUSICOS design \citep{Baudrand:1992}. The spectra were recorded using an Andor iKon-L DZ936N-BV CCD camera with a 2048 × 2048 array, 13.5 $\mu$m square pixels, 2.9e- readout noise, and a gain close to unity. The spectral range of the instrument is 4250–7375 Å (56 \'echelle orders) with a maximum resolving power of R = 38,000. The raw data were then reduced using IRAF package tasks, Linux shell scripts, and FORTRAN programs \citep{Pych:2004, Pribulla:2015, Garai:2017}.

These RVs were used to increase our confidence that TOI-1855 b is a bonafide exoplanet. Because there were only three observations from this telescope, they were not included in our final \exofast global fit. However, these observations were consistent with the $204^{+16}_{-15} \rm{ ms}^{-1}$ RV semi-amplitude measured by CHIRON.


\subsection{High-resolution Imaging}

Close stellar companions (bound or line of sight) can confound exoplanet discoveries in a number of ways. The detected transit signal might be a false positive due to a nearby eclipsing binary and even real planet discoveries will yield incorrect stellar and exoplanet parameters if a close companion exists and is unaccounted for \citep{Ciardi:2015, Furlan:2017b}. Additionally, the presence of a close companion star leads to the non-detection of small planets residing in the same exoplanetary system \citep{Lester:2021}. Given that nearly one-half of solar-like stars are in binary or multi-star systems \citep{Matson:2018}, high-resolution imaging (HRI) provides crucial information toward our understanding of exoplanet formation, dynamics, and evolution \citep{Howell:2021}.

In order to understand and properly account for the light contribution of nearby field stars and stellar companions, we observed all nine of the host stars in our sample using adaptive optics (AO) and speckle imaging. Both of these techniques are capable of high angular resolution and allow us to place upper limits on the brightness of a nearby star and therefore the potential of any nearby star to contaminate the signal in the photometric and spectroscopic data.

\begin{figure}
    \centering
    \includegraphics[width=0.95\linewidth]{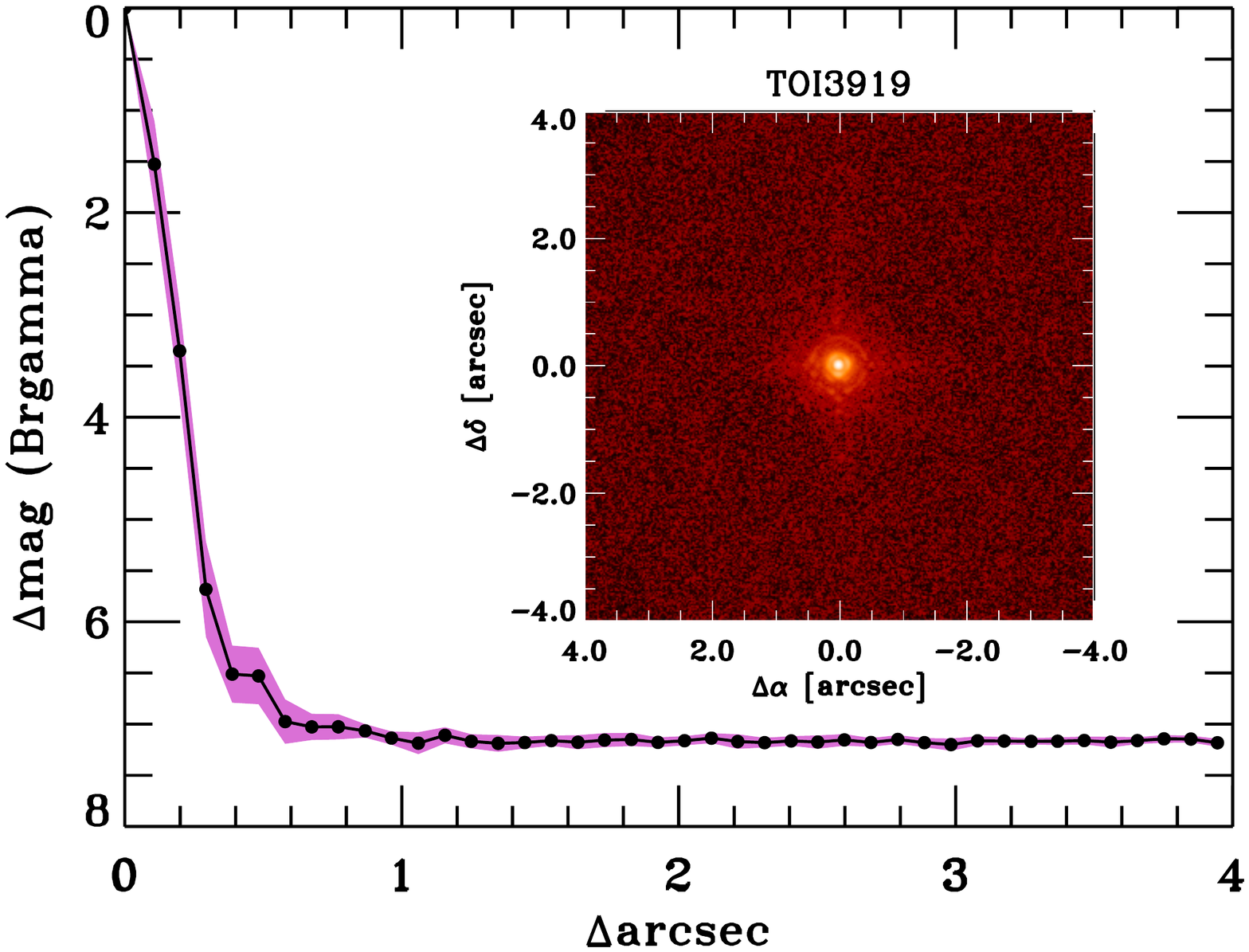}
    \includegraphics[width=0.95\linewidth]{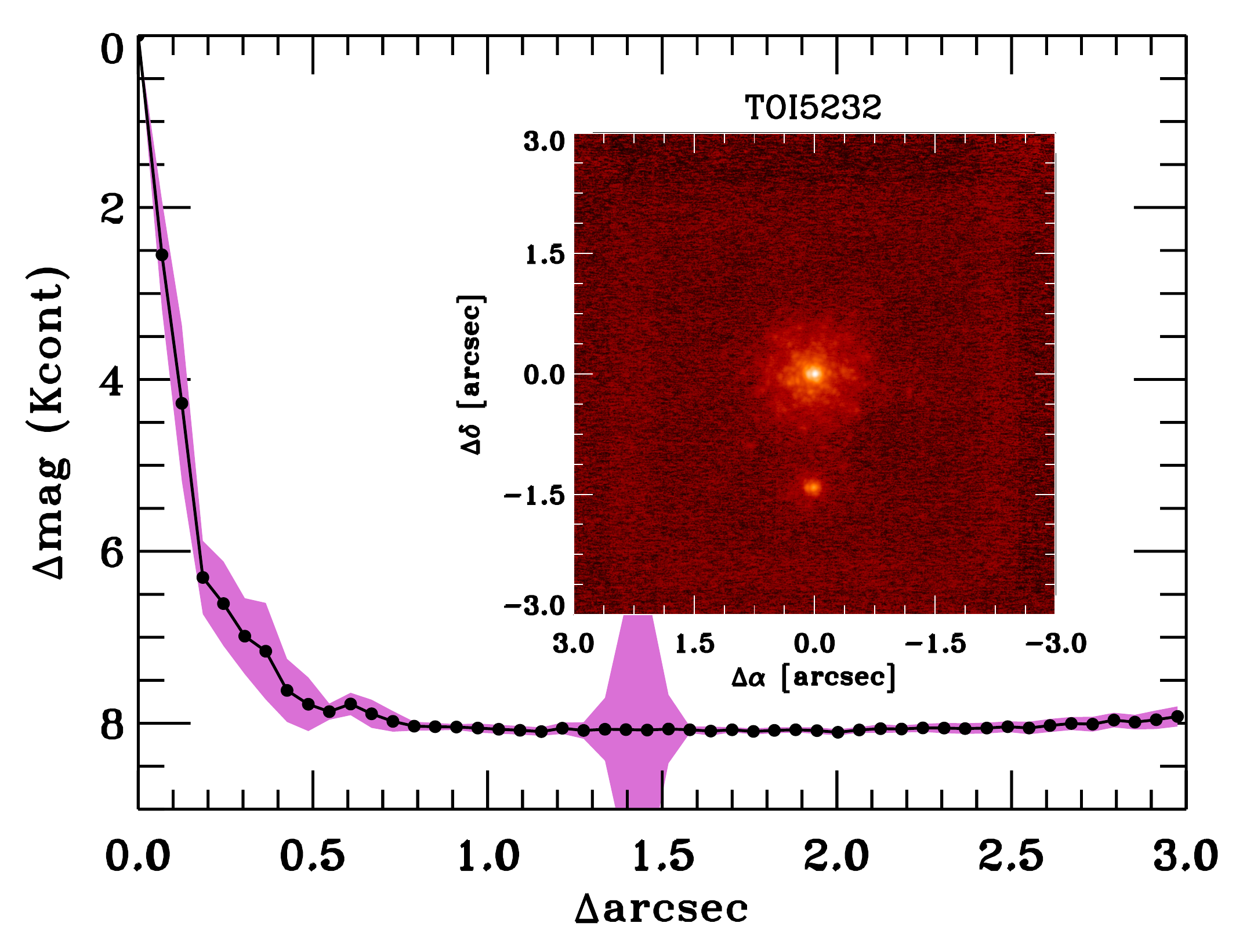}
    \caption{Adaptive optics images and sensitivity curves for TOI-3919 and TOI-5301. 
    \textbf{Top:} Palomar PHARO AO image and $5\sigma$ sensitivity curve of TOI-3919 showing no detection of a companion star, down to the detection limits of the instrument. The magenta-shaded region represents the uncertainty on the contrast curve.
    \textbf{Bottom:} Keck2 NIRC2 AO image and $5\sigma$ sensitivity curve of TOI-5232, showing a clear detection of a stellar companion at a separation of $1.417 \arcsec$ with $\Delta$mag = 2.952 in the Kcont filter.}
    \label{fig:hri_ex}
\end{figure}

Twenty AO and speckle observations from eight different instruments were used to find or rule out nearby contaminants to the host stars in this paper. The results of these observations are presented in the following sections, organized by the instrument used. An example of a non-detection (TOI-3919) and positive detection (TOI-5232) of a nearby stellar companion is presented in Figure \ref{fig:hri_ex}. The remainder of the high-resolution images and sensitivity curves are available on ExoFOP and are described in \S \ref{subsubsec:ao} and \S \ref{subsubsec:speckle} and presented in Table \ref{tab:hri}.

\begin{deluxetable*}{l l l l l l l l}[bt]
\tabletypesize{\scriptsize}
\tablecaption{Summary of High-Resolution Imaging Observations \label{tab:hri}}
\tablewidth{0pt}
\tablehead{
\colhead{Target} & \colhead{Telescope} & \colhead{Instrument} & \colhead{Image Type}  & \colhead{Filter} & \colhead{Contrast} & \colhead{Observation Date (UT)} & \colhead{Detection?$^\dagger$}
}
\startdata
TOI-1855 & SAI (2.5~m) & Speckle Polarimeter & Speckle & $I_c$ & $\Delta$ 6.2 mag at 1$\arcsec$ & 2020 Dec 29 & No \\
--- & Palomar (5~m) & PHARO & AO & Br$\gamma$ & $\Delta$ 6.732 mag at 0.5$\arcsec$ & 2021 Feb 24 & No \\
--- & SOAR (4.1~m) & HRCam & Speckle & $I_c$ & $\Delta$ 6.4 mag at 1$\arcsec$ & 2021 Feb 27 & No \\
--- & Shane (3~m) & ShARCS & AO & $K_s$ & --- & 2021 Mar 28 & No \\
--- & Shane (3~m) & ShARCS & AO & $J$ & --- & 2021 Mar 28 & No \\
--- & WIYN (3.5~m) & NESSI & Speckle & 562 nm & --- & 2021 Apr 1 & No \\
--- & WIYN (3.5~m) & NESSI & Speckle & 832 nm & --- & 2021 Apr 1 & No \\
--- & WIYN (3.5~m) & NESSI & Speckle & 832 nm & --- & 2021 Apr 18 & No \\
TOI-2107 & SOAR (4.1~m) & HRCam & Speckle & $I_c$ & $\Delta$ 6.0 mag at 1$\arcsec$ & 2021 Jul 14 & No \\
TOI-2368 & SOAR (4.1~m) & HRCam & Speckle & $I_c$ & $\Delta$ 5.8 mag at 1$\arcsec$ & 2020 Dec 29 & No \\
TOI-3321 & SOAR (4.1~m) & HRCam & Speckle & $I_c$ & $\Delta$ 7.0 mag at 1$\arcsec$ & 2021 Jul 14 & Yes \\
--- & Gemini (8~m) & Zorro & Speckle & 832 nm & $\Delta$ 6.23 mag at 0.5$\arcsec$ & 2021 Jul 23 & Yes \\
TOI-3894 & WIYN (3.5~m) & NESSI & Speckle & 832 nm & --- & 2022 Apr 17 & No \\
--- & Palomar (5~m) & PHARO & AO & Br$\gamma$ & $\Delta$ 7.278 mag at 0.5$\arcsec$ & 2022 Jul 15 & No \\
TOI-3919 & Palomar (5~m) & PHARO & AO & Br$\gamma$ & $\Delta$ 6.529 mag at 0.5$\arcsec$ & 2022 Jul 15 & No \\
--- & SAI (2.5~m) & Speckle Polarimeter & Speckle & $I_c$ & $\Delta$ 5.1 mag at 1$\arcsec$ & 2023 Jan 19 & No \\
TOI-4153 & SAI (2.5~m) & Speckle Polarimeter & Speckle & $I_c$ & $\Delta$ 4.9 mag at 1$\arcsec$ & 2021 Jul 19 & No \\
TOI-5232 & Keck2 (10~m) & NIRC2 & AO & $H$cont & $\Delta$ 7.017 mag at 0.5$\arcsec$ & 2022 Nov 8 & Yes \\
--- & Keck2 (10~m) & NIRC2 & AO & $K$cont & $\Delta$ 7.779 mag at 0.5$\arcsec$ & 2022 Nov 8 & Yes \\
--- & Gemini (8~m) & 'Alopeke & Speckle & 562 nm & $\Delta$ 5.92 mag at 0.5$\arcsec$ & 2022 Nov 14 & Yes \\
--- & Gemini (8~m) & 'Alopeke & Speckle & 832 nm & $\Delta$ 7.0 mag at 0.5$\arcsec$ & 2022 Nov 14 & Yes \\
TOI-5301 & SAI (2.5~m) & Speckle Polarimeter & Speckle & $I_c$ & $\Delta$ 6.1 mag at 1$\arcsec$ & 2022 Nov 18 & No \\
\hline
\enddata
\begin{flushleft}
    \textbf{NOTE:} All images and contrast curves are available on ExoFOP.\\
    $\dagger$ Detection refers to a positive detection of a star within the field of view of the AO or speckle instrument, subject to the maximum contrast possible with the instrument in question. TOI-1855's companion was outside of the field of view of each instrument that observed it.
\end{flushleft}
\end{deluxetable*}

\subsubsection{Adaptive Optics} \label{subsubsec:ao}

Three AO instruments were employed to observe the stars studied in this work: the ShARCS camera on the Shane 3-meter telescope at Lick Observatory (\citealt{2012SPIE.8447E..3GK, 2014SPIE.9148E..05G, 2014SPIE.9148E..3AM}; Dressing et al. in preparation), the PHARO instrument on the Hale 5-meter telescope at Palomar Observatory \citep{Hayward:2001}, and the NIRC2 instrument on the Keck II 10-meter telescope at Keck Observatory \citep{Wizinowich:2000}.

The ShARCS observations of TOI-1855 were taken with the Shane adaptive optics (AO) system in natural guide star mode. We collected sequences of observations using a $K_s$ filter ($\lambda_0 = 2.150$ $\mu$m, $\Delta \lambda = 0.320$ $\mu$m) and a $J$ filter ($\lambda_0 = 1.238$ $\mu$m, $\Delta \lambda = 0.271$ $\mu$m). We reduced the data using the publicly available \texttt{SImMER} pipeline \citep{2020AJ....160..287S, 2022PASP..134l4501S}.\footnote{\url{https://github.com/arjunsavel/SImMER}} This observation ruled out any bright stellar companions other than TIC 81247738, the source with a separation of $12.352\arcsec$ already identified by Gaia.

Additionally, we observed TOI-1855, TOI-3894, and TOI-3919 with the PHARO instrument \citep{Hayward:2001} on the Palomar Hale (5m) behind the P3K natural guide star AO system \citep{Dekany:2013} in the narrowband Br-$\gamma$ filter $(\lambda_0 = 2.1686; \Delta\lambda = 0.0326~\mu$m). The PHARO pixel scale is $0.025\arcsec$ per pixel. A standard 5-point quincunx dither pattern with steps of 5\arcsec\ was repeated twice with each repeat separated by 0.5\arcsec. The final resolutions of the combined dithers were determined from the full-width half-maximum (FWHM) of the point spread functions: 0.09\arcsec.

Observations of TOI-5232 were made with the NIRC2 instrument behind the natural guide star AO system \citep{Wizinowich:2000} in the standard 3-point dither pattern that is used with NIRC2 to avoid the left lower quadrant of the detector which is typically noisier than the other three quadrants. The dither pattern step size was $3\arcsec$ and was repeated twice, with each dither offset from the previous dither by $0.5\arcsec$. NIRC2 was used in the narrow-angle mode with a full field of view of $\sim10\arcsec$ and a pixel scale of approximately $0.0099442\arcsec$ per pixel. The NIRC2 observations were made in the $K$cont filter $(\lambda_o = 2.2706; \Delta\lambda = 0.0296~\mu$m) and the 
$H$cont filter $(\lambda_o = 1.5804; \Delta\lambda = 0.0232~\mu$m). The final resolutions of the combined dithers were once again determined from the FWHM of the point spread functions: 0.067\arcsec.  

For both the PHARO and the NIRC2 images, the sensitivities of the final combined AO image were determined by injecting simulated sources azimuthally around the primary target every $20^\circ $ at separations of integer multiples of the central source's FWHM \citep{Furlan:2017a}. The brightness of each injected source was scaled until standard aperture photometry detected it with $5\sigma $ significance. The final $5\sigma $ limit at each separation was determined from the average of all of the determined limits at that separation and the uncertainty on the limit was set by the rms dispersion of the azimuthal slices at a given radial distance. Only TOI-5232 shows a close companion - also detected in the optical speckle data and by Gaia.

\subsubsection{Speckle Imaging} \label{subsubsec:speckle}

In addition to our AO observations, five speckle instruments were used to observe the stars in our sample: the speckle polarimeter on the 2.5-m telescope at the Caucasian Mountain Observatory of the Sternberg Astronomical Institute (SAI) at Lomonosov Moscow State University, the HRCam instrument on the 4.1-meter Southern Astrophysical Research (SOAR) telescope \citep{Tokovinin:2018} at CTIO, the NN-EXPLORE Exoplanet Stellar Speckle Imager \citep[NESSI;][]{Scott:2018} on the WIYN 3.5-meter telescope at Kitt Peak Observatory, and the Zorro and 'Alopeke instruments on the Gemini-South and Gemini-North 8-meter telescopes, respectively \citep{Scott:2021}.

The HRCam observations were made of TOI-1855, TOI-2107, TOI-2368, and TOI-3321 in the Cousins $I$-band, a similar visible bandpass as TESS. The four observations were typically sensitive to a 5-magnitude fainter star at an angular distance of 1 arcsec from the target. More details of the speckle observations from the SOAR TESS survey are available in \cite{Ziegler:2020}. A 4.6 magnitude fainter star was detected at a separation of $0.69\arcsec$ from TOI-3321. No nearby stars were detected within 3$\arcsec$ of TOI-1855, TOI-2107, or TOI-2368 in the SOAR observations.

TOI-1855, TOI-3919, TOI-4153, and TOI-5301 were observed with the SAI speckle polarimeter. The speckle polarimeter used an Electron Multiplying CCD (EMCCD) Andor iXon 897 as its main detector prior to August 2022 \citep{Safonov:2017}. Since then, it has used the high–speed, low–noise CMOS detector Hamamatsu ORCA–quest \citep{Strakhov:2023}. The atmospheric dispersion compensator was active in all cases. Observations were carried out in the $I_c$ band, where the respective angular resolution is $0.083^{\prime\prime}$. No nearby stars were detected in any of the SAI images.

The NN-EXPLORE Exoplanet Stellar Speckle Imager \citep{Scott:2018} was used to obtain speckle imaging of TOI-1855 and TOI-3894. NESSI is a dual-channel speckle imager at the WIYN 3.5~m telescope on Kitt Peak, Arizona. TOI-1855 was observed with two filters having central wavelengths of $\lambda_c=562$ and 832~nm. It was then re-observed using only the 832~nm filter. TOI-3894 was observed using only the 832~nm filter. For each observation, the star was observed by taking 9 sets of 1000 40~ms exposures in a $256\times256$ pixel section of the EMCCD for a $4.6\times4.6$~arcsecond field of view centered on the target. A nearby single star was observed in close temporal proximity to each science target using a set of 1000 exposures to serve as a measure of the point spread function. The speckle pipeline data reduction followed the description given in \cite{Howell:2011}. No companion sources were detected for either TOI-1855 or TOI-3894.

\begin{figure}
    \centering
    \includegraphics[width=0.95\linewidth]{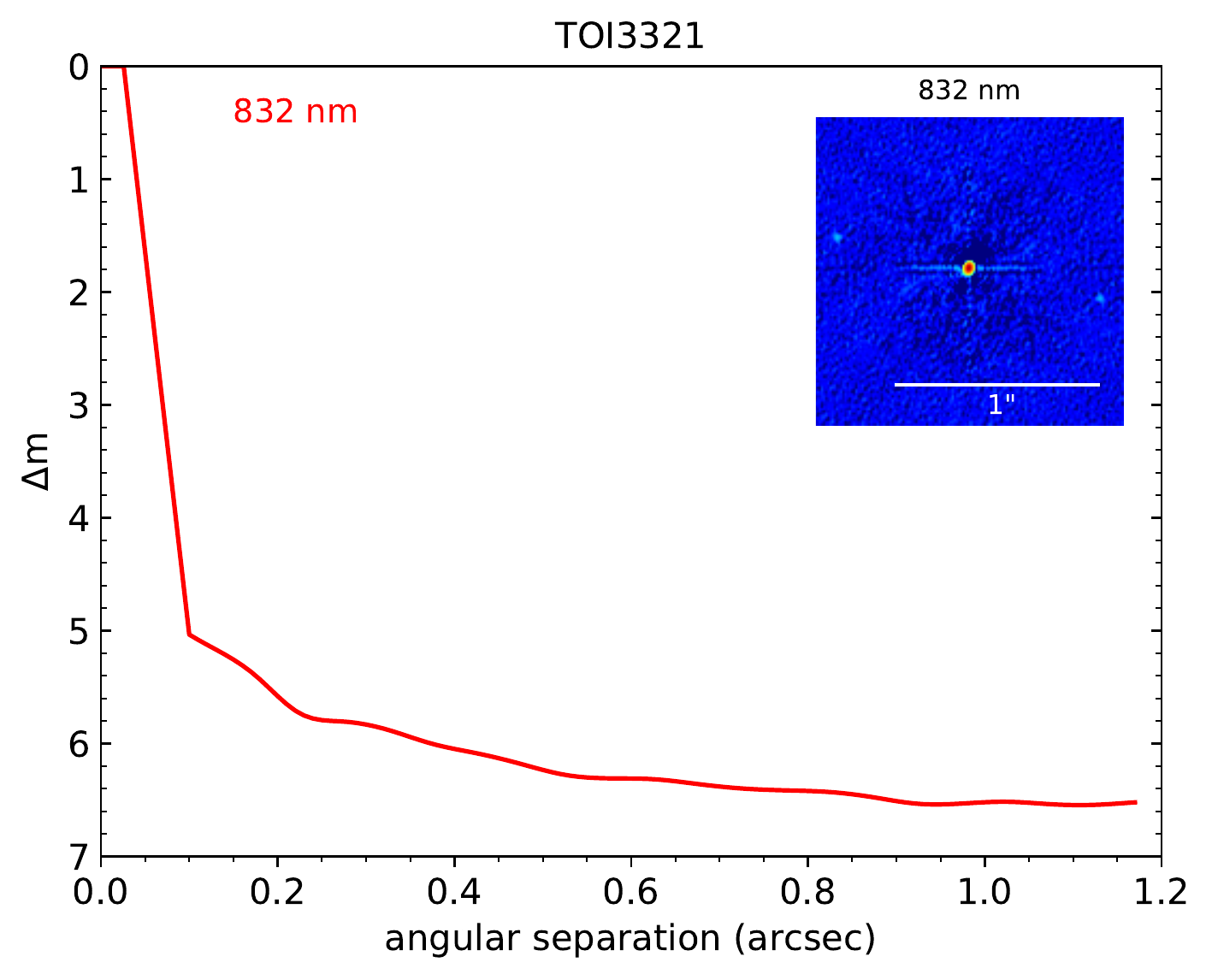}
    \caption{Speckle high-resolution image and $5\sigma$ sensitivity curve from Gemini's Zorro speckle imager. The observation reveals a source $79.1^{\circ}$ E of N and $0.663 \arcsec$ away from TOI-3321 that is 4.89 magnitudes fainter in Zorro's 832 nm narrowband filter.}
    \label{fig:hri_toi3321}
\end{figure}

Finally, TOI-3321 and TOI-5232 were observed using the Zorro and 'Alopeke speckle instruments mounted on the Gemini South/North 8-m telescopes \citep{Scott:2021}. Zorro and 'Alopeke both provide simultaneous speckle imaging in two bands (562 nm and 832 nm) with output data products including Fourier analysis and reconstructed images with robust contrast limits on companion detections. Seven sets of 1000 X 0.06 second images were obtained for TOI- 3321 and ten sets of 1000 X 0.06 second images were obtained for TOI-5232. Each of the observations was processed using our standard reduction pipeline (see \citealt{Howell:2011}). Figure \ref{fig:hri_toi3321} shows our final 5$\sigma$ contrast curve and the reconstructed speckle image from the TOI-3321 observation. We find that TOI-3321 has a close companion with a separation of 0.663$\arcsec$ at a Position Angle of 79.1 degrees and is only detected in the 832 nm filter.
The delta magnitude in 832 nm filter, that is the magnitude of the companion relative to the primary star at 832 nm, is 4.89 magnitudes.
TOI-5232 also reveals a companion star at a separation of $1.23 \arcsec$ and Position Angle of 5.2 degrees. This companion was detected in both filters and has delta magnitudes of $5.10 \pm 0.5$ (562 nm) and $4.64 \pm 0.48$ (832 nm) with both delta magnitude values being fairly uncertain due to the large separation ($>$1.0 arcsec) for which the speckles are decorrelated. At the distance of TOI-3321 ($d = 286$ pc) and TOI-5232 ($d = 609$ pc) the angular limits from the diffraction limit (20 mas) out to 1.2 arcsec correspond to spatial limits of 5.7 to 343 AU and 12 to 727 AU respectively.

\subsubsection{Possible Stellar Companions}

Of the nine planet-hosting stars in our sample, three (TOI-1855, TOI-3321, and TOI-5232) have nearby stars that are likely associated with our targets. The likely companions of TOI-1855 and TOI-5232 were both observed by Gaia and have distances and proper motions that are consistent with their probable host stars. TOI-3321's nearby star is faint, with $\Delta$mag $\sim 4.6-4.9$, and is fairly close to TOI-3321, with a separation of $\sim 0.679 \arcsec$. This source doesn't appear to be associated with any of the nearby Gaia DR3 \citep{GaiaDR3} detections. The Gaia Renormalized Unit Weight Error (RUWE), a metric that is used to determine whether or not the astrometric solution of a star is well-behaved for a single source, of this target is 1.283, which is less than the upper threshold of RUWE = 1.4, indicating that the Gaia solution is a well-behaved solution for a single source. A speckle image of TOI-3321 and the corresponding sensitivity curve from Gemini's Zorro speckle imager are shown in Figure \ref{fig:hri_toi3321}, illustrating the detection of TOI-3321's faint companion.

TOI-1855 and its tentative stellar companion have a $\Delta$mag of 1.94 in the Br$\gamma$ filter, which is small enough to influence TOI-1855's lightcurve; however, the $12.352\arcsec$ separation is distant enough that the Bergeron, MuSCAT, MuSCAT2, LCO-CTIO, and SOAR observations were all able to separate the stars and create uncontaminated lightcurves, confirming that the transit signal is not the consequence of a blended eclipsing binary. Additionally, the companion is distant enough that the spectral energy distribution of TOI-1855 (see Figure~\ref{fig:toi1855}) is unaffected by the light from the companion. TOI-3321's tentative stellar companion is fainter than TOI-3321 by 4.89 magnitudes in Gemini's 832 nm filter, which is too faint to significantly influence our lightcurve. Similarly, TOI-5232's tentative stellar companion is fainter than TOI-5232 by 4.64 magnitudes in the same filter and is also too faint to significantly influence the photometry.

\subsection{Archival Multi-Band Observations} \label{subsec:gaia}

In order to constrain the spectral energy distribution (SED) of the stars in our sample, we queried VizieR \citep{Ochsenbein:2000} to obtain broadband observations from Gaia EDR3 \citep{GaiaEDR3}, 2MASS \citep{Cutri:2003, Skrutskie:2006}, and WISE \citep{Wright:2010, Cutri:2012}. All of our targets were observed in the Gaia G, Bp, Rp, 2MASS J, H, Ks, and WISE W1, W2, and W3 bandpasses, which provide wavelength coverage from $\sim 0.3 - 12 \mu$m. In addition, astrometric measurements from Gaia EDR3 were obtained in order to place a Bayesian prior on the parallax of each target (see \S \ref{subsec:priors} for more details). These archival observations were then used in our \exofast global fits to determine relevant stellar parameters (see \S \ref{sec:exofast}). These photometric and astrometric parameters are reported in Table \ref{tab:lit}.

\section{Exofastv2 Global Fits} \label{sec:exofast}

We used the publicly available code \exofast\footnote{\url{https://github.com/jdeast/EXOFASTv2}} \citep{Eastman:2019} to determine all of the system parameters for all nine of the targets in this work. \exofast is a Differential Evolution Markov Chain Monte Carlo (MCMC) code, written in $\tt{IDL}$, designed to globally fit both the star(s) and planet(s) in each system together, ensuring a self-consistent set of parameters describing each system. These fits combine the TESS data (discussed in \S \ref{subsec:TESS}), follow-up transit observations from the ground (discussed in Section \ref{subsec:followup}) and the archival multi-band photometric observations (discussed in Section \ref{subsec:gaia}) to generate a SED model using MESA Isochrones and Stellar Tracks (MIST) \citep{Paxton:2011, Paxton:2013} and a transit model simultaneously at each step of the MCMC. Additional details about the \exofast modeling suite can be found in \cite{Eastman:2019}. 

To test for long-term radial velocity trends, we ran an initial fit for each system, allowing for a linear trend in the radial velocities. If the fit found that the slope of this trend was more than $1\sigma$ away from zero, we fit for the linear trend in the final fit; otherwise, the slope was fixed at zero. There were only two cases in which a tentative long-term radial velocity trend was found: TOI-2368 and TOI-3919. In all of the fits, we allowed the planet's orbital eccentricity to be a free parameter. In doing this, we provide a measurement and estimated uncertainty for the orbital eccentricity of each of the systems in this study, as other derived parameters depend on eccentricity. However, because of the Lucy-Sweeney bias \citep{Lucy:1971}, eccentricities that are smaller than $2.45\sigma$ above zero should be considered consistent with a circular orbit \citep{Eastman:2019}.

The resulting median and 68\% confidence interval of all relevant planetary and stellar parameters for each system are presented in Table \ref{tab:median}. Two of the systems that we fit had bimodal posterior distributions in stellar mass and age. These systems are discussed in further detail in \S \ref{subsec:bimodal}.

\begin{rotatepage}
\movetableright=-1in
\movetabledown=2in
\begin{rotatetable}
\begin{deluxetable*}{l>{\centering}cccccc}
\tablecaption{Median Values and 68\% Confidence Intervals for Fitted Stellar and Planetary Parameters \label{tab:median}}
\tabletypesize{\scriptsize}
\providecommand{\bjdtdb}{\ensuremath{\rm {BJD_{TDB}}}}
\providecommand{\feh}{\ensuremath{\left[{\rm Fe}/{\rm H}\right]}}
\providecommand{\teff}{\ensuremath{T_{\rm eff}}}
\providecommand{\teq}{\ensuremath{T_{\rm eq}}}
\providecommand{\ecosw}{\ensuremath{e\cos{\omega_*}}}
\providecommand{\esinw}{\ensuremath{e\sin{\omega_*}}}
\providecommand{\msun}{\ensuremath{\,M_\Sun}}
\providecommand{\rsun}{\ensuremath{\,R_\Sun}}
\providecommand{\lsun}{\ensuremath{\,L_\Sun}}
\providecommand{\mj}{\ensuremath{\,M_{\rm J}}}
\providecommand{\rj}{\ensuremath{\,R_{\rm J}}}
\providecommand{\me}{\ensuremath{\,M_{\rm E}}}
\providecommand{\re}{\ensuremath{\,R_{\rm E}}}
\providecommand{\fave}{\langle F \rangle}
\providecommand{\fluxcgs}{10$^9$ erg s$^{-1}$ cm$^{-2}$}
\providecommand{\tess}{\textit{TESS}\xspace}
\tablecolumns{7}
\tablehead{&  & \colhead{TOI-1855} & \colhead{TOI-2107} & \colhead{TOI-2368} & \colhead{TOI-3321} & \colhead{TOI-3894}}
\startdata
\multicolumn{7}{l}{\textbf{Priors}:} \\
$\pi$ & Gaia Parallax (mas)& $\mathcal{G}$[5.6565, 0.04552] & $\mathcal{G}$[4.2477, 0.034] & $\mathcal{G}$[4.7622, 0.0262] & $\mathcal{G}$[3.5126, 0.0398] & $\mathcal{G}$[2.4301, 0.0227] \\
$[{\rm Fe/H}]$ & Metallicity (dex)& $\mathcal{G}$[0.4055, 0.085] & $\mathcal{G}$[0.0492, 0.1384] & $\mathcal{G}$[0.0461, 0.2318] & $\mathcal{G}$[-0.3127, 0.1593] & $\mathcal{G}$[-0.1229, 0.1743] \\
$A_V$ & V-band extinction (mag)& $\mathcal{U}$[0, 0.05518] & $\mathcal{U}$[0, 0.15159] & $\mathcal{U}$[0, 5.7474] & $\mathcal{U}$[0, 0.20491] & $\mathcal{U}$[0, 0.03782] \\
$D_T$ & Dilution in \tess& $\mathcal{G}$[0, 0.0060132] & $\mathcal{G}$[0, 0.081363] & $\mathcal{G}$[0, 0.006256] & $\mathcal{G}$[0, 0.00477] & $\mathcal{G}$[0, 0.000040424] \\
\hline
\multicolumn{7}{l}{\textbf{Stellar Parameters}:} \\
$M_*$ & Mass (\msun) & $0.987^{+0.058}_{-0.045}$ & $0.961^{+0.049}_{-0.054}$ & $0.897^{+0.048}_{-0.049}$ & $1.041^{+0.084}_{-0.068}$ & $1.138^{+0.089}_{-0.075}$ \\
$R_*$ & Radius (\rsun) & $1.041^{+0.031}_{-0.033}$ & $0.932^{+0.027}_{-0.024}$ & $0.846^{+0.031}_{-0.026}$ & $1.549^{+0.067}_{-0.062}$ & $1.502^{+0.055}_{-0.052}$ \\
$L_*$ & Luminosity (\lsun) & $0.804^{+0.031}_{-0.03}$ & $0.786^{+0.035}_{-0.036}$ & $0.531^{+0.1}_{-0.065}$ & $2.54^{+0.15}_{-0.16}$ & $2.624^{+0.092}_{-0.083}$ \\
$\rho_*$ & Density (cgs) & $1.23^{+0.15}_{-0.11}$ & $1.68^{+0.12}_{-0.16}$ & $2.1^{+0.19}_{-0.23}$ & $0.396^{+0.07}_{-0.059}$ & $0.475^{+0.065}_{-0.059}$ \\
$\log{g}$ & Surface gravity (cgs) & $4.397^{+0.04}_{-0.031}$ & $4.484^{+0.024}_{-0.034}$ & $4.538^{+0.027}_{-0.038}$ & $4.076^{+0.057}_{-0.053}$ & $4.143^{+0.043}_{-0.046}$ \\
$T_{\rm eff}$ & Effective temperature (K) & $5359.0^{+89}_{-83}$ & $5627.0^{+94}_{-93}$ & $5360.0^{+230}_{-170}$ & $5850.0\pm 140$ & $6000.0\pm 110$ \\
$[{\rm Fe/H}]$ & Metallicity (dex) & $0.399^{+0.069}_{-0.079}$ & $0.07^{+0.12}_{-0.11}$ & $0.1^{+0.18}_{-0.15}$ & $-0.103^{+0.047}_{-0.062}$ & $0.021^{+0.085}_{-0.06}$ \\
$[{\rm Fe/H}]_{0}^\ddagger$ & Initial metallicity & $0.39^{+0.064}_{-0.074}$ & $0.06\pm 0.11$ & $0.09^{+0.17}_{-0.15}$ & $-0.026^{+0.054}_{-0.067}$ & $0.096^{+0.073}_{-0.059}$ \\
Age & Age (Gyr) & $8.2^{+3.5}_{-4}$ & $4.0^{+4}_{-2.7}$ & $4.2^{+5.1}_{-3}$ & $8.1\pm 2.4$ & $5.7^{+2.3}_{-1.8}$ \\
EEP & Equivalent evolutionary phase & $393.0^{+20}_{-43}$ & $343.0^{+32}_{-31}$ & $337.0^{+28}_{-37}$ & $447.2^{+8.3}_{-16}$ & $427.0^{+18}_{-30}$ \\
$A_V$ & V-band extinction (mag) & $0.03^{+0.018}_{-0.02}$ & $0.095^{+0.04}_{-0.056}$ & $0.34^{+0.23}_{-0.2}$ & $0.147^{+0.042}_{-0.073}$ & $0.02^{+0.012}_{-0.013}$ \\
$d$ & Distance (pc) & $176.8\pm 1.4$ & $235.4^{+1.9}_{-1.8}$ & $210.0\pm 1.2$ & $284.8^{+3.3}_{-3.2}$ & $411.9^{+3.9}_{-3.8}$ \\
\multicolumn{7}{l}{\textbf{Planetary Parameters}:} \\
$P$ & Period (days) & $1.36414864\pm 0.00000051$ & $2.4545467\pm 0.0000013$ & $5.1750073\pm 0.0000016$ & $3.6525145^{+0.0000027}_{-0.0000028}$ & $4.3345401\pm 0.0000022$ \\
$R_{\rm P}$ & Radius (\rj) & $1.65^{+0.52}_{-0.37}$ & $1.211^{+0.035}_{-0.032}$ & $0.967^{+0.036}_{-0.031}$ & $1.388^{+0.07}_{-0.064}$ & $1.358^{+0.054}_{-0.052}$ \\
$M_{\rm P}$ & Mass (\mj) & $1.133^{+0.096}_{-0.091}$ & $0.83\pm 0.11$ & $0.65\pm 0.18$ & $0.554^{+0.076}_{-0.075}$ & $0.85\pm 0.15$ \\
$T_C$ & Time of conjunction (\bjdtdb) & $2458929.44324^{+0.00031}_{-0.00032}$ & $2458654.61784^{+0.0004}_{-0.00039}$ & $2459297.19485^{+0.00026}_{-0.00028}$ & $2458681.48898^{+0.00044}_{-0.00041}$ & $2458912.41363^{+0.00041}_{-0.00042}$ \\
$T_0$ & Optimal conjunction time (\bjdtdb) & $2459243.19743^{+0.00028}_{-0.0003}$ & $2459228.98176\pm 0.00024$ & $2459752.59549^{+0.00022}_{-0.00023}$ & $2458940.81751^{+0.00039}_{-0.00036}$ & $2459250.50775^{+0.00036}_{-0.00038}$ \\
$a$ & Semi-major axis (AU) & $0.02398^{+0.00046}_{-0.00037}$ & $0.03515^{+0.00059}_{-0.00067}$ & $0.05649\pm 0.001$ & $0.047^{+0.0012}_{-0.001}$ & $0.0543^{+0.0014}_{-0.0012}$ \\
$i$ & Inclination (Degrees) & $78.1\pm 1$ & $89.44^{+0.39}_{-0.58}$ & $86.61^{+0.19}_{-0.29}$ & $86.33^{+1.3}_{-0.95}$ & $84.86^{+0.43}_{-0.48}$ \\
$e$ & Eccentricity & $0.033^{+0.038}_{-0.023}$ & $0.033^{+0.038}_{-0.023}$ & $0.061^{+0.066}_{-0.042}$ & $0.054^{+0.057}_{-0.038}$ & $0.048^{+0.052}_{-0.034}$ \\
$\omega_*$ & Argument of periastron (Degrees) & $-100.0^{+120}_{-110}$ & $153.0^{+100}_{-97}$ & $121.0^{+63}_{-91}$ & $12.0^{+98}_{-100}$ & $-190.0^{+110}_{-100}$ \\
$\teq$ & Equilibrium temperature (K) & $1701.0\pm 20$ & $1397.0^{+20}_{-18}$ & $1000.0^{+44}_{-31}$ & $1616.0^{+29}_{-28}$ & $1519.0^{+22}_{-21}$ \\
$\tau_{\rm circ}$ & Tidal circularization timescale (Gyr) & $0.00105^{+0.0029}_{-0.00079}$ & $0.0452^{+0.0092}_{-0.0094}$ & $2.43^{+0.98}_{-0.89}$ & $0.085^{+0.032}_{-0.024}$ & $0.332^{+0.1}_{-0.088}$ \\
$K$ & RV semi-amplitude (m/s) & $204.0^{+16}_{-15}$ & $128.0^{+17}_{-16}$ & $81.0\pm 23$ & $71.2^{+9.1}_{-9.2}$ & $96.0\pm 16$ \\
$R_{\rm P}/R_*$ & Radius of planet in stellar radii  & $0.162^{+0.049}_{-0.035}$ & $0.1335^{+0.0012}_{-0.0011}$ & $0.11742^{+0.00069}_{-0.0007}$ & $0.0921\pm 0.0015$ & $0.0929^{+0.00078}_{-0.0008}$ \\
$a/R_*$ & Semi-major axis in stellar radii  & $4.95^{+0.2}_{-0.15}$ & $8.13^{+0.2}_{-0.27}$ & $14.37^{+0.42}_{-0.55}$ & $6.54^{+0.36}_{-0.34}$ & $7.79\pm 0.34$ \\
Depth & \tess flux decrement at mid-transit & $0.00707^{+0.00028}_{-0.0003}$ & $0.02131^{+0.00054}_{-0.00053}$ & $0.01283\pm 0.00011$ & $0.00966^{+0.00029}_{-0.00028}$ & $0.008974^{+0.000089}_{-0.000091}$ \\
$\tau$ & Ingress/egress transit duration (days) & $0.02489^{+0.00025}_{-0.00027}$ & $0.01288^{+0.00029}_{-0.00017}$ & $0.02452^{+0.00088}_{-0.0009}$ & $0.0182^{+0.0025}_{-0.0022}$ & $0.0232^{+0.0019}_{-0.0018}$ \\
$T_{14}$ & Total transit duration (days) & $0.04977^{+0.00051}_{-0.00053}$ & $0.10815^{+0.00078}_{-0.00076}$ & $0.08439^{+0.00062}_{-0.00064}$ & $0.1796^{+0.0026}_{-0.0023}$ & $0.1495^{+0.0017}_{-0.0016}$ \\
$b$ & Transit impact parameter & $1.024^{+0.061}_{-0.049}$ & $0.078^{+0.081}_{-0.055}$ & $0.8239^{+0.0056}_{-0.0062}$ & $0.42^{+0.097}_{-0.15}$ & $0.695^{+0.025}_{-0.028}$ \\
$\rho_{\rm P}$ & Density (cgs) & $0.31^{+0.37}_{-0.18}$ & $0.577^{+0.093}_{-0.086}$ & $0.88^{+0.27}_{-0.26}$ & $0.255^{+0.057}_{-0.049}$ & $0.418^{+0.093}_{-0.083}$ \\
$\log{g_{\rm P}}$ & Surface gravity (cgs) & $3.01^{+0.23}_{-0.24}$ & $3.145^{+0.059}_{-0.065}$ & $3.23^{+0.11}_{-0.15}$ & $2.851^{+0.074}_{-0.078}$ & $3.056^{+0.078}_{-0.088}$ \\
$M_{\rm P}/M_*$ & Mass ratio  & $0.001092^{+0.000088}_{-0.000085}$ & $0.00082^{+0.00011}_{-0.0001}$ & $0.00069\pm 0.0002$ & $0.000507^{+0.000066}_{-0.000067}$ & $0.00071\pm 0.00012$ \\
$d/R_*$ & Separation at mid-transit  & $4.97^{+0.36}_{-0.26}$ & $8.08^{+0.36}_{-0.53}$ & $13.95^{+0.84}_{-1.1}$ & $6.51^{+0.57}_{-0.56}$ & $7.75^{+0.58}_{-0.62}$ \\
\enddata

\end{deluxetable*}
\end{rotatetable}
\addtocounter{table}{-1} 
\end{rotatepage}

\begin{rotatepage}
\movetableright=-1in
\movetabledown=1.2in
\begin{rotatetable}
\begin{deluxetable*}{l>{\centering}cccccc}
\tablecaption{\textit{(Continued)}}
\tabletypesize{\scriptsize}
\providecommand{\bjdtdb}{\ensuremath{\rm {BJD_{TDB}}}}
\providecommand{\feh}{\ensuremath{\left[{\rm Fe}/{\rm H}\right]}}
\providecommand{\teff}{\ensuremath{T_{\rm eff}}}
\providecommand{\teq}{\ensuremath{T_{\rm eq}}}
\providecommand{\ecosw}{\ensuremath{e\cos{\omega_*}}}
\providecommand{\esinw}{\ensuremath{e\sin{\omega_*}}}
\providecommand{\msun}{\ensuremath{\,M_\Sun}}
\providecommand{\rsun}{\ensuremath{\,R_\Sun}}
\providecommand{\lsun}{\ensuremath{\,L_\Sun}}
\providecommand{\mj}{\ensuremath{\,M_{\rm J}}}
\providecommand{\rj}{\ensuremath{\,R_{\rm J}}}
\providecommand{\me}{\ensuremath{\,M_{\rm E}}}
\providecommand{\re}{\ensuremath{\,R_{\rm E}}}
\providecommand{\fave}{\langle F \rangle}
\providecommand{\fluxcgs}{10$^9$ erg s$^{-1}$ cm$^{-2}$}
\providecommand{\tess}{\textit{TESS}\xspace}
\tablecolumns{6}
\tablehead{&  & \colhead{TOI-3919} & \colhead{TOI-4153} & \colhead{TOI-5232} & \colhead{TOI-5301}}
\startdata
\multicolumn{6}{l}{\textbf{Priors}:} \\
$\pi$ & Gaia Parallax (mas)& $\mathcal{G}$[1.6518, 0.0242] & $\mathcal{G}$[2.3867, 0.0201] & $\mathcal{G}$[1.6415, 0.0231] & $\mathcal{G}$[1.6998, 0.0501] \\
$[{\rm Fe/H}]$ & Metallicity (dex)& $\mathcal{G}$[0.2386, 0.1956] & $\mathcal{G}$[0.287, 0.1772] & $\mathcal{G}$[0.0365, 0.332] & $\mathcal{G}$[-0.1322, 0.1806] \\
$A_V$ & V-band extinction (mag)& $\mathcal{U}$[0, 0.02666] & $\mathcal{U}$[0, 0.8091] & $\mathcal{U}$[0, 0.434] & $\mathcal{U}$[0, 0.18414] \\
$D_T$ & Dilution in \tess& $\mathcal{G}$[0, 0.00047478] & $\mathcal{G}$[0, 0.00080667] & $\mathcal{G}$[0, 0.028318] & $\mathcal{G}$[0, 0.00040193] \\
\hline
\multicolumn{6}{l}{\textbf{Stellar Parameters}:} \\
$M_*$ & Mass (\msun) & $1.208^{+0.067}_{-0.07}$ & $1.572^{+0.064}_{-0.071}$ & $1.389^{+0.068}_{-0.075}$ & $1.483^{+0.081}_{-0.14}$ \\
$R_*$ & Radius (\rsun) & $1.319^{+0.052}_{-0.048}$ & $1.605^{+0.048}_{-0.046}$ & $1.785^{+0.067}_{-0.057}$ & $2.19^{+0.12}_{-0.11}$ \\
$L_*$ & Luminosity (\lsun) & $2.168^{+0.079}_{-0.077}$ & $5.17^{+0.42}_{-0.55}$ & $5.16^{+0.46}_{-0.53}$ & $6.55^{+0.55}_{-0.5}$ \\
$\rho_*$ & Density (cgs) & $0.742^{+0.1}_{-0.093}$ & $0.537^{+0.044}_{-0.046}$ & $0.344^{+0.038}_{-0.042}$ & $0.197^{+0.036}_{-0.032}$ \\
$\log{g}$ & Surface gravity (cgs) & $4.279^{+0.042}_{-0.044}$ & $4.224^{+0.024}_{-0.029}$ & $4.077^{+0.034}_{-0.042}$ & $3.926^{+0.052}_{-0.061}$ \\
$T_{\rm eff}$ & Effective temperature (K) & $6100.0\pm 110$ & $6860.0^{+150}_{-180}$ & $6500.0^{+180}_{-190}$ & $6240.0\pm 160$ \\
$[{\rm Fe/H}]$ & Metallicity (dex) & $0.175^{+0.096}_{-0.094}$ & $0.317^{+0.1}_{-0.13}$ & $-0.104^{+0.11}_{-0.086}$ & $0.03^{+0.13}_{-0.11}$ \\
$[{\rm Fe/H}]_{0}^\ddagger$ & Initial metallicity & $0.208^{+0.08}_{-0.078}$ & $0.39^{+0.072}_{-0.098}$ & $0.04^{+0.093}_{-0.084}$ & $0.11^{+0.12}_{-0.1}$ \\
Age & Age (Gyr) & $3.1^{+1.9}_{-1.7}$ & $0.4^{+0.53}_{-0.28}$ & $2.37^{+0.77}_{-0.56}$ & $2.43^{+1.1}_{-0.47}$ \\
EEP & Equivalent evolutionary phase & $368.0^{+40}_{-35}$ & $308.0^{+24}_{-38}$ & $378.0^{+24}_{-23}$ & $401.0^{+49}_{-18}$ \\
$A_V$ & V-band extinction (mag) & $0.0142^{+0.0086}_{-0.0094}$ & $0.686^{+0.083}_{-0.12}$ & $0.357^{+0.055}_{-0.097}$ & $0.136^{+0.036}_{-0.066}$ \\
$d$ & Distance (pc) & $605.2^{+8.8}_{-8.7}$ & $419.1\pm 3.5$ & $609.0^{+8.6}_{-8.3}$ & $587.0^{+18}_{-17}$ \\
\multicolumn{6}{l}{\textbf{Planetary Parameters}:} \\
$P$ & Period (days) & $7.433234\pm 0.000014$ & $4.6174141\pm 0.0000015$ & $4.0966692\pm 0.0000067$ & $5.858858\pm 0.000016$ \\
$R_{\rm P}$ & Radius (\rj) & $1.099^{+0.052}_{-0.05}$ & $1.438^{+0.045}_{-0.042}$ & $1.14^{+0.051}_{-0.045}$ & $1.177^{+0.079}_{-0.072}$ \\
$M_{\rm P}$ & Mass (\mj) & $3.88\pm 0.23$ & $1.15\pm 0.18$ & $2.34\pm 0.16$ & $3.65^{+0.38}_{-0.39}$ \\
$T_C$ & Time of conjunction (\bjdtdb) & $2458954.374\pm 0.0013$ & $2459026.05484\pm 0.00022$ & $2459440.0678^{+0.00086}_{-0.00087}$ & $2459494.1476^{+0.0011}_{-0.0012}$ \\
$T_0$ & Optimal conjunction time (\bjdtdb) & $2459519.29976^{+0.00081}_{-0.0008}$ & $2459557.05746\pm 0.00015$ & $2459857.92806^{+0.00052}_{-0.00053}$ & $2459277.3698\pm 0.001$ \\
$a$ & Semi-major axis (AU) & $0.0795^{+0.0014}_{-0.0016}$ & $0.06311^{+0.00084}_{-0.00096}$ & $0.05593^{+0.0009}_{-0.001}$ & $0.0726^{+0.0013}_{-0.0024}$ \\
$i$ & Inclination (Degrees) & $86.98^{+0.2}_{-0.23}$ & $88.75^{+0.76}_{-0.6}$ & $88.3^{+1.1}_{-1.3}$ & $83.1^{+1.8}_{-1.2}$ \\
$e$ & Eccentricity & $0.259^{+0.033}_{-0.036}$ & $0.039^{+0.047}_{-0.028}$ & $0.035^{+0.034}_{-0.024}$ & $0.33^{+0.11}_{-0.1}$ \\
$\omega_*$ & Argument of periastron (Degrees) & $-65.8^{+4.8}_{-3.4}$ & $-165.0^{+110}_{-100}$ & $46.0^{+61}_{-77}$ & $75.0^{+13}_{-14}$ \\
$\teq$ & Equilibrium temperature (K) & $1198.0^{+15}_{-14}$ & $1669.0^{+29}_{-39}$ & $1772.0^{+40}_{-45}$ & $1655.0^{+39}_{-34}$ \\
$\tau_{\rm circ}$ & Tidal circularization timescale (Gyr) & $22.5^{+6.3}_{-5}$ & $0.56\pm 0.13$ & $2.03^{+0.5}_{-0.47}$ & $3.8^{+3.6}_{-2.2}$ \\
$K$ & RV semi-amplitude (m/s) & $367.0^{+18}_{-17}$ & $103.0\pm 16$ & $239.0^{+15}_{-14}$ & $337.0^{+35}_{-33}$ \\
$R_{\rm P}/R_*$ & Radius of planet in stellar radii  & $0.0856\pm 0.0017$ & $0.0921^{+0.00039}_{-0.00033}$ & $0.0656\pm 0.0014$ & $0.0552^{+0.0015}_{-0.0014}$ \\
$a/R_*$ & Semi-major axis in stellar radii  & $12.95\pm 0.56$ & $8.46^{+0.22}_{-0.25}$ & $6.74^{+0.24}_{-0.28}$ & $7.11\pm 0.41$ \\
Depth & \tess flux decrement at mid-transit & $0.00695\pm 0.00026$ & $0.009323^{+0.000071}_{-0.000069}$ & $0.0048\pm 0.00021$ & $0.003254^{+0.000095}_{-0.000094}$ \\
$\tau$ & Ingress/egress transit duration (days) & $0.0371^{+0.0036}_{-0.0033}$ & $0.01642^{+0.0008}_{-0.0005}$ & $0.01282^{+0.0012}_{-0.00054}$ & $0.0134^{+0.0044}_{-0.0038}$ \\
$T_{14}$ & Total transit duration (days) & $0.1606^{+0.0031}_{-0.0029}$ & $0.18801^{+0.00074}_{-0.0006}$ & $0.1993^{+0.0016}_{-0.0014}$ & $0.1645^{+0.0047}_{-0.0042}$ \\
$b$ & Transit impact parameter & $0.835^{+0.014}_{-0.016}$ & $0.184^{+0.091}_{-0.11}$ & $0.19^{+0.14}_{-0.13}$ & $0.61^{+0.11}_{-0.23}$ \\
$\rho_{\rm P}$ & Density (cgs) & $3.63^{+0.6}_{-0.53}$ & $0.477^{+0.087}_{-0.085}$ & $1.96^{+0.28}_{-0.27}$ & $2.77^{+0.68}_{-0.57}$ \\
$\log{g_{\rm P}}$ & Surface gravity (cgs) & $3.901^{+0.048}_{-0.051}$ & $3.137^{+0.067}_{-0.08}$ & $3.649^{+0.043}_{-0.048}$ & $3.814^{+0.072}_{-0.076}$ \\
$M_{\rm P}/M_*$ & Mass ratio  & $0.00307^{+0.00016}_{-0.00015}$ & $0.0007\pm 0.00011$ & $0.00161^{+0.00011}_{-0.0001}$ & $0.00238\pm 0.00023$ \\
$d/R_*$ & Separation at mid-transit  & $15.8^{+1}_{-1.1}$ & $8.48\pm 0.46$ & $6.64^{+0.37}_{-0.47}$ & $4.81^{+0.86}_{-0.84}$ \\
\enddata

\begin{flushleft}
 \footnotesize{\textbf{NOTES:}\\
 The priors listed at the top of the table are labeled as $\mathcal{G}$[mean, standard deviation] if they are Gaussian priors and $\mathcal{U}$[lower limit, upper limit] if they are uniform priors.\\
 $\dagger$ TOI-1855's transit is grazing, and as a consequence, the radius measurement is highly uncertain.\\
 $\ddagger$ Initial metallicity represents the metallicity of the star at formation.\\
}
\end{flushleft}
\end{deluxetable*}
\end{rotatetable}
\end{rotatepage}

\subsection{Priors and Constraints} \label{subsec:priors}

In all nine fits, we adopted Gaussian priors for the Gaia parallax measurements from Early Data Release 3 \citep{GaiaEDR3}. We also placed wide Gaussian priors on the stellar metallicity, using averaged metallicity measurements from the TRES and CHIRON spectra, with a prior width equal to twice the standard deviation of the spectroscopic metallicities. Finally, we account for the possible contamination of the \TESS target pixel by fitting a dilution term and placing a Gaussian prior centered at 0\% with a standard deviation that is 10\% of the contamination ratio from the TESS Input Catalog (TIC) v8.2 \citep{Stassun:2018, Stassun:2019}. This contamination is already corrected by both the QLP and SPOC pipelines (see \S \ref{subsec:TESS}), so this prior acts as a conservative assumption that the contamination correction had a precision better than 10\%. Finally, we placed an upper limit on the V-band extinction of each target using the \cite{Schlegel:1998} and \cite{Schlafly:2011} dustmaps. Each of these priors is included at the top of Table \ref{tab:median}.

In addition to the supplied Gaussian priors, we adopt starting values for the stellar mass, stellar radius, and stellar effective temperature from the TIC. To provide a constraint on the stellar radius, we apply an upper limit on the V-band extinction from \cite{Schlafly:2011}. Finally, the starting values for the transit epoch $T_C$, orbital period $P$, and the ratio of planetary and stellar radii $R_P/R_*$ for our initial fit are retrieved from the \TESS mission catalog on ExoFOP\footnote{\url{https://exofop.ipac.caltech.edu/tess/}}.

\begin{figure}
    \centering
    \includegraphics[width=0.95\linewidth]{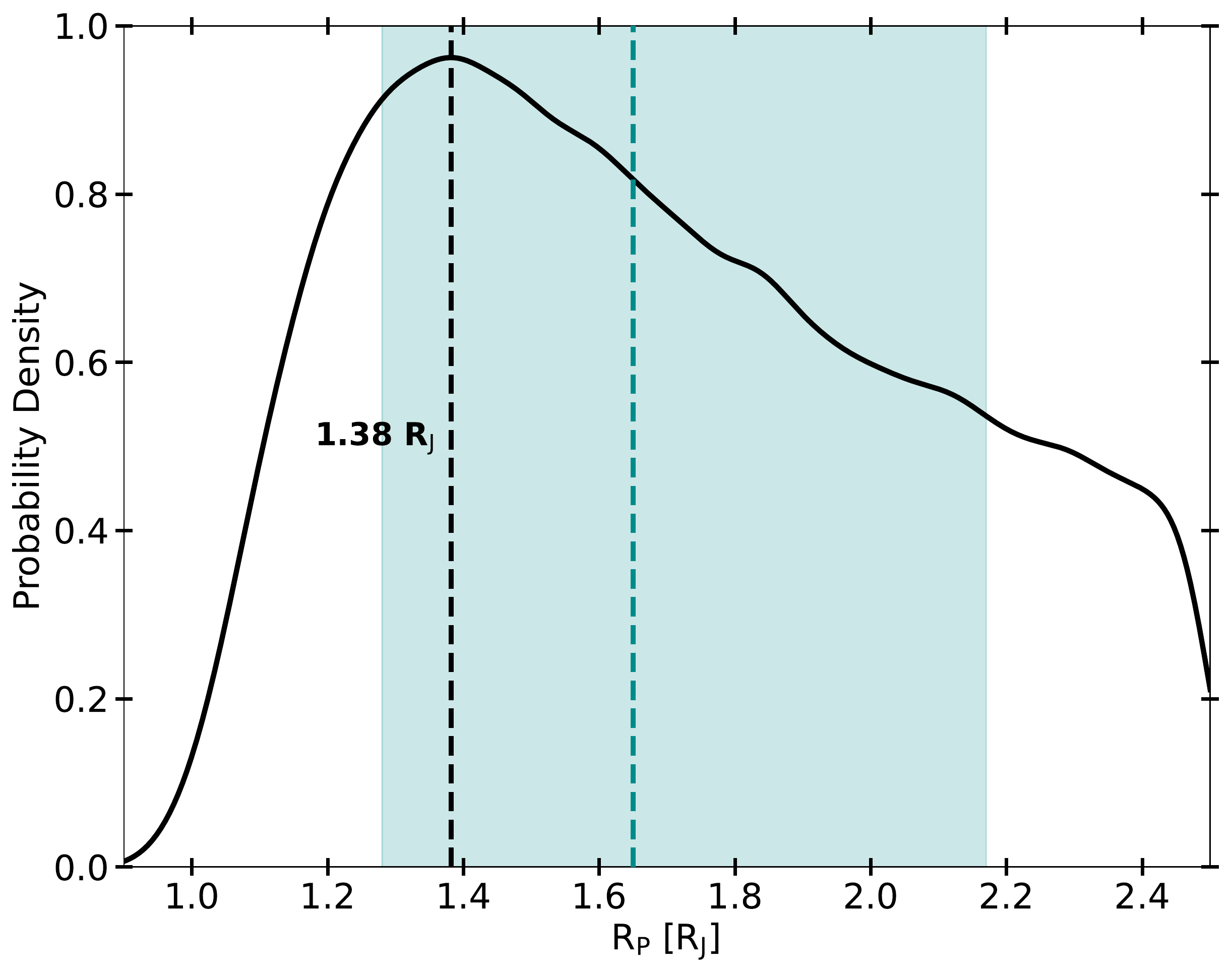}
    \caption{Posterior probability density function (PDF) of TOI-1855 b's radius. The black, labeled, vertical dashed line represents the posterior mode, 1.38 \rj. The cyan vertical dashed line and shaded region represent the median and 68\% confidence interval, 1.65$^{+0.52}_{-0.37}$ \rj. The allowed planetary radius was constrained by an upper limit of 2.5 \rj.}
    \label{fig:grazing_pdf}
\end{figure}

One of our targets, TOI-1855, has a grazing transit configuration, in which the planet incompletely transits the limb of the star. As a result, the radius of this planet is poorly constrained. To account for the effectively flat radius posterior and to reduce the computational cost of the fit, we placed a generous physical upper limit on the radius of the planet, equal to $2.5 \rm{R}_{\rm J}$. This upper limit is significantly larger than the largest nongrazing, transiting planet \citep{Smalley:2012}, ensuring that our interpretation is not affected by the placement of a strict upper limit. The upper limit does, however, likely affect the median of the R$_{\rm P}$ posterior listed in Table \ref{tab:median}, and so for that reason, we also highlight the posterior mode, 1.38 \rj, as the most likely radius of the planet, illustrated with a vertical dashed line in Figure \ref{fig:grazing_pdf}. A more detailed discussion of TOI-1855 and our interpretation of the radius distribution is included in \S \ref{subsec:toi1855}.

\section{Results and Discussion} \label{sec:results}

In this first installment of the Migration and Evolution of giant ExoPlanets (MEEP) survey, we confirm nine new hot Jupiters (TOI-1855 b, TOI-2107 b, TOI-2368 b, TOI-3321 b, TOI-3894 b, TOI-3919 b, TOI-4153 b, TOI-5232 b, and TOI-5301 b) detected by \TESS, orbiting FGK stars. These planets range in mass between 0.554 \mj and 3.88 \mj and range in size between 0.967 \rj and 1.438 \rj. The stars they orbit are all brighter than $G = 12.7$, which makes them suitable for further follow-up from a variety of ground-based facilities.

The transit photometry, radial velocity, spectral energy distribution, and MIST evolutionary plots for each system investigated within this work are organized by system and presented in Figures \ref{fig:toi1855}-\ref{fig:toi5301}. The relevant stellar and planetary parameters from our \exofast global fits are presented in Table \ref{tab:median}.

\subsection{TOI-1855 b's grazing transit} \label{subsec:toi1855}

One of our planets, TOI-1855 b, exhibits a V-shaped transit caused by the planet only grazing the limb of the star. This transit shape is similar to what is usually seen for an eclipsing binary. Because of this, we paid special attention to TOI-1855 to ensure that our interpretation of it as a transiting giant planet is correct. TOI-1855 was observed by three different speckle instruments with four different filters and by two different AO instruments in three different filters. These observations detected a stellar companion, previously identified by Gaia (TIC 81247738), at a separation of $12.352\arcsec$ with a \TESS magnitude of 13.524 ($\Delta T = 2.82$ mag). No other stellar companions were detected by any of the six HRI instruments. While these sources are blended in the \TESS observations, the Bergeron, MuSCAT, MuSCAT2, LCO-CTIO, and SOAR observations used uncontaminated apertures and confirmed the transit signal to be on target. Additionally, the \TESS data are corrected for contamination by both the SPOC and QLP pipelines (see \S \ref{subsec:TESS}) and we leave the dilution term for the \TESS lightcurves as a free parameter in our \exofast global fits (see \S \ref{sec:exofast}) to account for possible errors in the dilution correction.

Another indicator of a blended eclipsing binary that is visible in the \TESS data is a centroid offset, where the position of the target star appears to change as the stars enter an eclipse \citep{Batalha:2010}. We find from the data validation report\footnote{\url{https://tev.mit.edu/data/delivered-signal/i244634/}} for the SPOC-processed \TESS Sector 50 data that the measured centroid offset is $0.408 \pm 2.55\arcsec$, well within the expected variance and therefore consistent with no centroid offset. Finally, we searched the \TESS data for evidence of a secondary eclipse, another sign of an eclipsing binary, by phasing the \TESS data on the optimal ephemeris from our \exofast global fit and binning the phased lightcurve. We find no evidence of a secondary eclipse in either sector of data. The combination of an RV orbit, HRI non-detection, secondary eclipse non-detection, and the lack of a centroid offset imply that it is highly unlikely that TOI-1855 b's transit signal is the consequence of an eclipsing binary.

The other difficulty raised by the grazing configuration of TOI-1855 b's transit is that we are unable to precisely constrain its radius from the transit alone. In \S \ref{subsec:priors}, we discussed the R$_{\rm P}$ upper limit placed on TOI-1855's fit and the resulting posterior mode, 1.38 \rj. In order to get a different constraint on the radius of TOI-1855 b, we used the code \texttt{Forecaster}\footnote{\url{https://github.com/chenjj2/forecaster}} from \cite{ChenKipping:2017} to generate a PDF of the radius of TOI-1855 given its mass distribution from our \exofast global fit. From this mass-radius relation alone, we find a median planetary radius of $1.21 \pm 0.21$ \rj, which is consistent within 1$\sigma$ with the posterior mode (1.38 \rj) and the median ($1.65^{+0.52}_{-0.37}$) from the fit. \texttt{Forecaster} is built from a sample of planets with a wide range of isolations, but it is possible that TOI-1855 b's size could be underestimated by using only \texttt{Forecaster} if its radius is inflated by insolation.

\subsection{Bimodal Solutions} \label{subsec:bimodal}

\begin{figure*}[ht!]
    \centering
    \subfigure{\includegraphics[width=0.49\textwidth]{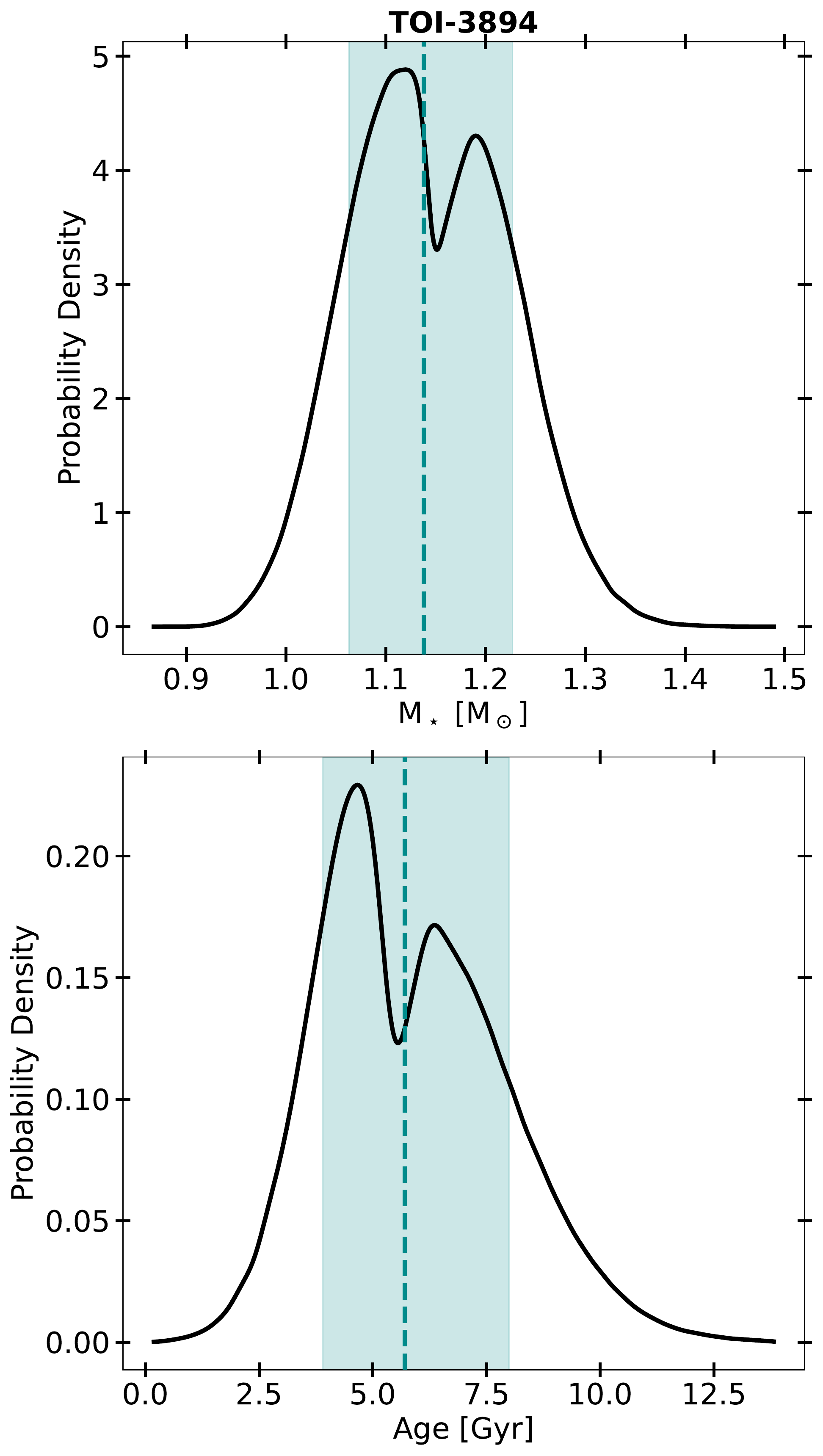}} 
    \subfigure{\includegraphics[width=0.49\textwidth]{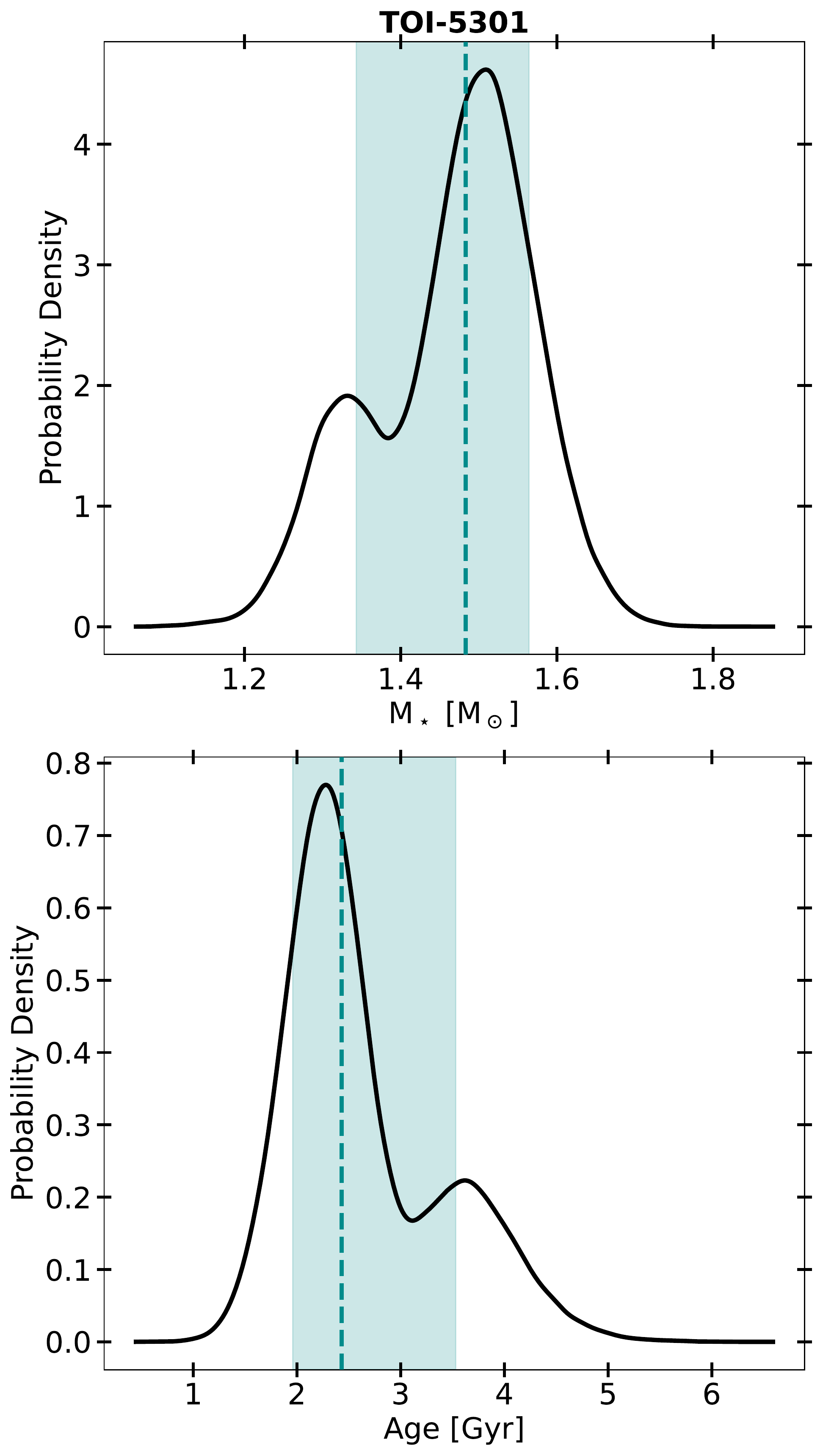}} 
    \caption{Gaussian kernel density estimations of TOI-3894 \textbf{(left)} and TOI-5301's \textbf{(right)} stellar mass and age posterior distributions from our \exofast global fits, illustrating a bimodal distribution in both parameters. The cyan dashed line indicates the median value of each parameter, while the shaded region represents the 68\% confidence region. The normalized probability densities were estimated using the Gaussian kernel density estimator in the Python package \texttt{scipy.stats}. TOI-3894's fit slightly favored the lower mass solution with a probability of 53.0\% while TOI-5301's fit favored the higher mass solution with a probability of 78.3\%.}
    \label{fig:bimodal}
\end{figure*}

Two of our \exofast global fits converged on solutions that had bimodal posterior distributions in stellar mass and age: TOI-3894 and TOI-5301. In order to characterize each individual solution from both fits, we split the probability density functions of TOI-3894's and TOI-5301's fits, identifying a solution with a higher stellar mass and a separate solution with a lower stellar mass for each (see Figure \ref{fig:bimodal}). For TOI-3894, if we separate the PDF using a stellar mass of 1.145 \msol (the center of the valley between the two peaks in Figure \ref{fig:bimodal}), the \exofast global fit slightly favors the lower mass solution (M$_\star = 1.089^{+0.038}_{-0.051}$ \msol, age = $7.2^{+1.7}_{-1.1}$ Gyr) with a probability of 53.0\%. In the case of TOI-5301, we split the PDF using a stellar mass of 1.375 \msol and find that the higher mass solution (M$_\star = 1.507^{+0.068}_{-0.067}$ \msol, age = $2.29^{+0.42}_{-0.4}$ Gyr) is favored with a probability of 78.3\%. Both of these uncertain fits are consequences of the stars' observed parameters indicating that they are near an evolutionary transition. To encapsulate both possibilities in both systems, we present the median and standard deviation of each parameter for the higher mass and lower mass solutions of both targets in Table \ref{tab:split_medians}.

\begin{deluxetable*}{l>{\centering}cccccc}
\tablecaption{Median Values and 68\% Confidence Intervals for Bimodal Fits \label{tab:split_medians}}
\tabletypesize{\scriptsize}
\providecommand{\bjdtdb}{\ensuremath{\rm {BJD_{TDB}}}}
\providecommand{\feh}{\ensuremath{\left[{\rm Fe}/{\rm H}\right]}}
\providecommand{\teff}{\ensuremath{T_{\rm eff}}}
\providecommand{\teq}{\ensuremath{T_{\rm eq}}}
\providecommand{\ecosw}{\ensuremath{e\cos{\omega_*}}}
\providecommand{\esinw}{\ensuremath{e\sin{\omega_*}}}
\providecommand{\msun}{\ensuremath{\,M_\Sun}}
\providecommand{\rsun}{\ensuremath{\,R_\Sun}}
\providecommand{\lsun}{\ensuremath{\,L_\Sun}}
\providecommand{\mj}{\ensuremath{\,M_{\rm J}}}
\providecommand{\rj}{\ensuremath{\,R_{\rm J}}}
\providecommand{\me}{\ensuremath{\,M_{\rm E}}}
\providecommand{\re}{\ensuremath{\,R_{\rm E}}}
\providecommand{\fave}{\langle F \rangle}
\providecommand{\fluxcgs}{10$^9$ erg s$^{-1}$ cm$^{-2}$}
\providecommand{\tess}{\textit{TESS}\xspace}
\tablecolumns{6}
\tablehead{& & \multicolumn{2}{c}{\textbf{TOI-3894}} & \multicolumn{2}{c}{\textbf{TOI-5301}} \\ &  & \colhead{Low-mass solution} & \colhead{High-mass solution} & \colhead{Low-mass solution} & \colhead{High-mass solution} \\ & & (53.0\% probability) & (47.0\% probability) & (21.7\% probability) & (78.3\% probability)}
\startdata
\hline
\multicolumn{6}{l}{\textbf{Stellar Parameters}:} \\
$M_*$ & Mass (\msun) & $1.089^{+0.038}_{-0.051}$ & $1.206^{+0.052}_{-0.039}$ & $1.316^{+0.039}_{-0.05}$ & $1.507^{+0.068}_{-0.067}$ \\
$R_*$ & Radius (\rsun) & $1.506^{+0.057}_{-0.052}$ & $1.498^{+0.053}_{-0.054}$ & $2.23^{+0.12}_{-0.11}$ & $2.18\pm 0.11$ \\
$L_*$ & Luminosity (\lsun) & $2.621^{+0.094}_{-0.083}$ & $2.627^{+0.091}_{-0.082}$ & $6.48^{+0.59}_{-0.52}$ & $6.57^{+0.54}_{-0.49}$ \\
$\rho_*$ & Density (cgs) & $0.447^{+0.054}_{-0.051}$ & $0.507^{+0.062}_{-0.05}$ & $0.167^{+0.026}_{-0.024}$ & $0.205^{+0.034}_{-0.028}$ \\
$\log{g}$ & Surface gravity (cgs) & $4.117^{+0.035}_{-0.039}$ & $4.169^{+0.036}_{-0.031}$ & $3.859^{+0.042}_{-0.046}$ & $3.939^{+0.046}_{-0.042}$ \\
$T_{\rm eff}$ & Effective temperature (K) & $5990.0\pm 110$ & $6010.0\pm 110$ & $6170.0\pm 160$ & $6260.0^{+160}_{-150}$ \\
$[{\rm Fe/H}]$ & Metallicity (dex) & $0.01^{+0.073}_{-0.054}$ & $0.038^{+0.092}_{-0.069}$ & $-0.02^{+0.1}_{-0.11}$ & $0.05^{+0.13}_{-0.11}$ \\
$[{\rm Fe/H}]_{0}$ & Initial metallicity & $0.08^{+0.064}_{-0.055}$ & $0.117^{+0.075}_{-0.062}$ & $0.036^{+0.085}_{-0.095}$ & $0.13^{+0.12}_{-0.099}$ \\
$Age$ & Age (Gyr) & $7.2^{+1.7}_{-1.1}$ & $4.27^{+0.74}_{-1}$ & $3.78^{+0.49}_{-0.39}$ & $2.29^{+0.42}_{-0.4}$ \\
$EEP$ & Equal evolutionary phase & $440.8^{+7.9}_{-9.3}$ & $405.0^{+13}_{-23}$ & $453.7^{+4.1}_{-6.3}$ & $397.0^{+11}_{-17}$ \\
$A_V$ & V-band extinction (mag) & $0.019\pm 0.013$ & $0.021^{+0.012}_{-0.014}$ & $0.113^{+0.051}_{-0.068}$ & $0.141^{+0.032}_{-0.062}$ \\
$d$ & Distance (pc) & $411.6^{+3.9}_{-3.8}$ & $412.1^{+3.9}_{-3.8}$ & $585.0^{+17}_{-16}$ & $588.0^{+18}_{-17}$ \\
\multicolumn{6}{l}{\textbf{Planetary Parameters}:} \\
$P$ & Period (days) & $4.33454\pm 0.0000022$ & $4.3345401\pm 0.0000022$ & $5.858857^{+0.000016}_{-0.000017}$ & $5.858858\pm 0.000016$ \\
$R_P$ & Radius (\rj) & $1.363^{+0.055}_{-0.051}$ & $1.352^{+0.053}_{-0.052}$ & $1.207^{+0.081}_{-0.073}$ & $1.168^{+0.076}_{-0.07}$ \\
$M_P$ & Mass (\mj) & $0.82\pm 0.14$ & $0.88\pm 0.15$ & $3.4\pm 0.33$ & $3.72\pm 0.37$ \\
$T_C$ & Time of conjunction (\bjdtdb) & $2458912.41365^{+0.00041}_{-0.00043}$ & $2458912.41361^{+0.0004}_{-0.00042}$ & $2459494.1476^{+0.0012}_{-0.0013}$ & $2459494.1476^{+0.0011}_{-0.0012}$ \\
$T_0$ & Optimal conjunction time (\bjdtdb) & $2459250.50777^{+0.00037}_{-0.00038}$ & $2459250.50773^{+0.00035}_{-0.00038}$ & $2459277.3699^{+0.001}_{-0.0011}$ & $2459277.36981\pm 0.00099$ \\
$a$ & Semi-major axis (AU) & $0.05354^{+0.00061}_{-0.00085}$ & $0.05539^{+0.00079}_{-0.0006}$ & $0.06975^{+0.00068}_{-0.00089}$ & $0.073\pm 0.0011$ \\
$i$ & Inclination (Degrees) & $84.63^{+0.4}_{-0.46}$ & $85.08^{+0.37}_{-0.36}$ & $82.0^{+1.4}_{-1.1}$ & $83.3^{+1.9}_{-1}$ \\
$e$ & Eccentricity & $0.05^{+0.052}_{-0.035}$ & $0.046^{+0.052}_{-0.032}$ & $0.337^{+0.12}_{-0.099}$ & $0.33^{+0.11}_{-0.1}$ \\
$\omega_*$ & Argument of periastron (Degrees) & $108.0^{+89}_{-93}$ & $-140.0\pm 100$ & $76.0\pm 13$ & $75.0^{+13}_{-14}$ \\
$T_{eq}$ & Equilibrium temperature (K) & $1531.0^{+17}_{-15}$ & $1504.0\pm 15$ & $1682.0^{+38}_{-34}$ & $1648.0^{+34}_{-31}$ \\
$\tau_{\rm circ}$ & Tidal circularization timescale (Gyr) & $0.304^{+0.093}_{-0.081}$ & $0.365^{+0.1}_{-0.086}$ & $2.7^{+2.5}_{-1.7}$ & $4.2^{+3.8}_{-2.4}$ \\
$K$ & RV semi-amplitude (m/s) & $96.0\pm 16$ & $96.0\pm 16$ & $338.0^{+36}_{-33}$ & $336.0^{+34}_{-33}$ \\
$R_P/R_*$ & Radius of planet in stellar radii  & $0.09302^{+0.00077}_{-0.00079}$ & $0.09275^{+0.00077}_{-0.00078}$ & $0.0557^{+0.0015}_{-0.0016}$ & $0.0551^{+0.0015}_{-0.0014}$ \\
$a/R_*$ & Semi-major axis in stellar radii  & $7.63^{+0.29}_{-0.3}$ & $7.96^{+0.31}_{-0.27}$ & $6.72^{+0.33}_{-0.34}$ & $7.2^{+0.38}_{-0.34}$ \\
$Depth$ & \tess flux decrement at mid-transit & $0.008978^{+0.000089}_{-0.00009}$ & $0.008969^{+0.00009}_{-0.000091}$ & $0.003269^{+0.000093}_{-0.000096}$ & $0.00325^{+0.000095}_{-0.000093}$ \\
$\tau$ & Ingress/egress transit duration (days) & $0.0235\pm 0.0018$ & $0.0228^{+0.0018}_{-0.0017}$ & $0.0149^{+0.0045}_{-0.0043}$ & $0.013^{+0.0042}_{-0.0036}$ \\
$T_{14}$ & Total transit duration (days) & $0.1498\pm 0.0016$ & $0.1492^{+0.0016}_{-0.0015}$ & $0.166\pm 0.0047$ & $0.1641^{+0.0046}_{-0.0041}$ \\
$b$ & Transit impact parameter & $0.7^{+0.023}_{-0.028}$ & $0.689^{+0.025}_{-0.028}$ & $0.655^{+0.088}_{-0.19}$ & $0.6^{+0.11}_{-0.23}$ \\
$\rho_P$ & Density (cgs) & $0.4^{+0.087}_{-0.078}$ & $0.44^{+0.095}_{-0.084}$ & $2.39^{+0.55}_{-0.47}$ & $2.88^{+0.67}_{-0.54}$ \\
$logg_P$ & Surface gravity  & $3.039^{+0.076}_{-0.087}$ & $3.076^{+0.075}_{-0.085}$ & $3.76^{+0.067}_{-0.07}$ & $3.828^{+0.067}_{-0.069}$ \\
$M_P/M_*$ & Mass ratio  & $0.00073\pm 0.00012$ & $0.00069\pm 0.00012$ & $0.00248^{+0.00023}_{-0.00024}$ & $0.00236^{+0.00022}_{-0.00023}$ \\
$d/R_*$ & Separation at mid-transit  & $7.51^{+0.49}_{-0.64}$ & $8.0^{+0.61}_{-0.47}$ & $4.49^{+0.76}_{-0.84}$ & $4.91^{+0.86}_{-0.83}$ \\
\enddata
\end{deluxetable*}

\subsection{Early Trends in the Hot Jupiter Population}

The purpose of this survey is to generate a large, complete sample in order to statistically assess the origins of HJs. As such, it is too early to make conclusive claims that require robust statistics. Instead, we argue that there are several tentative trends arising in the present sample of HJs.

\begin{figure*}[ht]
    \centering
    \includegraphics[width=0.9\linewidth, keepaspectratio]{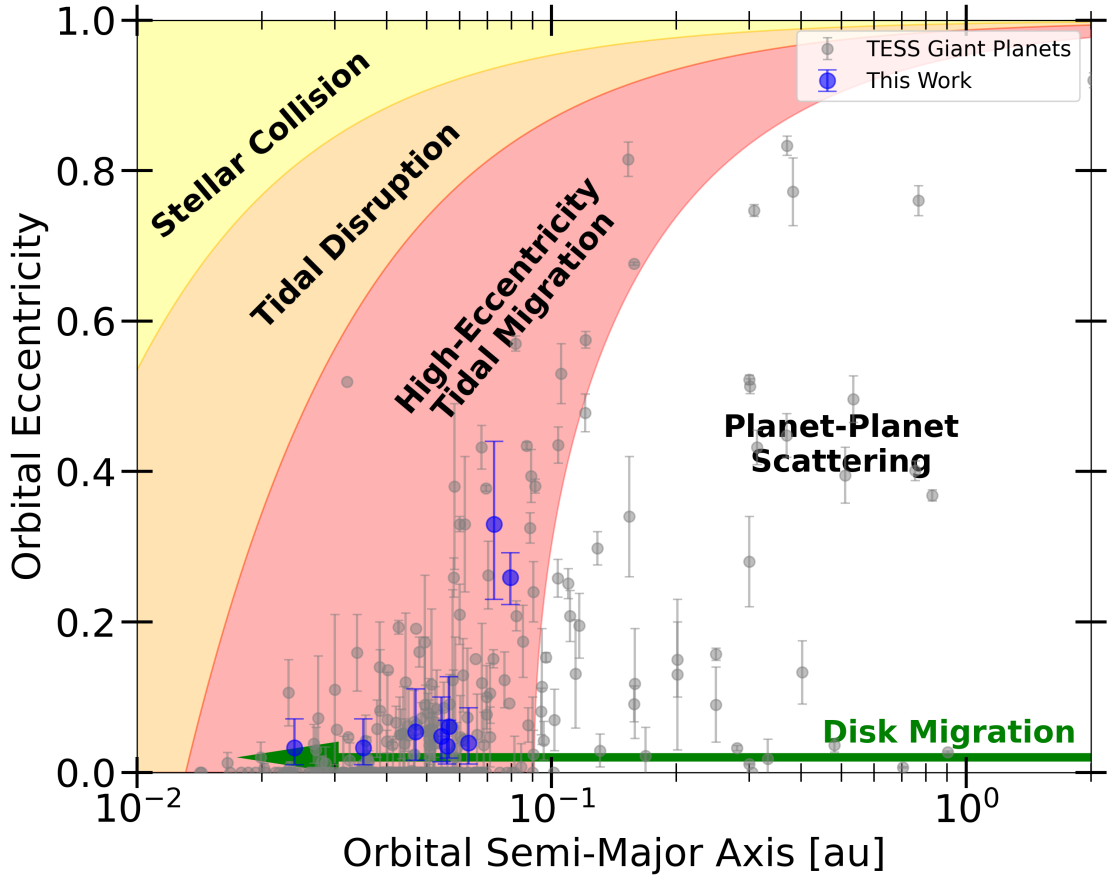}
    \caption{Eccentricity and semi-major axis distribution of the planets discovered in this work, compared to giant planets in the literature and several avenues and outcomes of giant planet migration (similar to Figure 4 from \citealt{Dawson:2018}). The region labeled "Stellar Collision" corresponds to a giant planet colliding with the star, assuming the star is 1 R$_\odot$ in size. The region labeled "Tidal Disruption" corresponds to a Jupiter-like planet falling within the Roche limit of the Sun. Finally, the red region corresponds to the tidal circularization of a highly eccentric giant planet around a Sun-like star. The blue circles represent the \exofast median semi-major axis and eccentricity of the planets in this work, while the gray circles represent substellar bodies with precise eccentricities and masses between 0.25 -- 13 \mj, obtained from the NASA Exoplanet Archive. Many of the eccentric planets, including those that fall outside of the three shaded regions, can be explained by planet-planet scattering that has not yet or will not be tidally circularized. The equations that describe each of these equations are shown in \cite{Dawson:2018}.}
    \label{fig:a_ecc}
\end{figure*}

First, it is notable that seven of our nine planets have orbits that are consistent with circular within $ 2.45 \sigma$ \citep{Lucy:1971} and are thus consistent with either migration mechanism: gas-disk migration or high-eccentricity tidal migration. Two of the planets in this sample, TOI-3919 b and TOI-5301 b, have significant eccentricities that cannot be explained by quiescent disk migration alone. TOI-3919's radial velocities show signs of a long-term trend that could be evidence of an outer companion and therefore a possible perturber to boost the eccentricity of TOI-3919 b and trigger high-eccentricity tidal migration. Both of these planets is shown in Figure \ref{fig:a_ecc} and compared to several different avenues and outcomes of planet migration. These two planets join the 39 other HJs with reported eccentricities $> 3\sigma$ above zero (11.2\% of the HJ population) per the NASA Exoplanet Archive\footnote{\url{https://exoplanetarchive.ipac.caltech.edu/index.html}} (date accessed: 2023 October 23). While the NASA Exoplanet Archive is a heterogeneous source of exoplanet parameters and should not be used to identify definitive trends, this is a significant minority of the population that must receive further scrutiny. Since many planets with periods less than ten days are expected to have very short tidal circularization timescales \citep{Adams:2006}, the fraction of HJs that once had boosted eccentricities could be much larger than 11.2\%. This implies, that at the very least, high-eccentricity tidal migration is an important migration mechanism that must be considered for a non-negligible fraction of the HJ population. \cite{Rodriguez:2023} and \cite{Zink:2023} both argue that the current eccentricity distribution of known HJs is consistent with high-eccentricity tidal migration being the dominant pathway for the evolution of HJs, an argument that can continue to be tested with a growing sample of HJs.

Additionally, two of the stars that we observed have long-term RV trends: TOI-2368 has a linear slope of $0.21 \pm 0.11$~m/s/day and TOI-3919 has a slope of $-0.79^{+0.31}_{-0.30}$~m/s/day. If further RV follow-up reveals a continuation of the trend consistent with a giant substellar companion on a wide orbit, then these systems would be further evidence that HJ-hosting systems are not always devoid of other planets, as RV surveys \citep[e.g.,][]{Knutson:2014, Bryan:2016, Zink:2023} have argued. \cite{Zink:2023} also argue that the properties of systems hosting a HJ and a distant giant companion favor a scenario in which the HJ migrated via coplanar high-eccentricity tidal migration.

Several issues with high-eccentricity tidal migration have been raised, however, that must be addressed. For example, \cite{Socrates:2012} argued that we should be able to detect a significant number of "super-eccentric" Jupiters that are actively migrating to become HJs, more than have yet been observed. Additionally, they argue that the tidal efficiency of HJs must be at least 10 times larger than Jupiter's. Further studies should address these issues and elucidate the importance of each mechanism of giant planet formation and evolution.

\subsection{Lithium in TOI-5301} \label{subsec:lithium}

While vetting the spectra of the stars in our sample, we discovered a discernible lithium absorption feature in the spectra of TOI-5301.
The strength of a star's lithium absorption feature serves as a valuable tool to determine stellar age \citep[e.g.,][]{1973stag.conf...38V}. 
This feature is also useful in identifying instances where a star may have ingested planets \citep{Montalban:2002}, particularly among stars that share similar mass and evolutionary stages as TOI-5301 \citep[e.g.,][]{2021AJ....162..273S}.

\textit{Is the lithium abundance of our target unusual when compared to a control sample of stars?}
To answer this question, we performed a lithium abundance analysis of the target by co-adding 19 TRES observations taken between 2022-09-01 and 2022-11-09. 
Before co-adding the spectra to build up the signal-to-noise ratio (SNR), each spectrum was continuum normalized, barycenter corrected, and Doppler corrected.
Following the procedures outlined in \cite{2023MNRAS.523..802J}, we measured the equivalent width (EW) of the lithium absorption doublet (Li I) at 6707.8\,\AA.
We measure an SNR=84 for the co-added spectra. 
Using \texttt{pymoog}\footnote{\url{https://github.com/MingjieJian/pymoog/}} and \texttt{pyMOOGi}\footnote{\url{https://github.com/madamow/pymoogi/}}, we measured a lithium abundance on the 12-point scale of $\mathrm{A(Li)}=2.47\pm 0.04$\,dex.

To determine if this measured abundance strength is unusual, we compared this value to the lithium abundances of an ensemble of 392 control sample stars with reported lithium abundance measurements in the GALAH DR3 database \citep{2021MNRAS.506..150B}.
We selected control stars that exhibited stellar properties akin to that of our target. 
This included stars with a surface gravity of $3.9 \pm 0.05$\,dex, an effective temperature of $6260 \pm 50$\,K, an iron abundance ([Fe/H]) of $-0.046 \pm 0.05$\,dex, and a Gaia renormalized unit weight error (RUWE) value $<1.4$, which indicates that the target is less likely to be part of a binary system \citep{2020MNRAS.496.1922B}.

The lithium abundances of our control population present a skewed lithium distribution.
Therefore, we computed a modified z-score, 
\begin{equation}
    Z=\frac{0.6745\times (X_{i}-\mathrm{Median}(X))}{\mathrm{MAD}},
\end{equation}
which incorporates the median and absolute deviation from the median (MAD) for the control sample lithium abundances, as opposed to the population mean and standard deviation. 
Our target's modified Z-score is measured at 0.003, suggesting that its lithium abundance is nearly identical to the median of our dataset. 
Therefore, we report no indication of abnormal lithium contamination in the spectra of TOI-5301.

\subsection{Building a Complete Sample of Hot Jupiters}

This paper is the first in a new series of papers aimed at constructing a homogeneously analyzed set of HJ parameters, built off of the works of \cite{Rodriguez:2019, Rodriguez:2021, Ikwut-Ukwa:2022, Yee:2022, Rodriguez:2023, Yee:2023}. Per \cite{Yee:2021}, roughly 300-400 transiting HJs around FGK stars with $G < 12.5$ mag should be discoverable with \TESS, a sample that is now $> \sim 60\%$ complete. Once this sample is complete, which is achievable within the next few years with the help of \TESS and TFOP, it will be possible to carefully consider the different possible migration mechanisms of HJs using detailed statistics with well-quantified biases. Models of other physical processes, such as the radius inflation of hot Jupiters subject to high instellation, will also be easier to test with this complete sample. 

Individual discoveries of HJ systems with confounding features are constantly being made that will weigh on the models that we use to describe the evolution of systems with HJs \citep[e.g.,][]{Becker:2015, Wittenmyer:2022, Yoshida:2023}. For example, since HJs that underwent high-eccentricity tidal migration are unlikely to have nearby companions due to the high likelihood of ejections and engulfment \citep{Mustill:2015}, the exceptions to this rule will be key systems to test models of gas-disk migration. Additionally, photometric \citep[e.g.,][]{De:2023} and chemical (\citealt{Montalban:2002}; see \S \ref{subsec:lithium}) signatures of planet engulfment may be found that provide insight into the stability of systems containing HJs. Throughout the construction of the sample, we will give greater consideration and care to benchmark systems in order to generate further constraints on models of migration. It is with the combination of a statistical sample of systems hosting HJs and the careful consideration of these benchmark systems that we hope to gain further insight into the origins and evolution of the enigmatic population of HJs.

\section{Summary} 
\label{sec:summary}

In this article, we present nine new hot Jupiters discovered by NASA's \TESS mission as part of an ongoing effort to discover and characterize all HJs orbiting FGK stars brighter than $G = 12.5$ mag. This article is the first in a series of articles titled the Migration and Evolution of giant ExoPlanets (MEEP) survey. This effort aims to use homogenous detection and analysis techniques to generate a population of HJs with well-quantified selection biases. In pursuit of this effort, we are obtaining precise eccentricities, along with a host of other important parameters (such as planetary and stellar mass and radius), to test theories of giant planet formation and evolution. 

The HJs we present in this work have masses ranging from $0.554^{+0.076}_{-0.075}$ \mj to $3.88 \pm 0.23$ \mj and radii ranging from $0.967^{+0.036}_{-0.031}$ \rj to $1.438^{+0.045}_{-0.042}$ \rj, excluding the one planet in our sample that exhibits a grazing transit and for which the radius cannot be precisely constrained. Two of the planets in our sample, TOI-3919 b and TOI-5301 b, exhibit significant orbital eccentricities that imply that they may have undergone high-eccentricity tidal migration. We investigated the lithium absorption feature of the spectra of one star, TOI-5301, and found no evidence of an anomalous lithium enrichment that could be associated with youth or planetary engulfment. Each of these planets, however, will be an important member of the rising population of HJs that, once complete, will help answer the longest-standing question of exoplanet science, \textit{how do hot Jupiters form and evolve?}

\makeatletter\onecolumngrid@push\makeatother
\clearpage
\begin{acknowledgments}

Many of the data used in this paper are available on the Mikulski Archive for Space Telescopes (MAST). This research has made use of the VizieR catalogue access tool, CDS, Strasbourg, France (DOI 10.26093/cds/vizier). The original description of the VizieR service was published in 2000, A\&AS 143, 23. This research has made use of the NASA Exoplanet Archive and the Exoplanet Follow-up Observation Program (ExoFOP; DOI: 10.26134/ExoFOP5) website, which is operated by the California Institute of Technology, under contract with the National Aeronautics and Space Administration under the Exoplanet Exploration Program. EP acknowledges funding from the Spanish Ministry of Economics and Competitiveness through project PGC2021-125627OB-C32. TRAPPIST-South is funded by the Belgian National Fund for Scientific Research (F.R.S.-FNRS) under grant PDR T.0120.21, with the participation of the Swiss National Science Fundation (SNF). M.G. and E.J. are FNRS Senior Research Associates. The postdoctoral fellowship of KB is funded by F.R.S.-FNRS grant T.0109.20 and by the Francqui Foundation. This work is partly supported by JSPS KAKENHI Grant Numbers JP17H04574, JP18H05439, JP20K14521, JP21K13955, and JST CREST Grant Number JPMJCR1761. This article includes observations made with the MuSCAT2 instrument, developed by ABC, at Telescopio Carlos S\'anchez operated on the island of Tenerife by the IAC in the Spanish Observatorio del Teide.
This paper contains data taken with the NEID instrument, which was funded by the NASA-NSF Exoplanet Observational Research (NN-EXPLORE) partnership and built by Pennsylvania State University. NEID is installed on the WIYN telescope, which is operated by the National Optical Astronomy Observatory, and the NEID archive is operated by the NASA Exoplanet Science Institute at the California Institute of Technology. NN-EXPLORE is managed by the Jet Propulsion Laboratory, California Institute of Technology under contract with the National Aeronautics and Space Administration.
Data presented herein were obtained at the WIYN Observatory from telescope time allocated to NN-EXPLORE through the scientific partnership of the National Aeronautics and Space Administration, the National Science Foundation, and NOIRLab.
This work was supported by a NASA WIYN PI Data Award (2022A-543544, PI: Yee), administered by the NASA Exoplanet Science Institute.
The authors are honored to be permitted to conduct astronomical research on Iolkam Du’ag (Kitt Peak), a mountain with particular significance to the Tohono O’odham.
The research of M.V. was supported by the Slovak Research and Development Agency, under contract No. APVV-20-0148.
Some of the observations in this paper made use of the High-Resolution Imaging instrument ‘Alopeke/Zorro and were obtained under Gemini LLP Proposal Number: GN/S-2021A-LP-105. ‘Alopeke/Zorro was funded by the NASA Exoplanet Exploration Program and built at the NASA Ames Research Center by Steve B. Howell, Nic Scott, Elliott P. Horch, and Emmett Quigley. Alopeke/Zorro was mounted on the Gemini North/South telescope of the international Gemini Observatory, a program of NSF’s OIR Lab, which is managed by the Association of Universities for Research in Astronomy (AURA) under a cooperative agreement with the National Science Foundation. on behalf of the Gemini partnership: the National Science Foundation (United States), National Research Council (Canada), Agencia Nacional de Investigaci\'on y Desarrollo (Chile), Ministerio de Ciencia, Tecnolog\'ia e Innovaci\'on (Argentina), Ministério da Ci\^encia, Tecnologia, Inovações e Comunicações (Brazil), and Korea Astronomy and Space Science Institute (Republic of Korea).
Based in part on observations obtained at the Southern Astrophysical Research (SOAR) telescope, which is a joint project of the Minist\'{e}rio da Ci\^{e}ncia, Tecnologia e Inova\c{c}\~{o}es (MCTI/LNA) do Brasil, the US National Science Foundation’s NOIRLab, the University of North Carolina at Chapel Hill (UNC), and Michigan State University (MSU).
This publication benefits from the support of the French Community of Belgium in the context of the FRIA Doctoral Grant awarded to MT.
We acknowledge financial support from the Agencia Estatal de Investigaci\'on of the Ministerio de Ciencia e Innovaci\'on MCIN/AEI/10.13039/501100011033 and the ERDF “A way of making Europe” through project PID2021-125627OB-C32, and from the Centre of Excellence “Severo Ochoa” award to the Instituto de Astrofisica de Canarias.
P.A.R. acknowledges support from the National Science Foundation under grant no. 1952545.
R.K. acknowledges the support by Inter-transfer grant no LTT-20015.
F.J.P acknowledges financial support from the grant CEX2021-001131-S
funded by MCIN/AEI/ 10.13039/501100011033 and through projects
PID2019-109522GB-C52 and PID2022-137241NB-C43.
We acknowledge the use of public \TESS data from pipelines at the \TESS Science Office and at the \TESS Science Processing Operations Center.
Resources supporting this work were provided by the NASA High-End Computing (HEC) Program through the NASA Advanced Supercomputing (NAS) Division at Ames Research Center for the production of the SPOC data products.
This paper made use of data collected by the TESS mission and are publicly available from the Mikulski Archive for Space Telescopes (MAST) operated by the Space Telescope Science Institute (STScI). Funding for the TESS mission is provided by NASA’s Science Mission Directorate.
HP acknowledges support by the Spanish Ministry of Science and Innovation with the Ramon y Cajal fellowship number RYC2021-031798-I.
CAC acknowledges that this research was carried out at the Jet Propulsion Laboratory, California Institute of Technology, under a contract with the National Aeronautics and Space Administration (80NM0018D0004).

\end{acknowledgments}
\makeatletter\onecolumngrid@pop\makeatother

\facilities{TESS, FLWO 1.5~m (TRES), SMARTS 1.5~m (CHIRON), SOAR 4.1~m (Goodman, HRCam), LCOGT, NAOJ-Okayama 1.88~m (MuSCAT2), TCS 1.52~m (MuSCAT), FLWO 1.2~m (KeplerCam), WIYN 3.5~m (NEID, NESSI), Shane 3~m (ShARCS), Palomar 5~m (PHARO), SAI 2.5~m, Gemini-North 8~m ('Alopeke), Gemini-South 8~m (Zorro), Keck 10~m (NIRC2), TRAPPIST-South 0.6~m}

\software{EXOFASTv2 \citep{Eastman:2019}, lightkurve \citep{LightkurveCollab:2018}, AstroImageJ \citep{Collins_2017}, TAPIR \citep{Jensen:2013}, astropy \citep{Astropy:2013, Astropy:2018}, numpy \citep{Harris:2020}, matplotlib \citep{Hunter:2007}, pandas \citep{pandas:2023}, scipy \citep{Virtanen:2020}}



\begin{figure*}[p]
    \centering
    \includegraphics[width=\textwidth,height=0.8\textheight,keepaspectratio]{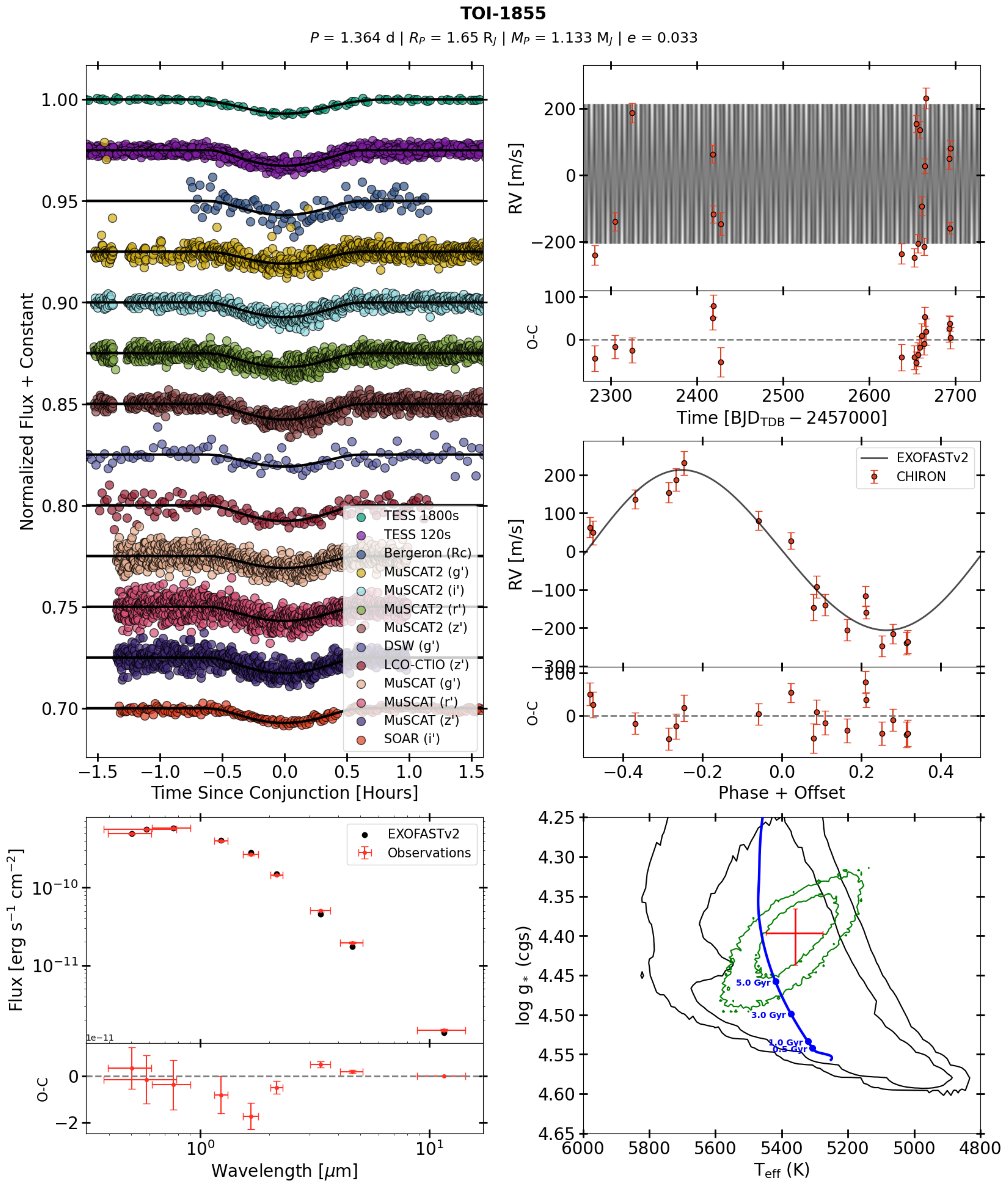}
    \caption{\TESS, Follow-up and archival observations of TOI-1855 as they compare to the \exofast results.
    \textbf{Upper left:} Unbinned \TESS and follow-up ground-based transits, phase-folded and shown in comparison to the best-fit time of conjunction with an arbitrary normalized flux offset. Multiple TESS sectors in the same cadence are stacked on top of each other.
    \textbf{Bottom left:} The spectral energy distribution of the target star compared to the best-fit \exofast model. Residuals are shown on a linear scale, using the same units as the primary y-axis.
    \textbf{Upper right:} RV observations versus time, including any significant long-term trend. The residuals are shown in the subpanel below in the same units.
    \textbf{Middle right:} RV observations phase-folded using the best-fit ephemeris from the \exofast global fit. The phase is shifted so that the transit occurs at Phase + Offset = 0. The residuals are shown in the subpanel below in the same units.
    \textbf{Bottom right:} The evolutionary track and current evolutionary stage of the planet according to the best-fit MESA Isochrones and Stellar Tracks (MIST) model. The blue line indicates the best-fit MIST track, while the black contours show the 1$\sigma$ and 2$\sigma$ constraints on the star's current \teff and log g from the MIST isochrone alone. The green contours represent the 1$\sigma$ and 2$\sigma$ constraints on the star's \teff and log g from the \exofast global fit, combining constraints from observations of the star and planet. The red cross indicates the median and 68\% confidence interval reported in Table \ref{tab:median}.}
    \label{fig:toi1855}
\end{figure*}

\begin{figure*}[p]
    \centering
    \includegraphics[width=\textwidth,height=\textheight,keepaspectratio]{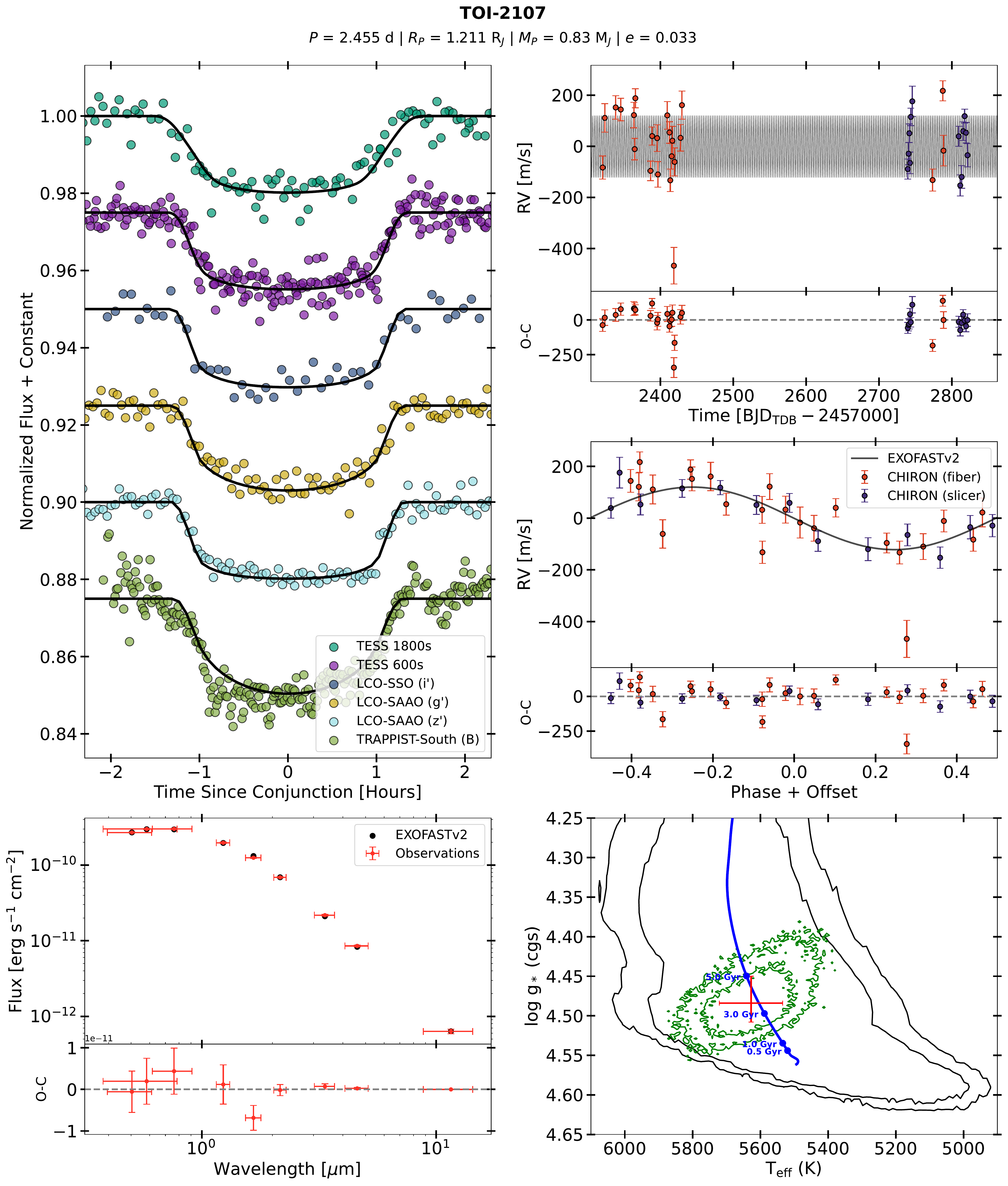}
    \caption{Same as Figure \ref{fig:toi1855}, but for TOI-2107.}
    \label{fig:toi2107}
\end{figure*}

\begin{figure*}[p]
    \centering
    \includegraphics[width=\textwidth,height=\textheight,keepaspectratio]{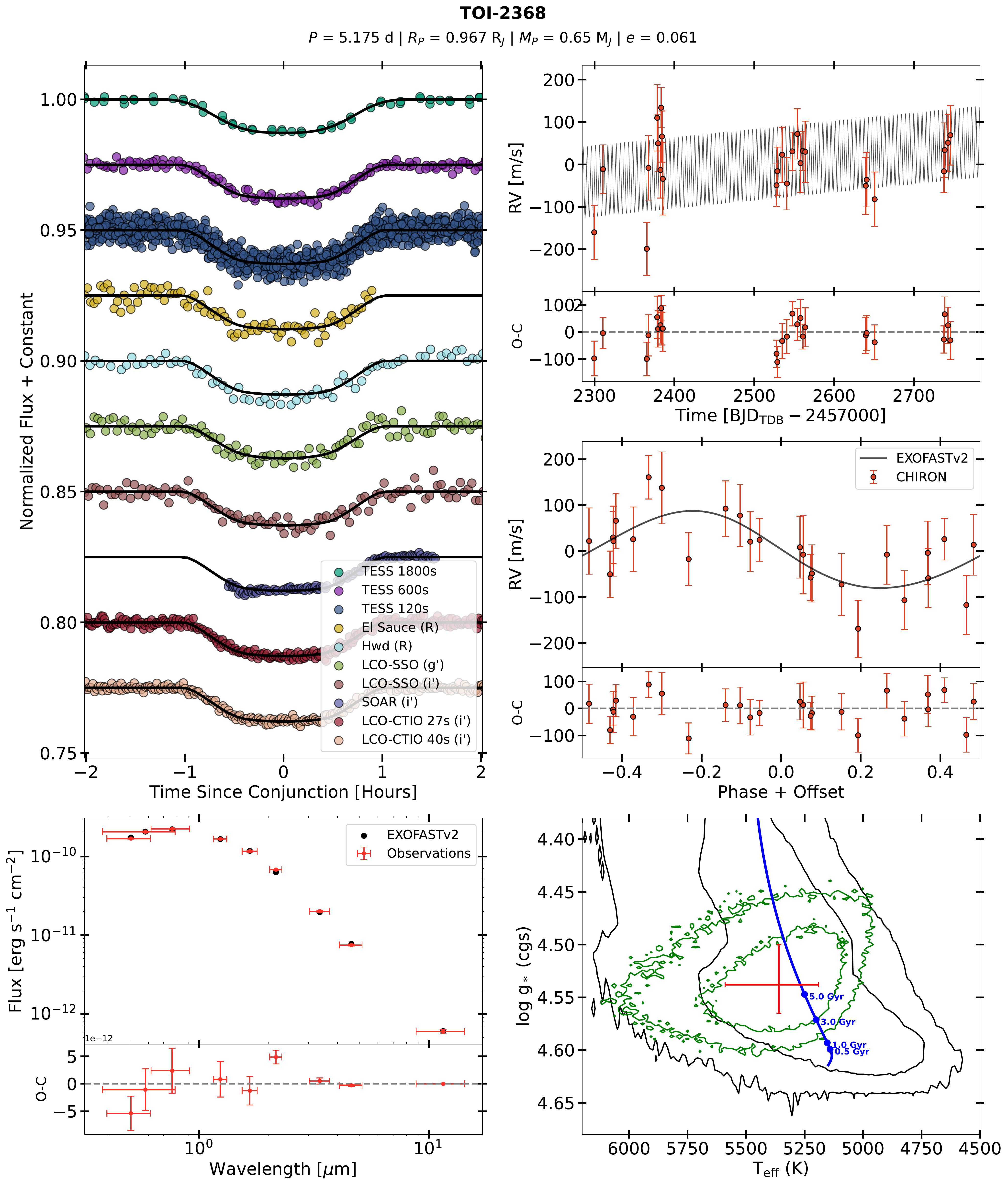}
    \caption{Same as Figure \ref{fig:toi1855}, but for TOI-2368. Notably, the best-fit RVs for TOI-2368 imply a linear RV trend of $0.21 \pm 0.11$ m/s/day, which was included in our fit per our prescription to fit for an RV trend if it is favored to greater than $1\sigma$.}
    \label{fig:toi2368}
\end{figure*}

\begin{figure*}[p]
    \centering
    \includegraphics[width=\textwidth,height=\textheight,keepaspectratio]{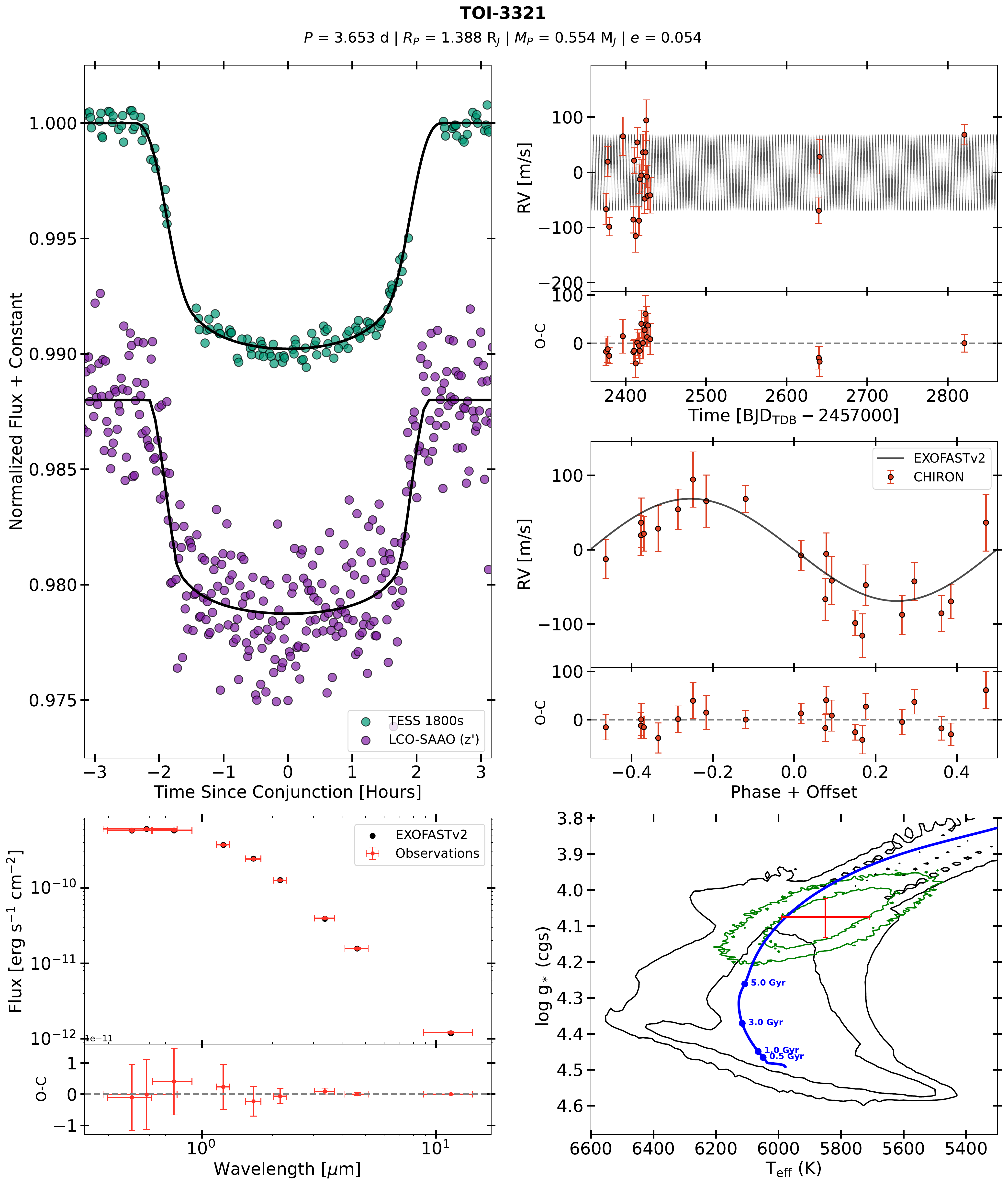}
    \caption{Same as Figure \ref{fig:toi1855}, but for TOI-3321.}
    \label{fig:toi3321}
\end{figure*}

\begin{figure*}[p]
    \centering
    \includegraphics[width=\textwidth,height=\textheight,keepaspectratio]{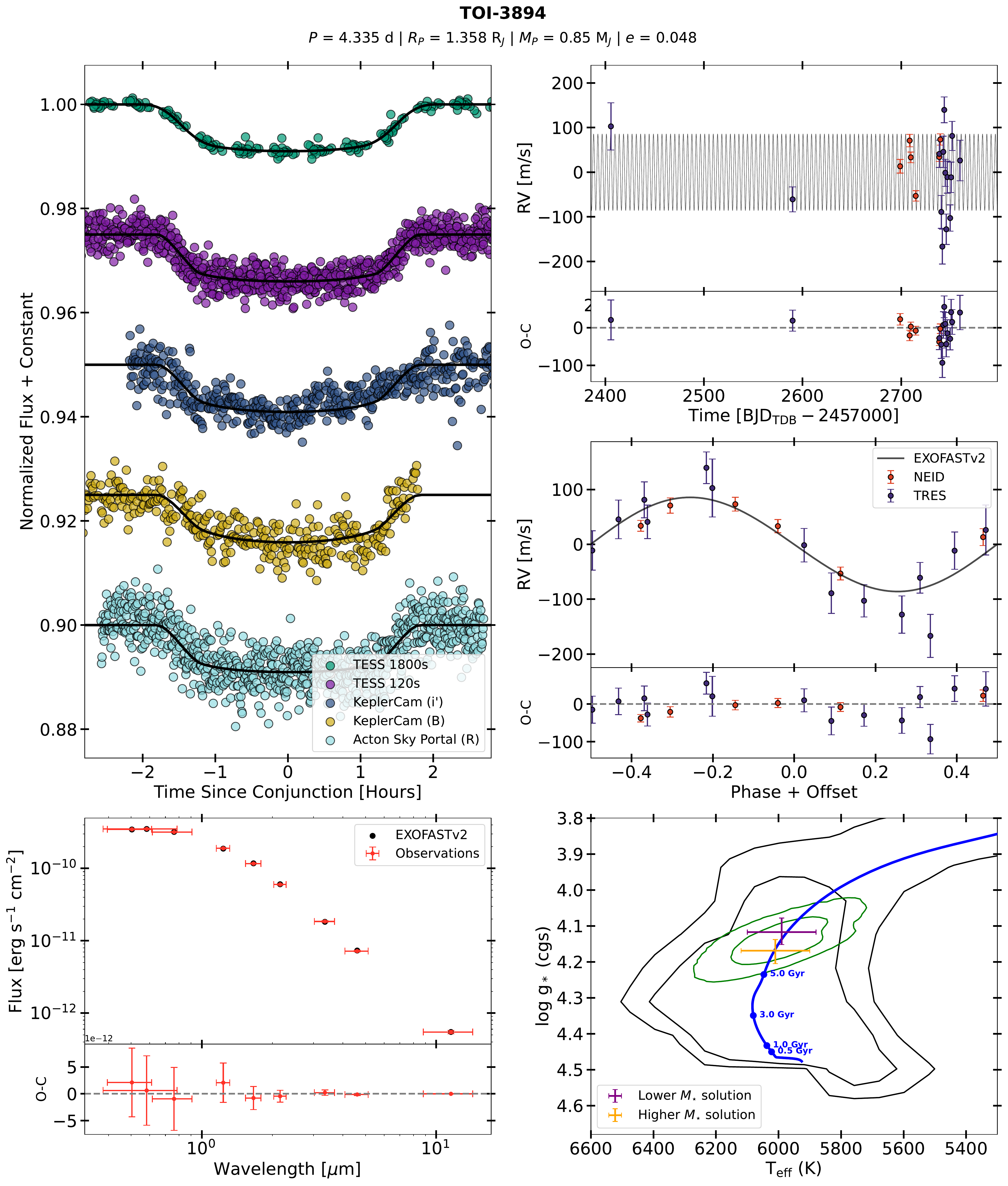}
    \caption{Same as Figure \ref{fig:toi1855}, but for TOI-3894. TOI-3894's \exofast global fit was bimodal in stellar mass and age. To represent both modes, we show median values and standard deviations from each stellar mass solution in the bottom right plot.}
    \label{fig:toi3894}
\end{figure*}

\begin{figure*}[p]
    \centering
    \includegraphics[width=\textwidth,height=\textheight,keepaspectratio]{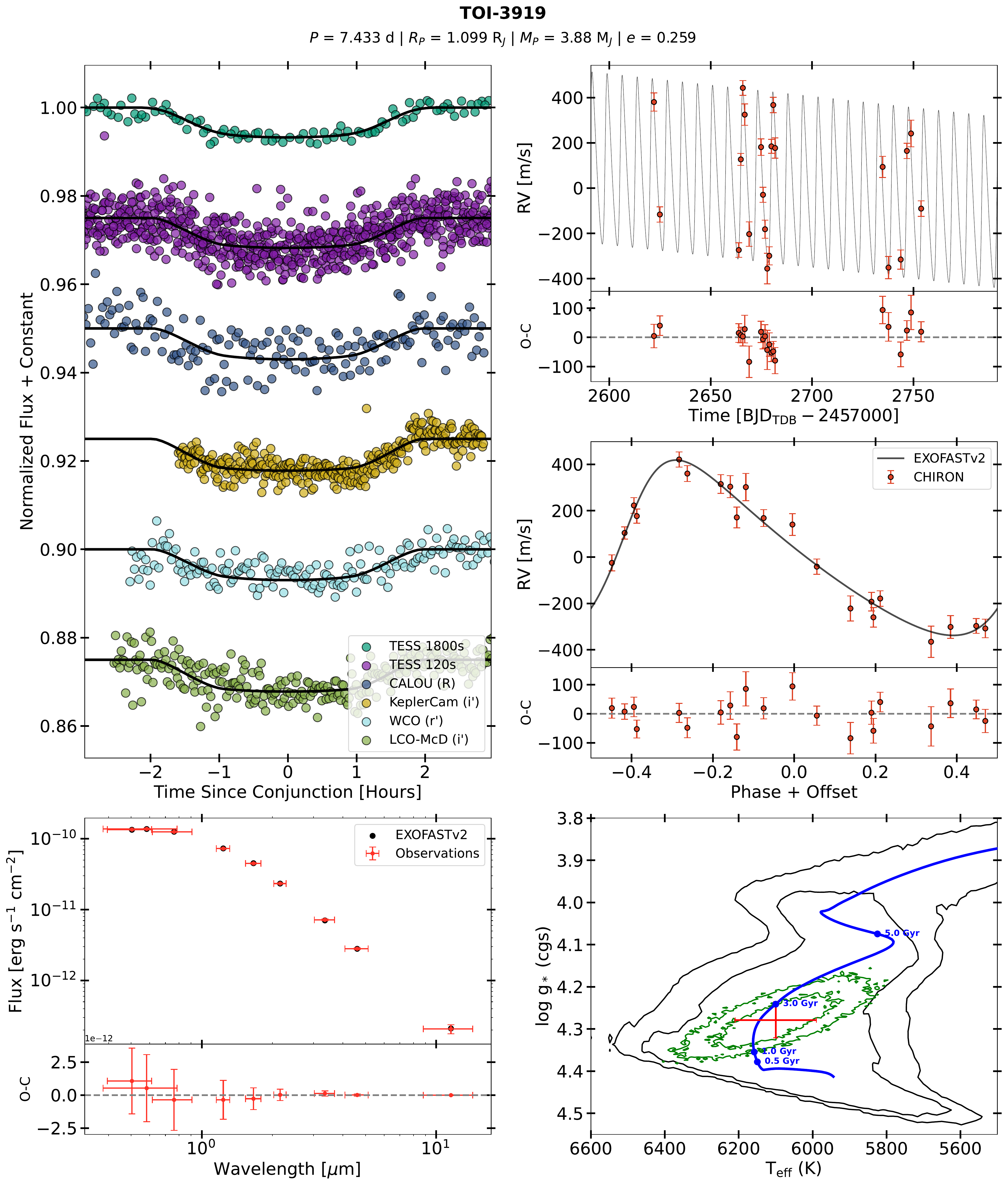}
    \caption{Same as Figure \ref{fig:toi1855}, but for TOI-3919. Notably, the best-fit RVs for TOI-3919 imply a linear RV trend of $-0.79^{+0.31}_{-0.30}$ m/s/day, which was included in our fit per our prescription to fit for an RV trend if it is favored to greater than $1\sigma$.}
    \label{fig:toi3919}
\end{figure*}

\begin{figure*}[p]
    \centering
    \includegraphics[width=\textwidth,height=\textheight,keepaspectratio]{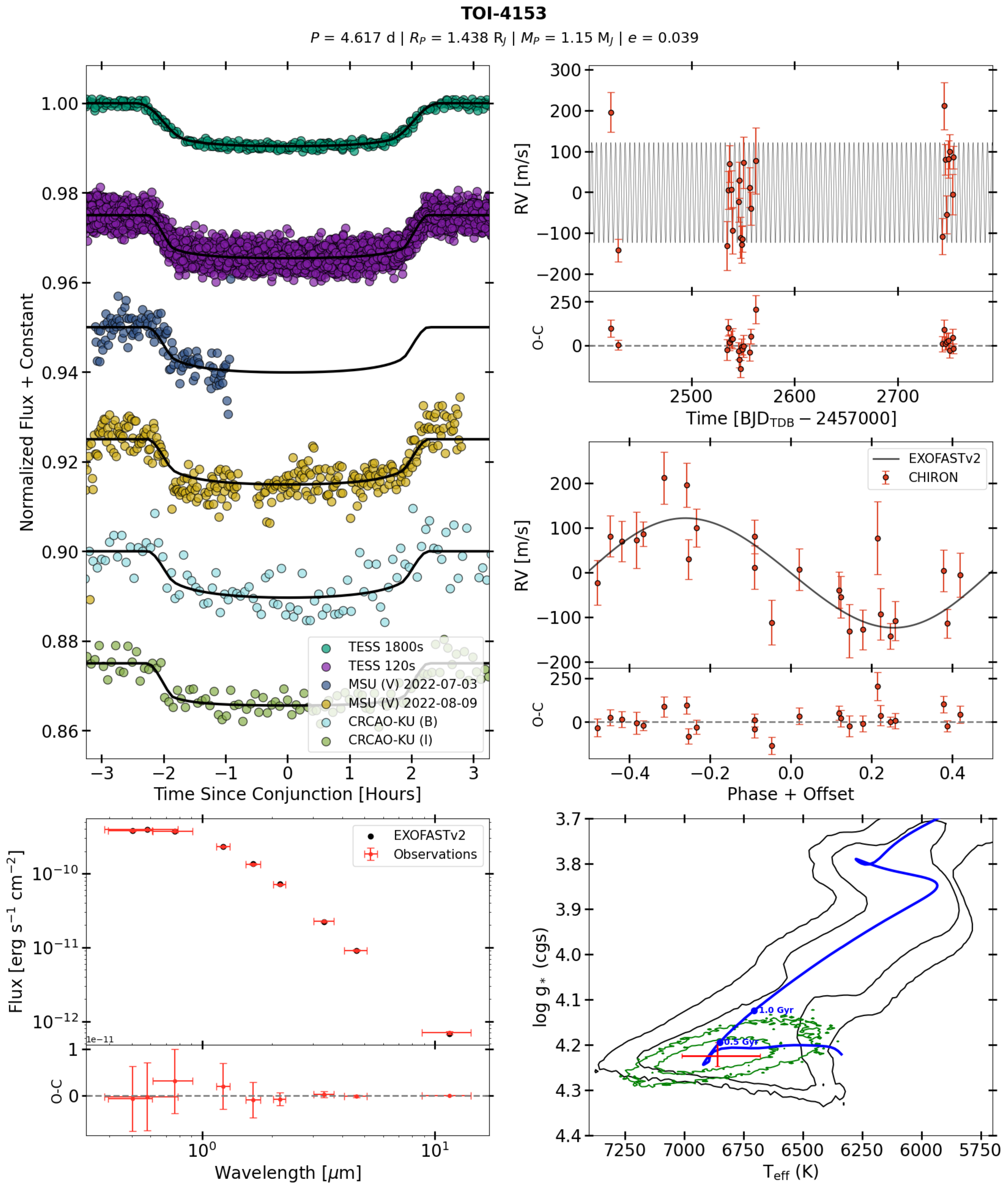}
    \caption{Same as Figure \ref{fig:toi1855}, but for TOI-4153.}
    \label{fig:toi4153}
\end{figure*}

\begin{figure*}[p]
    \centering
    \includegraphics[width=\textwidth,height=\textheight,keepaspectratio]{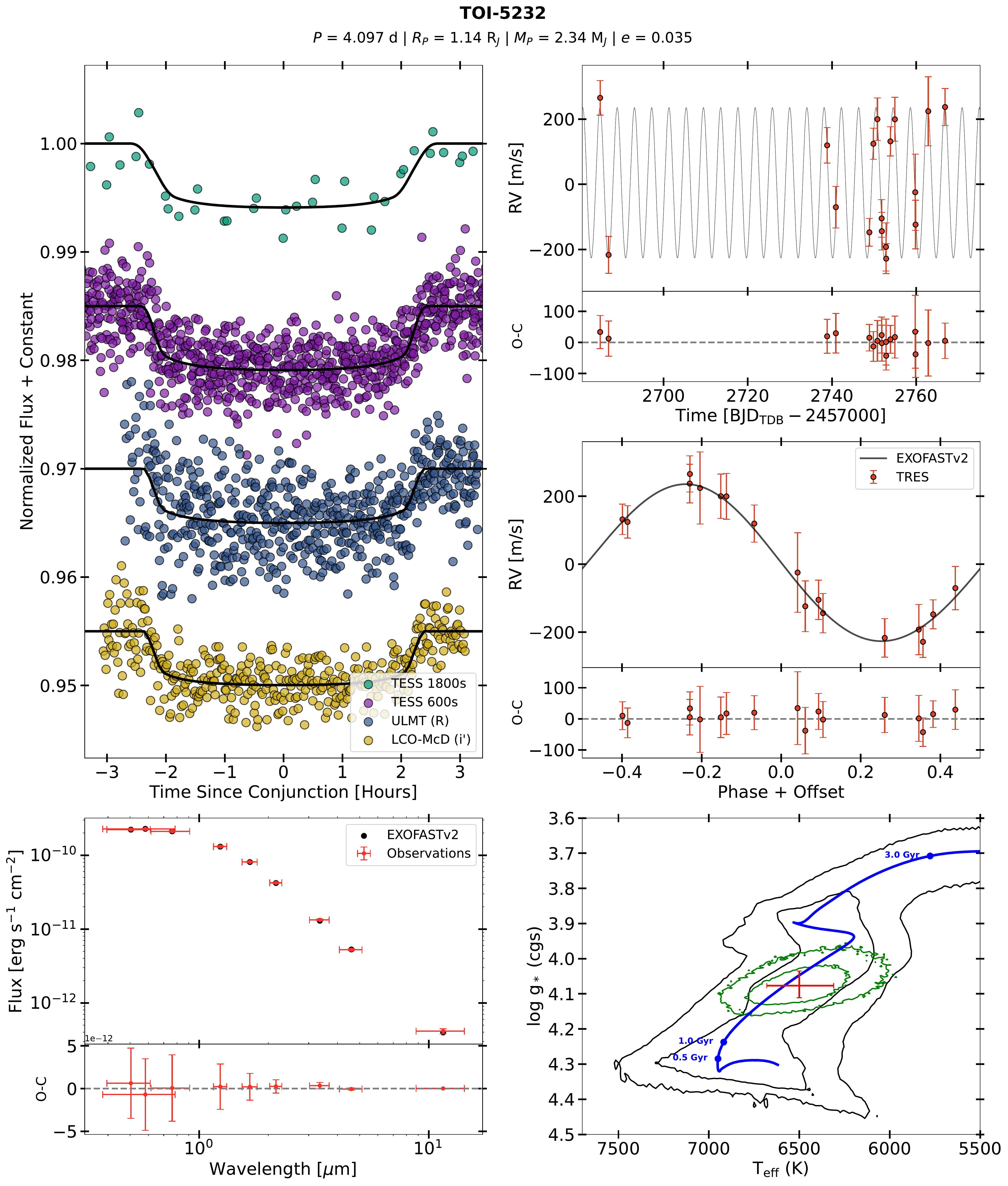}
    \caption{Same as Figure \ref{fig:toi1855}, but for TOI-5232.}
    \label{fig:toi5232}
\end{figure*}

\begin{figure*}[p]
    \centering
    \includegraphics[width=\textwidth,height=\textheight,keepaspectratio]{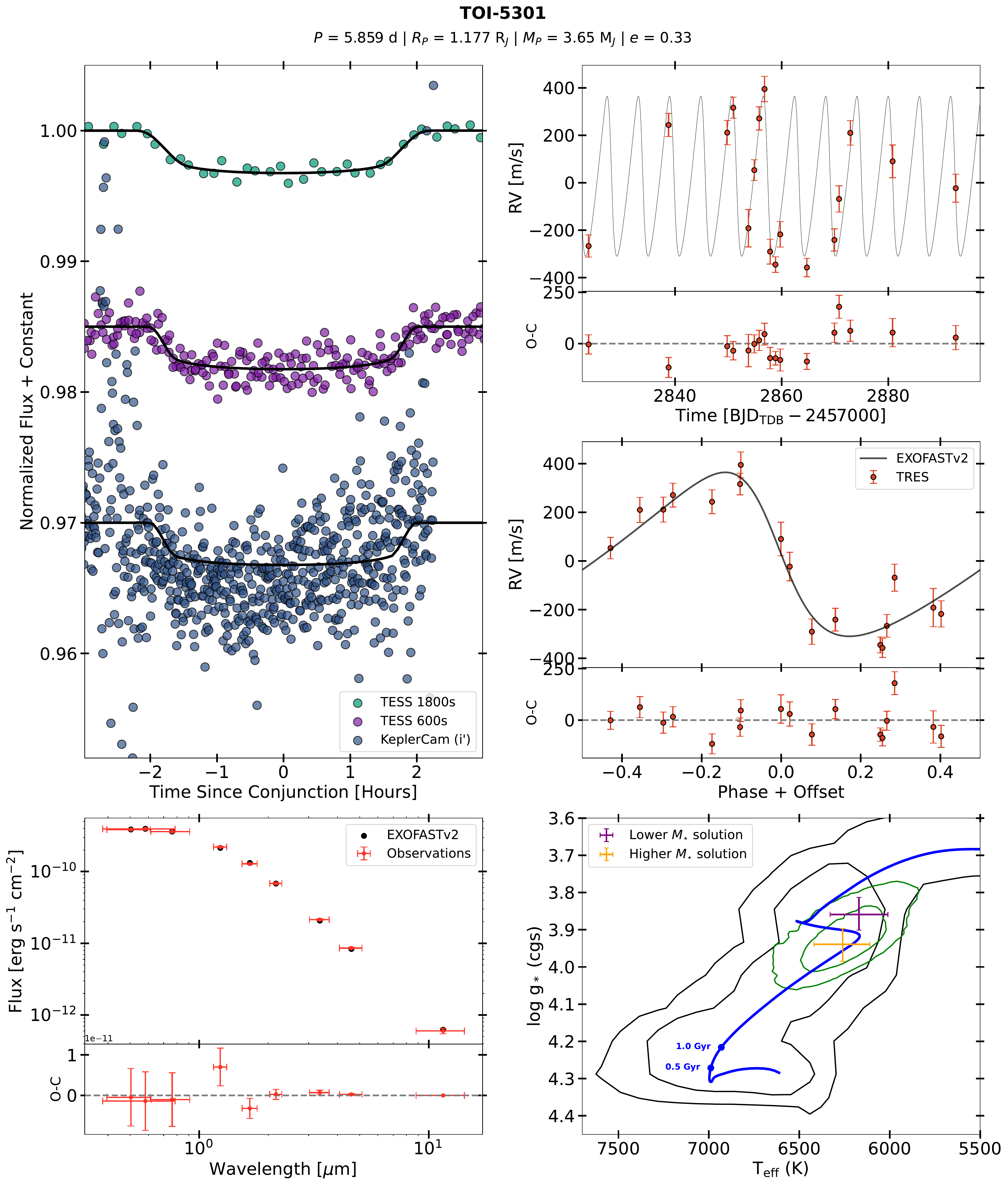}
    \caption{Same as Figure \ref{fig:toi1855}, but for TOI-5301. Similarly to Figure \ref{fig:toi3894}, TOI-5301 had a bimodal distribution of possible stellar masses, and we therefore represent the median and standard deviation of either solution in the bottom right plot.}
    \label{fig:toi5301}
\end{figure*}


\bibliography{MAIN.bib}
\bibliographystyle{apj}

\end{document}